# NONRELATIVISTIC THEORY OF ELECTROMAGNETIC FORCES ON PARTICLES AND NANOPROBES MOVING NEAR A SURFACE

G.V. Dedkov[1] and A.A.Kyasov

Kabardino-Balkarian State University, 360004, Nalchik, Russia

**Abstract**

Closed nonrelativistic (nonretarded) theory of conservative and dissipative electromagnetic forces and heat exchange between moving particles (nanoprobes) and a surface (flat and cylindrical) is reviewed. The formalism is based on methods of classical and fluctuating electrodynamics using minimum assumptions. The spatial dispersion effects are introduced via the surface response functions. The theory allows to treat various problems related with dynamic interactions of charged particles, dipole molecules, neutral atoms and nanoprobes in a unified manner. For the first time, a brief review of the recently obtained consistent relativistic results is also given. The corresponding formulae exactly reduce to the nonrelativistic ones in the limit $c \rightarrow \infty$. Applications to experiments with the scanning probe microscopes, quartz crystal microbalance technique and transmission of particle beams in the near field of surfaces (through nanochannels) are discussed.

**Contents**

1. Introduction
2. Physical processes and concepts related with interactions of mobile particles and surfaces
3. Problem statement and general equations
   3.1 Problem statement
   3.2 Some general equations
   3.3 Specular reflection model and surface response dielectric functions
4. Interactions of moving atomic and molecular particles with flat surface
   4.1 Charged particle
   4.2 Dipole molecule
   4.3 Neutral spherical particle (ground state atom)
   4.4 Heating effects
   4.5 A brief review of relativistic theory
   4.6 Numerical comparison of different approximations
   4.7 Nonlinear velocity - resonance effects
   4.8 Spatial dispersion effects
   4.9 Sructural effects

---

[1] Corresponding author e-mail: gv_dedkov@mail.ru

**1.Introduction**

How do loose energy nanoscale objects in relative motion ? What are the involved dependencies on the gap width, velocity, temperature and material parameters ? These are the key questions of the macro-/nanotribology and many other important topics, such as adsorption/desorption processes on surfaces and damping motion of adsorbates, physical properties of thin films, the Brownian motion and energy loss of slow and fast charged and neutral atomic particles moving in close vicinity to a surface, the friction of nominally flat surfaces, heating effects in nanostructures via evanescent fields, etc.

As the bodies in relative motion represent an example of the system with confined geometry of space, we find here variety of interesting phenomena like dynamic screening of ions, wake potentials and accompanied effects in the vicinity of a surface [1-3], fluctuation-induced forces [4-10] and the Casimir effect [11-14], cavity induced effects [15], quantum and thermal electromagnetic fluctuations [16, 17]. Two basic ingredients of these phenomena are: i) external bodies which response to the electromagnetic fields or suppress/modify the quantum and thermal fluctuations and ii) the electromagnetic vacuum, which looks like a medium acting upon the limiting surfaces and particles.

Nowadays, great practical significance of these issues is stimulated by needs of the scanning probe microscopy [18-20], the quartz-crystal microbalance [21] and the surface force apparatus technique [22], frictional drag experiments between (2D) electron systems [23], possibilities of transmission and manipulation of thin particle beams in nanochannels [24], energy modulation of particles in evanescent fields [25, 26], etc.

Due to a great complexity of the corresponding problems we are going to review only three main aspect of the particle-surface interactions: i) conservative dynamical forces ; ii) dissipative forces ; iii) heating effects. We aim at developing general nonrelativistic formalism based on electromagnetic and fluctuation electromagnetic theory, making it possible to determine necessary quantities for different kind of particles like bare charges, dipole and quadrupole molecules, ground-state atoms and nanoprobes. Moreover, using an additional assumption, the theory allows to treat conservative and dissipative interactions between nano- (mesoscopic) objects in relative motion. Of this kind is the problem of friction and heat transfer between two smooth flat surfaces with a narrow gap of 1-100 $nm$ width. A brief discussion of the recently developed relativistic theory is also given, and we show that the corresponding formulae exactly reduce to the nonrelativistic ones in the limit $c \rightarrow \infty$, $c$ being the speed of light. This is an important theoretical achievement, because until now a lot of authors failed to get such an agreement between nonrelativistic and relativistic results.

Generally speaking, in the discussed problem the relativistic and retardation effects become negligibly small at $V/c << 1$ and $\omega_0 z_0 /c << 1$, where $V$ is the particle velocity, $\omega_0$ is the characteristic frequency of the absorption spectra, $z_0$ is the particle distance from the surface. For normal metals $\omega_0 \approx 2\pi\sigma \approx 10^{16} \sec^{-1}$ ($\sigma$ is the conductance), the second condition is fulfilled at $z_0 \leq 30$ $nm$. For poor conductors and dielectrics, validity of nonrelativistic (and nonretarded) approximation is broader.

Anyhow, just this range of separations is of major significance in processes related with evanescent fields.

It should be pointed out that theoretical discussion of the problems mostly involved with fluctuation dissipative forces (FDF), has led to many controversial issues in the past [28-36] and present day [37-59] publications of different authors, which have used various theoretical approaches. Among them are the quantum electrodynamic field theory propagator methods (virtual photon exchange) [28, 29, 34, 37], a zero-point energy approach [7, 30] and self-energy formalism [2, 32, 33], the quantum perturbation theory and Hamiltonian approach [3, 31, 38-41, 43, 45], the density-functional methods [2, 40], the general theory of the fluctuating fields and a generalized Kirchhoff's law [44, 45, 47, 53 - 55, 59]. Our method [48, 49, 51, 52, 56 - 58] is based on direct statistical averaging of electrodynamic expressions for multipole forces, and energy dissipation integral of fluctuating field using the fluctuation –dissipation relations.

As a matter of fact, the canonical result for the molecule –metal system ( the dependence of the friction parameter, $\eta \sim z_0^{-10}$ at $T = 0$ ) obtained by Schaich and Harris (1981) has been recognized by many authors from the very beginning. However, the next important generalization of the theory at $T \neq 0$ has been done by Tomassone and Widom only in 1997, while the most general expressions for the fluctuation dissipative forces have been obtained quite recently [55, 56, 58]. Obviously, the necessity to compare differing theoretical results and experimental data available is mandatory.

We guess that most of yet existing ambiguities in these problems were caused by misunderstanding of the ground relations between the involved physical quantities, such as tangential force on a moving particle, the rate of energy dissipation of fluctuating electromagnetic field, the heat flow, a role of spontaneous and induced components of the electric fields and currents, etc. These important points and other ones, due to the spatial dispersion effects, material properties, temperature and nonlinear velocity effects were scarcely touched upon in the current literature, and their reviewing must be interesting to the readers.

The organization of this paper is as follows. In Section 2 we discuss basic physical processes and concepts related with interactions of mobile particles and surfaces. Section 3 is devoted to a problem statement, some reference results and dielectric formalism which brings necessary links between the electromagnetic forces and fluctuation electromagnetic ones. The spatial dispersion effects are introduced in the theory in a simple manner via the specular reflection model. Section 4 represents the most important part of the paper, where some general expressions for normal and tangential forces for different kind particles and heat flow are given. In Section 5 the same formalism is applied to interactions with cylindrical surfaces, where the particles move parallel to generatrix of concave (convex) cylindrical surface. Section 6 contains the results relevant to the viscous –type friction forces (without of wear and atomic exchange) in sliding contacts nanotip –flat surface and between two flat surfaces. Section 7 is devoted to several experiments where the dynamical fluctuation forces could be observed. The main conclusions are summarized in Section 8 and, finally, Appendixes A-F (Section 9) present important mathematical details.

The Gauss units system is used throughout the paper except for the Sections 3.3, 4.8 devoted to the spatial dispersion effects, where the atomic units $e = \hbar = m = 1$ are used, too. Also, when discussing energy –loss effects for charged particles and dipole molecules, we denote the ion charge by $Ze$, irrespectively of the units system used. If not indicated, the integrations over components of the planar wave-vectors $k_x, k_y$ $(q_x, q_y)$ and frequency $\omega$ are performed in the interval $(-\infty, \infty)$, except for the final formulae like (4.19), (4.20), (4.27), (4.28), (4.32), (4.34) – (4.36) , where the interval $(0, \infty)$ is assumed. The primed and double primed functions $\alpha(\omega)$, $\Delta(\omega)$, $\tilde{\Delta}(\omega)$, etc. denote the corresponding real and imaginary parts. In several formulae presented in Section 5, where we summarize the Bessel functions in the range $0 \leq n \leq \infty$, the term $n = 0$ is taken with the numerical factor $1/2$ (see, for instance, Eqs.(5.3), (5.4), (5.6) – (5.8) , (5.10) - (5.12)). In general case, the particle temperature is assumed to be $T_1$ ($T_1 = 0$ for a bare charge and an atom), the surface temperature is $T_2$.

## 2. Physical processes and concepts related to the interactions of mobile particles and surfaces

A particle approaching a polarizable body may engender collective excitations which act back upon the particle. Both virtual and real excitations occur, and the resulting interaction thereby has both conservative and dissipative components [2, 32, 33, 43, 60]. It is possible to treat these excitations using the general self –energy expression for a particle outside the surface [2, 32, 33]. In the case of neutral atom this expression gives the van der Waals (vdW) energy as its real part and the inelastic scattering rate as its imaginary part. In the case of ion (dipole molecule) the real part of the proper expression gives the image potential. Excitations may include surface plasmons [61, 62], surface optical phonons [63], polaritons [64, 65], surface excitons, helicons, ripplons [66, 67], electron –hole pairs [19, 31, 38], etc. Relative importance of the electron –hole pair excitation as opposed to phonon creation has been the subject of considerable interest with respect to the frictional damping of adsorbates [19, 38, 68 - 71]. There are strong indications that during sliding a crucial role might play a few layers of lubrication molecules which could drastically change the interface properties. Thus, in the case of insulators, the microscopic sliding friction necessarily involves excitation of phonons, while in metals the excitation of phonons and low –energy electron –hole pairs is feasible [18 - 22, 40].

Consider, for example, the case of a metal substrate and a moving particle above it. For a moving bare charge (Fig.1a), it produces the moving "image charge" on the metal surface, while in the case of moving dipole (Fig.1b) – the moving image dipole. This motion of the induced charges is accompanied by Ohm's law heating within the metal. Therefore, the friction is due to the conduction electrons, and the heat generated inside the metal is the source of this friction force on the particles outside the metal surface [39, 43].

However, strictly speaking, yet in the case of moving dipole molecule one has to account for an additional channel of energy loss caused by the dipole moment reorientation [72]. Due to this fact, an intuitively clear relation between rate of work of friction force and rate of work performed by electric field in the volume of the particle (rate of energy absorption) needs to be correctly substantiated. To greater extent this concerns a neutral particle slowing down [55- 57, 73]. Another example when the rate of work of electromagnetic field does not exactly determine the energy absorption, is related with longitudinal plasmon wave –field at a vacuum –metal boundary [74].

As we strictly show in what follows, the rate of work of the fluctuating field over a moving neutral particle (in the laboratory reference frame) is spent on change of the kinetic energy of its center of mass (an origin of friction) and on excitation (deexcitation) of the inner degrees of freedom. The corresponding excitation (heating) of a neutral ground state atom can be associated with some kind of Lamb –shift due to the atom –surface coupling [56, 73]. In the opposite case (deexcitation) the heat is transferred to the surface. The balance can be expressed in terms of energy conservation ($W$ is the field energy, $Q$ is the heat released, and $F_x$ is the force of friction):

$$-\frac{dW}{dt} = \frac{dQ}{dt} + F_x V \tag{2.1}$$

It is necessary to note that all the quantities in Eq.(2.1) are taken in the laboratory frame (the substrate surface is in rest). Alternatively, in the particle rest frame (the nonrelativistic statement implies $t = t', Q = Q'$) , one has to write

$$-\frac{dW'}{dt} = \frac{dQ}{dt} \tag{2.2}$$

Despite a very clear meaning, Eqs.(2.1),(2.2) have been brought into practice only quite recently [55, 56, 73] ( see also [35]), where it has been considered friction between semi –infinite media.

It is possible to give a little bit different explanation of origin of the friction force (energy loss) on moving particles [26, 31, 40, 43]. Fig.2 shows the case of a moving charge. If the charge moves with a constant velocity along classical trajectory $\mathbf{R}(t)$, the screening charge will not follow adiabatically but will lag behind due to a finite response time of the electron gas. If $x$ is the direction of motion, the friction force is given by

$$-F_x = (Ze) E_x \tag{2.3}$$

where $E_x$ is the reaction electric field at the particle location point . Alternatively, Eq.(2.3) determines the involved stopping power of the particle (energy loss per unit path length traveled) via the identity $dE / dx = F_x$ , $E$ being the particle energy [26].

The case of fluctuating dipole is shown in Fig.3, illustrating a thermal or a quantum fluctuation, which gives rise to a temporal charge imbalance and an electric field .The electric field penetrates into the solid where it creates excitations [43]. The same situation would be expected for two surfaces in relative motion, or a small micro-(nanoparticle) above the surface. In general case, certainly, there are generated the dipole and higher order multipole fluctuating moments.

At thermal equilibrium for two stationary bodies, being situated in close vicinity to one another, there is no net energy transfer between them, but during sliding a net energy and momentum transfer will occur from the moving body to the resting one, leading to the friction force and surface heating [43]. When the bodies have different temperatures, the situation can be much more complex [35, 56, 73]. In particular, the force acting on a moving body may either slow it down or accelerate it. Moreover, the direction of heat flow can be "anomalous" : from a "cold" body to a "hot" one. This can be compared with the situation realized in a refrigerator. The additional work is produced by fluctuating electromagnetic field (see Eq.(2.1)).

It is necessary to draw attention to a fundamental difference between the quantum and thermal fluctuations [43]. So, the quantum fluctuations contribute to the linear in velocity $V$ sliding friction only in the second and higher orders of the perturbation theory [31, 38 - 40, 43], while thermal fluctuations –already in the lowest order (in theelectric field) [39, 55, 56]. This explains large difference in the distance dependence of the friction force : for a small body above a flat surface the thermal-induced drag fluctuation force is scaled to be $z_0^{-5}$ [39, 55, 56], while the quantum-induced one - as $z_0^{-10}$ [31, 38 - 40, 68]. Referring to the conservative fluctuation –induced forces [10, 12, 75], we do not observe such a difference (see below).

The role of quantum effects in stopping (friction) of low-velocity charges (dipole particles) is not so crucial, despite that the field created by charged particle represents a very strong perturbation [2]. The corresponding screening effects of the surrounding electron gas can be taken into account if we go beyond the lowest –order perturbation theory, and use is made of the nonlinear density functional formulation [40, 76]. However, yet the linear –response approximation allows to reproduce correct distance dependence of the friction force and its order of magnitude. It was realized that for charged and dipole particles the linear –response dielectric theory is very good at high velocities, $ZV_B/V << 1$ , and satisfactory at $V << V_F$ (see [2] and references therein), where $V_B$ and $V_F$ are the Bohr and Fermi velocities, respectively. On the contrary, a coupling between a neutral particle and a surface is weak from the very start , and therefore, the second- (high-order) perturbation theory is principally needed in evaluation of the friction force at $T = 0$ [31, 38 - 40, 43, 68].

The improved calculations of the friction forces within the time –dependent local-density approximation reveals an important role of exchange –correlation terms in the local potential, which make the surface electronic density more polarizable. This leads to larger numerical values of the friction coefficients for both ions and atoms [40].

The order of magnitude of attractive and drag friction forces for different particles above conducting surface can be determined from a simple analysis of physical dimensions. The needed quantities are: the height of the particle above the surface $z_0 \sim (cm)$; the charge $Ze \sim (gm^{0.5}cm^{1.5}\sec^{-1})$; the dipole moment $d \sim (gm^{0.5}cm^{2.5}\sec^{-1})$; the conductivity of substrate $\sigma \sim (\sec^{-1})$; the particle velocity $V \sim (cm/\sec)$; the (spherical) particle radius $R \sim (cm)$; the attractive and friction forces $F_z, F_x \sim (gm\,cm\,\sec^{-2})$. The resulting dependencies are:

i)      charge-surface interaction: $F_z \sim \dfrac{(Ze)^2}{z_0^2}$, $F_x \sim \dfrac{(Ze)^2}{z_0^3}\dfrac{V}{\sigma}$;

ii)      dipole molecule – surface interaction: $F_z \sim \dfrac{d^2}{z_0^4}$, $F_x \sim \dfrac{d^2}{z_0^5}\dfrac{V}{\sigma}$ ;

iii)      neutral particle – surface interaction (thermal–induced fluctuations):

$$F_z \sim \frac{(k_B T)R^3}{z_0^4}, \quad F_x \sim \frac{(k_B T)R^3}{z_0^5}\frac{V}{\sigma}$$

where $k_B$ is the Boltzmann constant and $T$ is the particle temperature. Obviously, the dependence $F_x \sim 1/\sigma$ is a simple consequence of the Ohm's law. Moreover, one sees that formulae iii) follow from ii) after the replacement $d^2 \to (k_B T)R^3$. For normal component of the quantum–induced fluctuation force $F_z$, one has to replace $k_B T$ by $\hbar \omega_0$, but the corresponding tangential component $F_x$ can not be obtained in such a simple manner.

From the above expressions it follows that $F_x / F_z = V / z_0\sigma$ for each type of the particle and, evidently for many practical cases, the relation $F_x / F_z \geq 1$ proves to be realistic one, especially for more resistive materials. As far as neutral particles is concerned, this is in contrast with the ordinary opinion [31] that FDF are always much smaller than vdW forces. This conclusion seems to be valid only at $T = 0$.

Also, from analysis of dimensions, one can construct an expression for the heat flow between a hot particle and a cold surface caused by evanescent field. Evidently, such a flow should exist even at $V = 0$. In this case, correspondingly,

$$\dot{Q} \sim k_B T \frac{R^3}{z_0^3} \sigma$$

This expression shows that heat exchange through evanescent modes of fluctuating electromagnetic field, where a small particle is placed in close vicinity to a surface, may significantly differ from that one due to black body radiation, being independent of distance and having another temperature dependence ($\dot{Q} \sim T^4$).

Furthermore, we want to draw attention to a method of treating of the involved friction problem as a problem of metal optics. Thus, the authors of [26] have computed the energy loss of low –energy electrons near metal surfaces, while in [41, 46] -the friction force and heat flow have been determined between two flat perfectly smooth featureless surfaces. Close to this approach was used in [44, 45]. The general idea is quite simple [41] : each component of fluctuating evanescent field created by moving body is reflected from the surface of resting body and vice versa. The corresponding waves experience Doppler shifts and, assuming the dispersion with frequency, this leads to a net momentum transfer between the bodies in relative motion. The resulting friction force depends on reflectivity coefficients (Fresnel coefficients) corresponding to different wave polarizations. In the same manner, it is possible to evaluate the heat flow through evanescent photon tunneling modes between the surfaces [46].

Finally, one has to mention the results following from complete theory of thermal fluctuating fields [16,17]. This theory finds various applications starting from vdW forces [5,75], the theory of heat transfer [47, 54, 75, 77, 79, 80 ], friction [44, 45, 50, 53, 55, 81, 82], and other phenomena [59, 82, 83].

For resting bodies, the results of different authors are in agreement, but a more general theory must include the fact of relative motion of the interracting bodies. This essentially changes the involved electrodynamical problem, because one needs to employ other material equations and boundary conditions, even in the nonrelativistic statement [53]. To date, there is no strict relativistic generalization of this theory and different authors have used different approaches in order to find, for example, the drag friction coefficient. So, the authors of [44, 45] have used a dynamical modification of the Lifshitz formulae [5], introducing the Doppler shifts in the corresponding amplitudes of the fluctuating fields via the Galileo transformation, and expanding them in $V/c$. The frictional stress between two semi-infinite solids was then calculated from $xz-$ component of the Maxwell stress tensor $\sigma_{ij}$. The obtained results agree with [31-33, 41] at $T=0$ (no friction in linear velocity order) and manifest the linear-velocity drag force at $T\neq 0$.

In [53] the authors have used an expression for losses $Q$ (from Ref. [16] ) of the electromagnetic field of electric and magnetic dipoles being situated above a flat surface. Furthermore, they have expanded $Q$ over the velocity of the moving dipole and restricted themselves to the first nonvanishing term. The obtained formula predicts the drag force $F_x \sim V/c^2$, which apparently vanishes in the limit $c \to \infty$. This is in sharp contrast with papers [31, 38 - 45, 55 - 58, 73] and is, possibly, due to a rejection of more important terms of the particle –surface interaction. The same shortcomings are relevant to papers [34, 35], where the authors have employed the quantum field –theoretical methods. However, a more detailed analysis of relativistic results is out of a scope of this article.

Concluding this primary discussion, we briefly summarize several important issues of the dissipative particle –surface interactions:

i)     for charged and dipole particles the linear –response dielectric theory is very good at high velocities, $ZV_B/V << 1$ , and satisfactory at $V << V_F$ ; in the last case the induced potential and electric field are both integrated quantities of the charge density fluctuations, and the results obtained do not significantly differ both in linear and nonlinear theories; at intermediate velocities, different charge states are present, and one has to account various processes leading to a change of charge, together with the interaction of different charge states with the surface;

ii)     several theoretical approaches allow to treat FDF with more or less degree of completeness, but special caution is needed when use is made of the corresponding relativistic generalizations, even at small velocity factor, $V/c$ ; of special importance are general (guide) relations like (2.1), (2.2) between the ground physical quantities;

ii)    at $T=0$ , the linear –velocity friction force (on neutral particle) has quantum origin and can not be determined from linear fluctuating electrodynamic; for metal and semiconductor substrates, the leading energy dissipation mechanism below the energy threshold for direct surface plasmon creation is related with generation of electron –hole pairs;

iii)   at $T \neq 0$ ,the friction force on neutral particles is dominated by thermal fluctuations and may be successfully calculated using the linear – response dielectric formalism; the needed nonlinear quantum corrections may be incorporated using the density functional theory; the thermal –induced fluctuation –dissipative forces exceed the conservative ones at $V > z_0 \sigma$ .

## 3.Problem statement and general equations

### 3.1 Problem statement

Let us consider a small moving particle with velocity $V$ in a vacuum region over a half –space (Fig.4). The half –space $z \leq 0$ is filled with material characterized by complex dielectric function, $\varepsilon_2 = \varepsilon_2{}' + i\varepsilon_2{}''$ , while the material of the particle, correspondingly, by $\varepsilon_1 = \varepsilon_1{}' + i\varepsilon_1{}''$ . A neutral atom (spherical particle) is also characterized by the dipole polarizability, $\alpha = \alpha' + i\alpha''$ . In general case, the functions $\varepsilon_{1,2}$ depend on frequency $\omega$ and wave –vector $\mathbf{k} = (\mathbf{q}, \mathbf{k}_z)$ , where $\mathbf{q}$ is the two –dimensional vector parallel to the surface. In the case of local (frequency) dispersion of the substrate material, we use notation $\mathbf{k} = (k_x, k_y)$ , i.e. vector $\mathbf{k}$ is assumed to be two –dimensional, too. Polarizability $\alpha$ depends on frequency $\omega$ . For simplicity, we consider nonmagnetic materials (if not specially mentioned). The spatial dispersion effects become important at $\lambda \leq z_0$ , with $\lambda$ being the wave –length of the involved surface excitation. In the range of separations $z_0 < 100\,nm$ , the local dielectric approximation is well justified in many cases of practical importance.

The own system of coordinates $K'$ is connected with the particle, and laboratory system of coordinates $K$ is located at the sample surface. The $z$ - axis of $K$ system is directed to the vacuum, being perpendicular to the surface, the $x-$ axis corresponds to the velocity direction, as the typical case of interest. However, we shall consider also normal (and arbitrary) direction of motion with respect to the surface.

Within the nonrelativistic and nonretarded statement we assume $V/c << 1$ and $r_0 << z_0 << c/\omega_0$ , where $r_0$ is the characteristic particle radius and $\omega_0$ is the characteristic absorption frequency. For ions, $r_0$ determines the core –radius and so the condition $z_0 >> r_0$ allows to consider them as bare charges. For dipole molecules and neutral atoms, respectively, it is possible to treat them as the point –like dipoles.

Relation $z_0 << c/\omega_0$, as we have already noted in Section 1, determines nanoscale distances in the range of 1 to 30 $nm$, which can be effectively probed by fast particle beams in the surface near field , or by moving nanotips of the scanning probe microscopes. Of course, the consistent relativistic formulation must be in accordance with the nonrelativistic results.

What we interested in, are : the tangential and normal forces ($F_x$, $F_z$), applied to a particle, and its rate of heating, $dQ/dt$. Alternatively, $-dQ/dt$ is the heating rate of the surface. In the case of lateral motion, $F_z$ defines the conservative component of the force (dynamic vdW force), while $F_x$ - the dissipative one. In the case of normal motion, $F_z$ includes both conservative and dissipative components. For the conservative force, we can also relate it with the potential, $F_z = -\partial U/\partial z$.

### 3.2 Some general equations

Now let us write down some general equations for the particle –surface interactions.

i)        charged particle – surface

An obvious generalization of Eq.(2.3) is

$$\mathbf{F} = (Ze)\mathbf{E}^{ind} \qquad (3.1)$$

where $\mathbf{E}^{ind}$ is the induced potential of the surface created by moving bare charge.

ii)        dipole molecule – surface

Introducing the dipole moment of the particle $\mathbf{d} = (d_x, d_y, d_z)$, the force applied from the surface reads

$$\mathbf{F} = (\mathbf{d}\nabla)\mathbf{E}^{ind} \qquad (3.2)$$

iii)        neutral particle –surface  (dipole approximation)

In this case, using a linear fluctuation electrodynamics, we must take into account both spontaneous and induced components of the fields and multipole moments. So, for the most important case of dipole approximation, the interaction potential is given by [8]

$$U = -\frac{1}{2}\langle \mathbf{d}\,\mathbf{E} \rangle = -\frac{1}{2}\langle \mathbf{d}^{sp}\,\mathbf{E}^{ind} \rangle - \frac{1}{2}\langle \mathbf{d}^{ind}\,\mathbf{E}^{sp} \rangle \qquad (3.3)$$

where the angular brackets imply total quantum and thermal averaging. Alternatively, the total fluctuation –induced  force applied to the particle is determined by [56]

$$\mathbf{F} = \left\langle (\mathbf{d}\nabla)\mathbf{E} \right\rangle = \left\langle (\mathbf{d}^{sp}\nabla)\mathbf{E}^{ind} \right\rangle + \left\langle (\mathbf{d}^{ind}\nabla)\mathbf{E}^{sp} \right\rangle \tag{3.4}$$

The force $F_z$ can be calculated either from (3.4), or from $F_z = -\partial U / \partial z$.

It must be emphasized that within the used formulation all vectorial quantities are assumed to be the Heisenberg operators.

In order to find $\mathbf{E}^{ind}$, we need to solve the corresponding Poisson's equation with the corresponding continuity conditions for the potential $\phi$ and normal component of the electric displacement at $z = 0$:

$$\nabla^2\phi = -4\pi\rho, \ \ \phi(+0) = \phi(-0), \ \ -\phi'(+0) = -\varepsilon\phi'(-0), \tag{3.5}$$

where the charge density $\rho$ is given by (for moving charges and dipoles, respectively)

$$\rho(x, y, z, t) = Ze\delta(x - Vt)\delta(y)\delta(z - z_0) \tag{3.6}$$

$$\rho(x, y, z, t) = -\text{div}\,\mathbf{P} \tag{3.7}$$

$$\mathbf{P}(x, y, z, t) = \delta(x - Vt)\delta(y)\delta(z - z_0)\mathbf{d} \tag{3.8}$$

In Eqs.(3.6)-(3.8) a particle is assumed to be moving in $x-$ direction. For normal motion (in $z-$ direction) the needed modification is trivial. In the case of fluctuating dipole moment, vector $\mathbf{d}$ in (3.8) depends on time $t$. Eq.(3.5) is solved in an analytical form using the Fourier transformations over the variables $x, y, t$ (see Appendix A). Then, obviously, $\mathbf{E}^{ind} = -\nabla\phi^{ind}$, with $\phi^{ind}$ being the induced part of the electric potential.

In addition, one has to employ the integral relation between the induced dipole moment $\mathbf{d}^{ind}(t)$ of the particle and the surface fluctuation field $\mathbf{E}^{sp}(t)$ [8] ($\alpha(\tau)$ is the particle polarizability, see also Appendix E)

$$\mathbf{d}^{ind}(t) = \int_0^\infty \alpha(\tau)\mathbf{E}^{sp}(t - \tau)d\tau \tag{3.9}$$

The foregoing calculations using (3.1) - (3.9), being clear in their main respects, take a lot of space and we are referring the readers to the results of Appendixes A, C - E and papers [48, 49, 51, 56, 73], where the needed mathematical details of statistical averaging are presented in more details .

At this moment, it is more appropriate to formulate the ground relationship between the rate of work of fluctuating field, tangential force on the particle and heat flow [56-58, 73]. In absence of radiation, the energy conservation law reads

$$-\frac{dW}{dt} = \int \langle \mathbf{j}\mathbf{E}\rangle \, d^3r \quad , \tag{3.10}$$

where in the left –hand side we have the rate of energy dissipation of fluctuating electromagnetic field, while in the right-hand side – the averaged work per unit time performed by fluctuation electromagnetic field over the moving particle. The density current $\mathbf{j}$ in (3.10) can be expressed via the polarization $\mathbf{P}(\mathbf{r},t)$ (see (3.8)) :

$$\mathbf{j} = \partial \mathbf{P}(\mathbf{r},t)/\partial t \tag{3.11}$$

By inserting (3.11) into (3.10) with account of (3.8) and $\mathbf{d} = \mathbf{d}(t)$ we get

$$\int \langle \mathbf{j}\mathbf{E}\rangle \, d^3r = V\left\langle \frac{\partial}{\partial x}(\mathbf{d}\mathbf{E})\right\rangle + \langle \dot{\mathbf{d}}\mathbf{E}\rangle \tag{3.12}$$

It must be specially emphasized that one has firstly to take derivative $\partial/\partial x$, and then to substitute the corresponding coordinates of the particle location point, $\mathbf{r} = (Vt, 0, z_0)$.

Transforming the first addend in (3.12)

$$\frac{\partial}{\partial x}(\mathbf{d}\mathbf{E}) = \frac{\partial}{\partial x}\left(d_x E_x + d_y E_y + d_z E_z\right) =$$

$$= \left(d_x \frac{\partial}{\partial x} + d_y \frac{\partial}{\partial y} + d_z \frac{\partial}{\partial z}\right)E_x + d_y\left(\frac{\partial E_y}{\partial x} - \frac{\partial E_x}{\partial y}\right) + d_z\left(\frac{\partial E_z}{\partial x} - \frac{\partial E_x}{\partial z}\right) = \tag{3.13}$$

$$= (\mathbf{d}\nabla)E_x + d_y(\mathrm{rot}\,\mathbf{E})_z - d_z(\mathrm{rot}\,\mathbf{E})_y = (\mathbf{d}\nabla)E_x + [\mathbf{d}\,\mathrm{rot}\,\mathbf{E}]_x$$

and substituting (3.13) into (3.12) yields

$$\int \langle \mathbf{j}\mathbf{E}\rangle \, d^3r \equiv V\left\{\langle(\mathbf{d}\nabla)E_x\rangle + \langle[\mathbf{d}\,\mathrm{rot}\,\mathbf{E}]_x\rangle\right\} + \langle\dot{\mathbf{d}}\mathbf{E}\rangle \tag{3.14}$$

In the nonrelativistic case, $\mathrm{rot}\,\mathbf{E} = -\frac{1}{c}\frac{\partial \mathbf{B}}{\partial t} = 0$ and (3.14) reduces to

$$\int \langle \mathbf{j}\mathbf{E}\rangle \, d^3r \equiv V\langle(\mathbf{d}\nabla)E_x\rangle + \langle\dot{\mathbf{d}}\mathbf{E}\rangle \tag{3.15}$$

The first term in the right –hand side of (3.15) defines tangential component of the "dipole force", $F_x = \langle(\mathbf{d}\nabla)E_x\rangle$, while the second one – the heat flow $\dot{Q} = \langle\dot{\mathbf{d}}\mathbf{E}\rangle$ (compare (3.15) with (2.1),(3.10)). In particular, $\dot{Q} > 0$ corresponds to particle heating and $\dot{Q} < 0$ - to cooling. We see that, generally speaking, $V^{-1}\int\langle\mathbf{j}\mathbf{E}\rangle\,d^3r \neq \langle(\mathbf{d}\nabla)E_x\rangle$, and these two quantities are identical only for particles with constant dipole moment $\mathbf{d}$.

The physical meaning of Eq. (3.15) is quite clear: a work performed by fluctuating electromagnetic field over a system of bounded charges is spent on change of kinetic energy of the center of mass (friction or acceleration the particle), and on change of its inner degrees of freedom. For a multiatomic particle, the second term of (3.15) is identified with the heat exchange. For a ground state atom, it may be attributed to a some kind of the Lamb –shift caused by particle –surface coupling [56, 73].

It is not difficult to get the corresponding relationship in relativistic case, too (see Appendix B). In the laboratory frame $K$ the result is given by [57]

$$\int \langle \mathbf{j}\mathbf{E} \rangle d^3 r = F_x V + \langle \dot{\mathbf{d}}\,\mathbf{E} \rangle - \frac{V}{c} \left\langle \left[ \dot{\mathbf{d}}\,\mathbf{H} \right]_x \right\rangle, \qquad (3.16)$$

$$\dot{Q} = \langle \dot{\mathbf{d}}\,\mathbf{E} \rangle - \frac{V}{c} \left\langle \left[ \dot{\mathbf{d}}\,\mathbf{H} \right]_x \right\rangle \qquad (3.17)$$

where $\mathbf{H}$ is the magnetic field and the particle is assumed to be nonmagnetic in its rest frame, $K'$. Also, Eq. (3.17) can be derived using the Planck's formulation of relativistic thermodynamics.

In the most general case, the force applied to a moving particle from the surface is given by

$$\mathbf{F} = \int \langle \rho \mathbf{E} \rangle d^3 r + \frac{1}{c} \int \langle [\mathbf{j}\mathbf{H}] \rangle d^3 r \qquad (3.18)$$

with $\rho$ and $\mathbf{j}$ being the fluctuating charge and current densities. By definition, $\rho = -\operatorname{div}\mathbf{P}$, $\quad \mathbf{j} = \partial \mathbf{P}/\partial t + c\operatorname{rot}\mathbf{M}$ , where $\mathbf{P}$ and $\mathbf{M}$ are vectors of electric and magnetic polarization created by the particle. Using these relations, the total force on the particle and heat flow can be written in more compact form [57, 58]

$$\mathbf{F} = \langle \nabla(\mathbf{d}\,\mathbf{E} + \mathbf{m}\,\mathbf{H}) \rangle \qquad (3.19)$$

$$\dot{Q} = \langle \dot{\mathbf{d}}\,\mathbf{E} + \dot{\mathbf{m}}\,\mathbf{H} \rangle \qquad (3.20)$$

where $\mathbf{d}$ and $\mathbf{m}$ are dipole and magnetic moments of the particle in the laboratory frame. It should be underlined, that a nonmagnetic particle ( in its rest frame) with the electric dipole moment will have both electric and magnetic moments in the laboratory frame (see Section 4.5 and Appendix B). Obviously, (3.19), (3.20) agree with their nonrelativistic equivalents, (3.4) and $\dot{Q} = \langle \dot{\mathbf{d}}\,\mathbf{E} \rangle$ .

### 3.3 Specular reflection model and surface response dielectric functions

In order to study dynamic problems of particle –surface interactions and properties of interface between two media using dielectric formalism, one needs to have the corresponding dielectric response functions. The media can be undispersive, and then the appropriate dielectric response function may depend only on frequency (local approximation), or the media can be dispersive, and the corresponding nonlocal effects (spatial dispersion effects) must be taken into account, too.

A simple way to introduce the corresponding medium response provides the specular reflection model (SRM), firstly proposed in [84, 85] and later successfully applied to many dynamic problems of the particle –surface interactions (see [1-3] and references therein). Using SRM allows to account for nonlocal effects through the dielectric function of bulk material, $\varepsilon(\mathbf{k}, \omega)$, for different kind of particles.

According to SRM, a border of the polarizable medium is assumed to be sharp, without smooth variation of electric properties across the surface $z = 0$. The induced potential is determined by the external charge, its image and a fictitious surface charges fixed by the boundary conditions. For a charged particle incident onto the surface (Fig.5) the Fourier –components of scalar electric potential are given by

$$\phi(\mathbf{k}, \omega) = \left(8\pi^2 Ze / k^2\right)\left[\delta(\omega - \mathbf{kV}) + \delta(\omega - \mathbf{kV'}) + \rho_s(\mathbf{q}, \omega)\right], \ z<0 \tag{3.21}$$

$$\phi(\mathbf{k}, \omega) = -8\pi^2 Ze \, \rho_s(\mathbf{q}, \omega) / k^2 \varepsilon(\mathbf{k}, \omega) \ , \ z > 0 \tag{3.22}$$

where $\mathbf{V} = (\mathbf{V}_{II}, \mathbf{V}_\perp)$ is the particle velocity with components $\mathbf{V}_{II}, \mathbf{V}_\perp$ parallel and normal to the surface, $\mathbf{V'} = (\mathbf{V}_{II}, -\mathbf{V}_\perp), \mathbf{k} = (\mathbf{q}, k_z)$ .The first term in (3.21) represents the image charge necessary to symmetrise the potential seen by the surface electrons and a fictitious surface distribution required to satisfy the boundary conditions (continuity of the potential and normal component of the electric displacement). The continuity of the potential at $z = 0$ fixes $\rho_s(\mathbf{q}, \omega)$ as

$$\rho_s(\mathbf{q}, \omega) = -\frac{q}{\pi + qI_0} \int_{-\infty}^{+\infty} \frac{dk_z}{q^2 + k_z^2} \left[\delta(\omega - \mathbf{kV}) + \delta(\omega - \mathbf{kV'})\right] \tag{3.23}$$

$$I_0 = \int_{-\infty}^{+\infty} \frac{dk_z}{k^2 \varepsilon(\mathbf{k}, \omega)} \tag{3.24}$$

The induced scalar potential, as obtained from (3.22) after Fourier decomposition, has the form (the proper generalizations for dipole molecules and spherical particles will be obtained in Section 4) [86]

$$\phi^{ind}(\mathbf{r},t) = -\frac{Ze}{2\pi^2}\int d^2\mathbf{q}\int d\omega \frac{\left|\mathbf{V}_\perp\right|\exp(-q|z|)}{(\omega - \mathbf{V}_{II}\cdot\mathbf{q})^2 + (V_\perp q)^2}\left(\frac{\pi - qI_0}{\pi + qI_0}\right)\exp(i(\mathbf{q}\mathbf{V}_{II} - \omega)t) \quad (3.25)$$

According to this, the nonlocal surface –response function (SRF) reads

$$\Delta(q,\omega) = \frac{\pi - qI_0}{\pi + qI_0} \quad (3.26)$$

In local case, using (3.24), (3.26) yields

$$\Delta(\omega) = \frac{\varepsilon(\omega) - 1}{\varepsilon(\omega) + 1} \quad (3.27)$$

In this notation, additionally, (3.27) is identical to the local reflection factor of electromagnetic waves, $R_P(\omega)$, corresponding to $P-$polarized waves, when the electric field vector lies in the plane containing the wave vector and surface normal. For another wave with $S-$polarization, the electric field vector is perpendicular to the corresponding plane, but in the nonretarded approximation $R_S(\omega) = 0$ [87].

There are known several analytical expressions for $\varepsilon(\mathbf{k},\omega)$ and $\Delta(q,\omega)$ incorporating important properties of polarizable medium. Some of them being useful in future analysis are given in Appendix C.

## 4.Interactions of moving atomic and molecular particles with flat surface

Now let us review the results of the calculations for different kind of particles. At first, we shall neglect nonlocal effects ( Sections 4.1 - 4.7).

### 4.1 Charged particle

Using (3.1) and results from Apendix A, the general expressions for the normal and tangential forces at parallel movement are given by [48,51,52]

$$F_z(z_0,V) = -\frac{2(Ze)^2}{\pi}\int_0^{\pi/2}d\phi\int_0^\infty qdq\exp(-2qz_0)\Delta'(qV\cos\phi) \quad (4.1)$$

$$F_x(z_0,V) = -\frac{2(Ze)^2}{\pi}\int_0^{\pi/2}\cos\phi d\phi\int_0^\infty qdq\exp(-2qz_0)\Delta''(qV\cos\phi) \quad (4.2)$$

Denoting $\omega = qV\cos\phi$ , (4.1) and (4.2) reduce to simpler expressions, coinciding with [88]

$$F_z(z_0,V) = -\frac{2(Ze)^2}{\pi V^2}\int_0^\infty d\omega\omega K_1(2\omega z_0/V)\Delta'(\omega) \quad (4.3)$$

$$F_x(z_0,V) = -\frac{2(Ze)^2}{\pi V^2}\int_0^\infty d\omega\omega K_0(2\omega z_0/V)\Delta''(\omega) \quad (4.4)$$

where $K_0(x)$ and $K_1(x)$ are the modified Bessel functions. More general (relativistic) expressions for $F_x$ were obtained in [26, 88, 89].

For Drude dielectric response (C7) without damping ($\tau \to \infty$) Eq. (4.4) reduces to [86, 90]

$$F_x(z_0, V) = -\frac{(Ze)^2 \omega_s{}^2}{V^2} K_0\left(\frac{2\omega_s z_0}{V}\right) \tag{4.5}$$

where $\omega_s = \omega_p / \sqrt{2}$ is the surface plasmon frequency.

Using the metal dielectric response (C10) and small –velocity approximation, Eqs.(4.3), (4.4) reduce to [86, 39]

$$F_z(z_0, V) = -\frac{(Ze)^2}{4z_0{}^2}\left(1 - \frac{3}{16\pi^2}\frac{V^2}{\sigma^2 z_0{}^2} + ...\right) \tag{4.6}$$

$$F_x(z_0, V) = -\frac{(Ze)^2}{16\pi\sigma z_0{}^3} V + ... \tag{4.7}$$

In order to see equivalence (4.7) and Eq. (32) from Ref. [86], one should account for the standard relations: $\gamma = 1/\tau$, $4\pi\sigma/\tau = \omega_p{}^2$, $\omega_s = \omega_p/\sqrt{2}$ .

In the case of perpendicular motion, the force $F_z$ contains both conservative and dissipative components simultaneously. The corresponding expression is given by (we assume $V > 0$, i.e the particle approachs the surface)

$$F_z(z_0, V) = \frac{\mathrm{i}(Ze)^2}{\pi V^2} \int\limits_{-\infty}^{\infty} d\omega\,\omega\,\Delta(\omega)\exp(-\mathrm{i}\omega z_0/V)\int\limits_0^{\infty}\frac{dq\,q\exp(-qz_0)}{q^2 + \omega^2/V^2} =$$
$$= -(Ze)^2 \int\limits_0^{\infty} dq\,q\,\Delta(-iqV)\exp(-2qz_0) \tag{4.8}$$

The conservative and dissipative components of the force are seen when use is made of (C10), when expanding (4.8) in powers in $V$ :

$$F_z(z_0, V) = -\frac{(Ze)^2}{4z_0{}^2}\left(1 + \frac{V}{2\pi\sigma z_0} + \frac{3V^2}{8\pi^2\sigma^2 z_0{}^2} + ...\right) \tag{4.9}$$

It follows from (4.9) that dissipative contribution to the attraction force contains an odd power of $V$, while the conservative one – an even power. The first term of (4.9) describes attraction of a charge to its image. Comparing (4.7) with (4.9) we see that the drag friction coefficient for the normal motion is twice that for the lateral one.

**4.2 Dipole molecule**

Now we use Eq.(3.2) and, again, results from Appendix A. The final formulae for the attraction and dissipative forces are [51, 52]:

i) lateral movement

$$F_z(z_0, V) = \frac{-1}{2\pi} \iint dq_x \, dq_y \exp(-2qz_0)(q_x^2 d_x^2 + q_y^2 d_y^2 + q^2 d_z^2) \Delta'(q_x V) \qquad (4.10)$$

$$F_x(z_0, V) = \frac{-1}{2\pi} \iint dq_x \, dq_y \, q_x \exp(-2qz_0)(q_x^2 d_x^2 + q_y^2 d_y^2 + q^2 d_z^2) \Delta''(q_x V)/q \quad (4.11)$$

The relativistic generalization for $F_x$ see in [89].

ii) perpendicular movement $(V > 0)$

$$F(z_0, V) = -\frac{d_x^2 + d_y^2 + 2d_z^2}{2} \int_0^\infty dq q^3 \Delta(-iqV) \exp(-2q z_0) \qquad (4.12)$$

Performing the small –velocity expansions in Eqs. (4.10) - (4.12) yields

$$F_z(z_0, V) = -\frac{3(d_x^2 + d_y^2 + 2d_z^2)}{16z_0^4} \left( \begin{array}{l} \Delta'(0) + \dfrac{5V^2}{8z_0^2} \dfrac{d^2 \Delta'(\omega)}{d\omega^2}\Big|_{\omega=0} \cdot \\[2mm] \cdot \dfrac{(3d_x^2 + d_y^2 + 4d_z^2)}{(d_x^2 + d_y^2 + 2d_z^2)} + ... \end{array} \right) \qquad (4.13)$$

$$F_x(z_0, V) = -\frac{3(3d_x^2 + d_y^2 + 4d_z^2)V}{32z_0^5} \frac{d\Delta''(\omega)}{d\omega}\Big|_{\omega=0} + ... \qquad (4.14)$$

$$F(z_0, V) = -\frac{(3d_x^2 + d_y^2 + 2d_z^2)}{16z_0^4} \left( \Delta(0) + \frac{2V}{z_0} \frac{d\Delta''}{d\omega}\Big|_{\omega=0} + ... \right) \qquad (4.15)$$

Eqs. (4.13), (4.14) correspond to parallel movement, Eq.(4.15) - to perpendicular one. The first term of (4.15) describes the well-known expression for the electrostatic attraction of a dipole to the surface. Using (C10), Eqs.(4.13)-(4.15) reduce to [39]

$$F_z(z_0, V) = -\frac{3(d_x^2 + d_y^2 + 2d_z^2)}{16z_0^4} \left( 1 - \frac{5V^2}{8\pi^2 \sigma^2 z^2} + ... \right) \qquad (4.16)$$

$$F_x(z_0, V) = -\frac{3(3d_x^2 + d_y^2 + 4d_z^2)V}{64\pi\sigma z_0^5} + ... \qquad (4.17)$$

$$F(z_0, V) = -\frac{3(3d_x^2 + d_y^2 + 2d_z^2)}{16z_0^4} \left( 1 + \frac{V}{\pi\sigma z_0} + ... \right) \qquad (4.18)$$

One can see that the ratio of the friction coefficients for normal and lateral motion is between 2 and 4 at different projections of the dipole moment.

### 4.3 Neutral spherical particle (ground state atom)

For perpendicular movement of the particle, the general method (see Appendix A) does not allow to separate the space ($x, y, z$) and time ($t$) variables in the Poisson's equation (3.5), therefore we present here only the results corresponding to lateral motion, the most important case in practical applications. The starting equations are (3.3), (3.4). The corresponding details of the calculations are given in Appendixes A, D-E.

The authors of [55] used the second –order perturbation theory and estimated the drag friction coefficients for both perpendicular ($\eta_z$) and lateral ($\eta_x$) motion which were found to be $\eta_z / \eta_x = 7 \div 8.4$ in the case of thermal –induced fluctuations. For the quantum –induced fluctuations, corespondingly, it was found $\eta_z / \eta_x \approx 5$ [40], thus we may accept the above estimates and those for constant dipoles ($\eta_z / \eta_x = 2 \div 4$) as the guide ones when addressing the corresponding FDF in general case at arbitrary direction of motion.

In the case of parallel motion, the most general expression for the vdW potential is given by [51, 52]

$$U(z_0, V) = -\frac{\hbar}{\pi^2} \iiint d\omega dk_x\, dk_y\, k \exp(-2kz_0) \cdot$$

$$\cdot \left\{ \begin{array}{l} \coth \dfrac{\omega \hbar}{2k_B T_1} \alpha''(\omega) \big[ \Delta'(\omega - k_x V) + \Delta'(\omega + k_x V) \big] + \\[2mm] + \coth \dfrac{\omega \hbar}{2k_B T_2} \Delta''(\omega) \big[ \alpha'(\omega - k_x V) + \alpha'(\omega + k_x V) \big] \end{array} \right\} \qquad (4.19)$$

Eq. (4.19) generalizes the well –known formula for the vdW interaction between a neutral spherical particle and a surface with account of different finite temperatures and arbitrary particle velocity. If we take $V$=0, $T_1 = T_2 = 0$, then (4.19) reduces to

$$U(z_0) = -\frac{2\hbar}{\pi^2} \cdot \iiint d\omega dk_x\, dk_y\, k \exp(-2kz_0) \big[ \alpha''(\omega)\Delta'(\omega) + \alpha'(\omega)\Delta''(\omega) \big] =$$

$$= -\frac{2\hbar}{\pi^2} \iint dk_x\, dk_y\, k \exp(-2kz_0) \cdot \mathrm{Im} \int_0^\infty d\omega \alpha(\omega)\Delta(\omega) \qquad (4.20)$$

The corresponding integration over $k_x, k_y$ is trivial , while the frequency integral is simplified by turning the integration contour over the angle $90^0$ , so that it coincides the upper imaginary half –axis . Then we get

$$\text{Im}\int\limits_0^\infty d\omega\alpha(\omega)\Delta(\omega)=\int\limits_0^\infty d\omega\alpha(\mathrm{i}\omega)\Delta(\mathrm{i}\omega),$$

and retrieve the well –known reference result [8]

$$U(z_0)=-\frac{\hbar}{4\pi z_0^{\,3}}\int\limits_0^\infty d\omega\alpha(\mathrm{i}\omega)\Delta(\mathrm{i}\omega) \tag{4.21}$$

The first nonvanishing velocity-correction to (4.21) is obtained from (4.19) after the velocity expansion of the integrand functions. The result is

$$\Delta U(z_0,V)=-\frac{3\hbar V^2}{16\pi z_0^{\,5}}\int_0^\infty d\omega\coth\frac{\omega\hbar}{2k_\mathrm{B}T}\left(\alpha''(\omega)\frac{\partial^2\Delta'(\omega)}{\partial\omega^2}+\Delta''(\omega)\frac{\partial^2\alpha'(\omega)}{\partial\omega^2}\right)$$

(4.22)

Eq.(4.22) has a more general form than that obtained in [32], where use is made of the surface –plasmon –pole approximation for $\Delta(\omega)$ , with no account of the surface temperature and specific material properties.

Let us consider, for instance, a movement of a small conducting particle (of radius $R$ ) above a conducting surface, assuming both the particle and surface material to be described by dielectric function (C10) with the same conductance parameter $\sigma$ . Then, with account of the known relation for polarizability of a spherical particle

$$\alpha(\omega)=R^3\frac{\varepsilon(\omega)-1}{\varepsilon(\omega)+2}, \tag{4.23}$$

in the limit $2\pi\sigma<<k_\mathrm{B}T/\hbar$ the corresponding integrations in (4.22) and (4.19) (at $V\to0,\ T_1=T_2=T$ ) yield (cf. with simple formulae given in Section 2):

$$U(z_0,V)\approx-\frac{k_BTR^3}{4z_0^{\,3}}\left(1-0.014\frac{V^2}{z_0^{\,2}\sigma^2}\right) \tag{4.24}$$

In the opposite limit, $2\pi\sigma>>k_\mathrm{B}T/\hbar$ , we obtain (neglecting the smaller terms of the order of $\delta$ and $\delta^2$ $(\delta=k_\mathrm{B}T/\hbar\,\sigma\,)$ )

$$U(z_0,V)\approx-0.2\frac{\hbar\sigma R^3}{z_0^{\,3}}\left(1+0.026\frac{V^2}{z_0^{\,2}\sigma^2}\right) \tag{4.25}$$

It follows from (4.24), (4.25) that at $z_0 = 1\,nm$, $\sigma = 10^{11}\,sec^{-1}$ the corresponding dynamic effect becomes large at $V \geq 5\,m/\sec$, i.e. at much smaller velocities, than could be expected in the case of atom–plasmon coupling [32]. Moreover, one sees that the proper dynamic correction may be both repulsive and attractive (compare (4.24) and (4.25); see also Eq.(26) in Ref. [32]). This example is specific one, but Eq.(4.19), evidently, manifests a lot of new features, which we leave aside at the moment.

The general expression for the dissipative force at parallel motion is calculated from (3.4) with account of the results given in Appendixes A, D-E. The final formula has the form [56, 58]

$$F_x = \left\langle (\mathbf{d}^{sp}\nabla) E^{ind}_x \right\rangle + \left\langle (\mathbf{d}^{ind}\nabla) E^{sp}_x \right\rangle = -\frac{2\hbar}{\pi^2}\iiint d\omega\, dk_x dk_y\, k_x\, k \exp(-2k\, z_0)\, \cdot$$
$$\cdot\left\{ \begin{array}{l} \coth\left(\dfrac{\omega\hbar}{2k_B T_1}\right)\alpha''(\omega)\left[\Delta''(\omega+k_x V) - \Delta''(\omega - k_x V)\right] + \\[2mm] \coth\left(\dfrac{\omega\hbar}{2k_B T_2}\right)\Delta''(\omega)\left[\alpha''(\omega+k_x V) - \alpha''(\omega - k_x V)\right] \end{array} \right\} \qquad (4.26)$$

At $T_1 = T_2 = T$ (4.26) reduces to [55]

$$F_x = -\frac{2\hbar}{\pi^2}\iiint d\omega\, dk_x dk_y\, k_x\, k \exp(-2k\, z_0) \coth\left(\frac{\omega\hbar}{2k_B T}\right)\cdot$$
$$\cdot\left\{\alpha''(\omega)\left[\Delta''(\omega+k_x V) - \Delta''(\omega - k_x V)\right] + \Delta''(\omega)\left[\alpha''(\omega+k_x V) - \alpha''(\omega - k_x V)\right]\right\} \qquad (4.27)$$

Moreover, to linear velocity order, after expanding the integrand functions in powers $k_x V$ and integrating over $k_x, k_y$ we retrieve the formula firstly obtained by Tomassone and Widom [39]

$$F_x = \frac{3\hbar V}{2\pi z_0^5}\int_0^\infty \alpha''(\omega)\Delta''(\omega)\frac{d}{d\omega}\frac{1}{(\exp(\omega\hbar/k_B T) - 1)}d\omega \qquad (4.28)$$

As it follows from (4.28), $F_x$ exponentially approachs zero at $T \to 0$: $F_x \propto \exp(-E_0/k_B T)$, where $E_0$ is the energy of the first excited state of an atom relative to the ground state [39]. Therefore, it is necessary to go to higher order of the perturbation theory in order to get the drag friction force on spherical atom at $E_0 >> k_B T$. The coresponding result is (we use here the atomic units) [38]

$$F_x = -\frac{\left[k_F^{\,3}\alpha(0)\right]^2}{(k_F d)^{10}}\frac{\omega_F}{\omega_p}k_F I(d)V \qquad (4.29)$$

where $d$ is the distance of an atom from the metal jellium edge, the function $I(d)$ is slowly –dependent function of the order of 1, $\alpha(0)$ is the static polarizability of the adatom, $k_F$ and $\omega_F$ are the Fermi wave –vector and frequency. Eq.(4.29) agrees with that obtained in pioneering paper by Schaich and Harris [31]. For more refined calculations of $I(d)$ see in [40] where the author used time –dependent –density - functional approach incorporating the exchange –correlation terms in local potential.

In general case, at $T_1 \to 0$, $T_2 \to 0$ Eqs. (4.27), (4.28) reduce to

$$F_x = \frac{4\hbar}{\pi^2} \int\limits_0^\infty k_x \, dk_x \int\limits_0^\infty dk_y \, k \exp(-2k\,z_0) \int\limits_0^{k_x V} d\omega \, \Delta''(\omega) \, \alpha''(\omega - k_x V) \,, \qquad (4.30)$$

while in the limit $V \to 0$ we get $F_x \sim V^3 / z_0^{\,7}$ (see also Section 6.5).

Coming back to Eq.(4.26), let us consider the linear –velocity friction force at $T_1 \neq T_2$. Performing the corresponding expansions under the integrand (this is valid at $V << z_0 \omega_0$, with $\omega_0$ being the characteristic frequency), and integrating over the wave-vectors yields

$$F_x = -\frac{3\hbar V}{2\pi z_0^{\,5}} \int\limits_0^\infty d\omega \left[ \frac{1}{\exp(\hbar\omega / k_B T_1) - 1} - \frac{1}{\exp(\hbar\omega / k_B T_2) - 1} \right] \alpha''(\omega) \frac{d\Delta''(\omega)}{d\omega} -$$
$$-\frac{3\hbar V}{2\pi z_0^{\,5}} \frac{\hbar}{k_B T_2} \int\limits_0^\infty d\omega \, \alpha''(\omega) \, \Delta''(\omega) \frac{\exp\left( \dfrac{\omega\hbar}{k_B T_2} \right)}{\left[ \exp\left( \dfrac{\omega\hbar}{k_B T_2} \right) - 1 \right]^2} \qquad (4.31)$$

We see that only the second term in (4.31) is always frictional $(F_x < 0)$. The first one may be also accelerative $(F_x > 0)$, either at $T_2 > T_1$, $d\Delta''/d\omega > 0$, or at $T_1 > T_2$, $d\Delta''/d\omega < 0$. Therefore, the particle may acquire energy from the surface excitations.

Another interesting situation is related with the nonlilinear –velocity regime, $V \geq z_0 \omega_0$, and we shall turn back to this point in Section 4.6.

## 4.4 Heating effects

The heat flow between a moving spherical particle and a surface, $\dot{Q} = \langle \dot{\mathbf{d}} \mathbf{E} \rangle$ (see Eq.(3.15)) is calculated similarly to the dipole force. The general expression is given by [56,73]

$$\dot{Q} = \langle \dot{\mathbf{d}} \mathbf{E} \rangle = \langle \dot{\mathbf{d}}^{sp} \mathbf{E}^{in} \rangle + \langle \dot{\mathbf{d}}^{in} \mathbf{E}^{sp} \rangle = -\frac{2\hbar}{\pi^2} \iiint d\omega\, dk_x\, dk_y\, k \exp(-2k\,z_0) \cdot$$

$$\cdot \left\{ \begin{array}{l} \coth\left(\dfrac{\omega\hbar}{2k_B T_1}\right)\alpha''(\omega)\left[\omega\Delta''(\omega+k_x V) + \omega\Delta''(\omega-k_x V)\right] - \\[2mm] \coth\left(\dfrac{\omega\hbar}{2k_B T_2}\right)\Delta''(\omega)\left[(\omega+k_x V)\alpha''(\omega+k_x V) + (\omega-k_x V)\alpha''(\omega-k_x V)\right] \end{array} \right\} \qquad (4.32)$$

Also, it is interesting to consider the small-velocity limit $V \ll z_0\omega_0$ . In this case from (4.32) we obtain

$$\dot{Q} = -\frac{\hbar}{\pi z_0^3}\int\limits_0^\infty d\,\omega\,\omega\,\alpha''(\omega)\,\Delta''(\omega)\left[\frac{1}{\exp(\hbar\omega/k_B T_1)-1} - \frac{1}{\exp(\hbar\omega/k_B T_2)-1}\right] -$$

$$- \frac{3\hbar V^2}{8\pi z_0^5}\int\limits_0^\infty d\omega \left\{ \begin{array}{l} \coth\left(\dfrac{\omega\hbar}{2k_B T_1}\right)\omega\,\alpha''(\omega)\dfrac{d^2\Delta''}{d\omega^2} - \coth\left(\dfrac{\omega\hbar}{2k_B T_2}\right)\Delta''(\omega)\cdot \\[3mm] \cdot\left(2\dfrac{d\alpha''(\omega)}{d\omega} + \omega\dfrac{d^2\alpha''(\omega)}{d\omega^2}\right) \end{array} \right\} \qquad (4.33)$$

The first term of (4.33) agrees with that obtained in [46, 54] at $V = 0$. The dynamic correction ( the second term) has the order of magnitude $(V/z_0\omega_0)^2 \ll 1$ and, at first sight, is small. However, it will not be correct to neglect this one in the general case when the dielectric functions have resonant structure and dynamic contribution significantly increases. In particular, in such a situation the heat flow may have "anomalous direction", when a "hot" particle acquires heat from a "cold" surface ($T_1 > T_2$) [73]. A similar possibility in the case of heat exchange between two flat surfaces was noted by Polevoi [35].

Quite recently, the authors of [54, 55] have drawn attention to a drastically strong influence of retardation effect on heat transfer [54] and friction force [55] between conducting nanostructures. As they claim, the involved effect comes from $S$-polarized waves having the reflection factor $R_S$ (see Section 3.3). Because the corresponding contribution is missed within the nonrelativistic approximation, it is worthwhile giving a brief review of the relativistic theory before discussing this question in more details.

## 4.5 A brief review of relativistic theory

Using the same formalism, we have calculated normal/tangential forces for charges and dipole molecules in [89], and in [57, 58] – for neutral spherical particles (ground state atoms). The statement of the problem is similar to the nonrelativistic one (see Section 3.1), the particle is assumed to have fluctuating electric dipole moment and no magnetic moment in the $K'$-frame, but, due to the Lorentz transformations, the corresponding moments in the $K$-frame are determined by

$$\mathbf{d} = (d'_x / \gamma, \, d'_y, \, d'_z) \;, \; \mathbf{m} = (0, \, \beta \, d'_z, \, \beta \, d'_y) \;, \; \gamma = (1 - \beta^2)^{-1/2} \;, \; \beta = V / c \;,$$

where the primed variables correspond to the $K'$-frame. In this case, instead of (3.4) and Eq. $\dot{Q} = \langle \dot{\mathbf{d}} \mathbf{E} \rangle$, the general expressions for the force and heat flow are given by Eqs.(3.19), (3.20).

The formulas presented in [57, 58] are reduced to the more compact ones after simple transformations of the frequency arguments in the integrand functions:

$$F_x = -\frac{\hbar}{\pi^2} \gamma \iiint_{k > \omega / c} d\omega \, dk_x \, dk_y \, k_x \, \frac{\exp(-2q_0 z_0)}{q_0} \cdot$$

$$\cdot \left\{ \begin{array}{l} \cdot \left( \coth \dfrac{\hbar \omega}{2k_B T_2} - \coth \dfrac{\hbar (\omega + k_x V) \gamma}{2k_B T_1} \right) \cdot \\ \cdot \alpha''\big((\omega + k_x V) \gamma\big) \Big[ \chi_e^{(+)}(\omega, \mathbf{k}) \Delta_e''(\omega) + \chi_m^{(+)}(\omega, \mathbf{k}) \Delta_m''(\omega) \Big] - \\ - \left( \coth \dfrac{\hbar \omega}{2k_B T_2} - \coth \dfrac{\hbar (\omega - k_x V) \gamma}{2k_B T_1} \right) \cdot \\ \cdot \alpha''\big((\omega - k_x V) \gamma\big) \Big[ \chi_e^{(-)}(\omega, \mathbf{k}) \Delta_e''(\omega) + \chi_m^{(-)}(\omega, \mathbf{k}) \Delta_m''(\omega) \Big] \end{array} \right\} -$$

$$- \frac{\hbar}{\pi^2} \gamma \iiint_{k < \omega / c} d\omega \, dk_x \, dk_y \, k_x \left[ -\frac{\sin(2\tilde{q}_0 z_0)}{\tilde{q}_0} \right] \cdot \left\{ \Delta_e, \Delta_m \to \tilde{\Delta}_e, \tilde{\Delta}_m \right\} + \dots$$

(4.34)

$$F_z = -\frac{\hbar}{\pi^2}\gamma \iiint_{k>\omega/c} d\omega\, dk_x\, dk_y \exp(-2q_0 z_0)$$

$$\cdot \left\{ \begin{array}{l} \cdot \coth\dfrac{\hbar(\omega+k_x V)\gamma}{2k_B T_1}\alpha''\big((\omega+k_x V)\gamma\big)\cdot \\[2mm] \cdot\Big[\chi_e^{(+)}(\omega,\mathbf{k})\Delta_e'(\omega)+\chi_m^{(+)}(\omega,\mathbf{k})\Delta_m'(\omega)\Big]+ \\[2mm] + \coth\dfrac{\hbar(\omega-k_x V)\gamma}{2k_B T_1}\alpha''\big((\omega-k_x V)\gamma\big)\cdot \\[2mm] \cdot\alpha''\big((\omega-k_x V)\gamma\big)\Big[\chi_e^{(-)}(\omega,\mathbf{k})\Delta_e'(\omega)+\chi_m^{(-)}(\omega,\mathbf{k})\Delta_m'(\omega)\Big]+ \\[2mm] + \coth\dfrac{\hbar\omega}{2k_B T_1}\alpha'\big((\omega+k_x V)\gamma\big)\cdot \\[2mm] \cdot\Big[\chi_e^{(+)}(\omega,\mathbf{k})\Delta_e''(\omega)+\chi_m^{(+)}(\omega,\mathbf{k})\Delta_m''(\omega)\Big]+ \\[2mm] + \coth\dfrac{\hbar\omega}{2k_B T_2}\alpha'\big((\omega-k_x V)\gamma\big)\cdot \\[2mm] \cdot\Big[\chi_e^{(-)}(\omega,\mathbf{k})\Delta_e''(\omega)+\chi_m^{(-)}(\omega,\mathbf{k})\Delta_m''(\omega)\Big] \end{array}\right\} -$$

$$-\frac{\hbar}{\pi^2}\gamma \iiint_{k<\omega/c} d\omega\, dk_x\, dk_y \cos(2\tilde{q}_0 z_0)\cdot\Big\{\Delta_e,\Delta_m\rightarrow\tilde{\Delta}_e,\tilde{\Delta}_m\Big\}+... \qquad (4.35)$$

$$\dot{Q} = -\frac{\hbar}{\pi^2}\gamma \iiint_{k>\omega/c} d\omega\, dk_x\, dk_y \frac{\exp(-2q_0 z_0)}{q_0}\cdot$$

$$\cdot \left\{ \begin{array}{l} \alpha''\big((\omega+k_x V)\gamma\big)(\omega+k_x V)\left(\coth\dfrac{\hbar(\omega+k_x V)\gamma}{2k_B T_1}-\coth\dfrac{\hbar\omega}{2k_B T_2}\right)\cdot \\[2mm] \cdot\Big[\chi_e^{(+)}(\omega,\mathbf{k})\Delta_e''(\omega)+\chi_m^{(+)}(\omega,\mathbf{k})\Delta_m''(\omega)\Big]+ \\[2mm] + \alpha''\big((\omega-k_x V)\gamma\big)(\omega-k_x V)\left(\coth\dfrac{\hbar(\omega-k_x V)\gamma}{2k_B T_1}-\coth\dfrac{\hbar\omega}{2k_B T_2}\right)\cdot \\[2mm] \cdot\Big[\chi_e^{(-)}(\omega,\mathbf{k})\Delta_e''(\omega)+\chi_m^{(-)}(\omega,\mathbf{k})\Delta_m''(\omega)\Big] \end{array}\right\} \qquad (4.36)$$

$$-\frac{\hbar}{\pi^2}\gamma \iiint_{k<\omega/c} d\omega\, dk_x\, dk_y\left[-\frac{\sin(2\tilde{q}_0 z_0)}{\tilde{q}_0}\right]\cdot\Big\{\Delta_e,\Delta_m\rightarrow\tilde{\Delta}_e,\tilde{\Delta}_m\Big\}+...$$

where :

$$q_0 = \left(k^2-\omega^2/c^2\right)^{1/2},\ \tilde{q}_0 = \sqrt{\omega^2/c^2-k^2}\ ,$$

$$q = \left(k^2-\varepsilon(\omega)\mu(\omega)\omega^2/c^2\right)^{1/2},\ \tilde{q} = \left(\varepsilon(\omega)\mu(\omega)\omega^2/c^2-k^2\right)^{1/2}$$

$$\chi_e^{(\pm)}(\omega,\mathbf{k}) = 2(k^2-k_x^2\beta^2)(1-\omega^2/k^2 c^2)+\frac{(\omega\pm k_x V)^2}{c^2},$$

$$\chi_m^{(\pm)}(\omega,\mathbf{k}) = 2k_y^2\beta^2(1-\omega^2/k^2 c^2)+\frac{(\omega\pm k_x V)^2}{c^2},$$

$$\Delta_{\mathrm{m}}(\omega) = \left( \frac{\mu(\omega) q_0 - q}{\mu(\omega) q_0 + q} \right),$$

$$\Delta_{\mathrm{e}}(\omega) = \left( \frac{\varepsilon(\omega) q_0 - q}{\varepsilon(\omega) q_0 + q} \right),$$

Moreover, the functions $\widetilde{\Delta}_{\mathrm{e}}, \widetilde{\Delta}_{\mathrm{m}}$ are similar to $\Delta_{\mathrm{e}}, \Delta_{\mathrm{m}}$ with the replacements $q, q_0 \to \widetilde{q}, \widetilde{q}_0$.

The first terms in Eqs.(4.34)-(4.36) describe the contributions from nonradiative modes of electromagnetic field, $k = \sqrt{k_x{}^2 + k_y{}^2} > \omega/c$, while the second ones – from radiative modes. In the limit $c \to \infty$ Eqs.(4.34)-(4.36) completely agree with (4.20), (4.27), (4.32).

For a resting particle ($V = 0$), Eq. (4.36) reduces to a simpler one, which, however, explicitly takes into account the retardation effect (the second term in figure brackets)

$$\dot{Q} = -\frac{2\hbar}{\pi} \int\limits_0^\infty d\omega \, \omega \left[ \frac{1}{\exp(\hbar\omega/k_B T_1) - 1} - \frac{1}{\exp(\hbar\omega/k_B T_2) - 1} \right] \cdot$$

$$\cdot \int\limits_{k > \omega/c} dk \, k^3 \, \frac{\exp(-2q_0 z_0)}{q_0} \cdot \left\{ \begin{array}{l} 2\alpha''(\omega) \Delta_e''(\omega) + \\ + \left( \dfrac{\omega}{kc} \right)^2 \left[ \alpha'(\omega) \Delta_m''(\omega) - \alpha''(\omega) \Delta_e''(\omega) \right] \end{array} \right\} + \dot{Q}_r \qquad (4.37)$$

where $\dot{Q}_r$ describes the contribution from radiative modes, $k < \omega/c$. At $c \to \infty$, $V = 0$, evidently, (4.37) reduces to (4.33).

With the replacements $\Delta_e(\omega) \to R_P(\omega), \Delta_m(\omega) \to R_s(\omega)$, Eq. (4.37) is similar, but not identical to that one recently obtained in [54, 80]. In this respect, we must note crucial difference between our relativistic expression for the density current, $\mathbf{j} = \partial \mathbf{P} / \partial t + c \operatorname{rot} \mathbf{M}$ ($\mathbf{M}$ is magnetization vector) and that one from Ref. [55] (Eq.(23)), where the term $c \operatorname{rot} \mathbf{M}$ is ignored, but the authors used the expression for the Lorentz force (Eq.(19)), being identical to ours [57, 58]. We return to this point in the next section when comparing numerical calculations, because the corresponding difference drastically influences the involved contribution from $S-$polarized waves.

Furthermore, it is very interesting to get from (4.34) the linear – velocity relativistic correction to the tangential force. Neglecting, for simplicity, the radiative contribution and other ones containing derivatives $\dfrac{d\alpha''}{d\omega}$, $\dfrac{d\varepsilon''}{d\omega}$ under the integrand sign

yields

$$\delta F_x \approx -\frac{\hbar V}{4\pi c^2}\frac{1}{z_0^{\ 3}}\int_0^\infty d\omega\, \omega\, \alpha''(\omega)\, \Delta''(\omega)\left[\coth\frac{\omega\hbar}{2k_B T_2}-\coth\frac{\omega\hbar}{2k_B T_1}\right] \qquad (4.38)$$

One sees that $\delta F_x = 0$ at $T_1 = T_2 = 0$ (and generally, at $T_1 = T_2 = T$), so the basic statement (Section 2) that the velocity – linear drag force is due to higher –order quantum effects, remains to be valid in relativistic treatment, too.

From (4.31), (4.38) we get $\delta F_x / F_x \propto (z_0\omega_0 / c)^2 << 1$, as long as $z_0 << c/\omega_0$, (again, $\omega_0$ is the characteristic frequency). This is within the range of separations corresponding to ordinary nonretarded approximation. An important thing is that $\delta F_x$ may have different sign. In order to see how large is correction (4.38), we must compare it with the nonretarded contributions to the friction force. On the other hand, it is interesting to compare the corresponding heating rates.

### 4.6 Numerical comparison of different approximations

At first, we are going to compare different approximations for the drag tangential force, given by Eqs. (4.29), (4.30) and (4.38). Let us consider the following thermal configurations for the thermal – induced FDF:

i) $T_1 = 0, T_2 = T$ ("cold" particle and "hot" surface), when (4.31) reduces to

$$F_x = -\frac{3\hbar V}{2\pi z_0^{\ 5}}\int_0^\infty d\omega\frac{1}{\exp(\hbar\omega / k_B T)-1}\Delta''(\omega)\frac{d}{d\omega}\alpha''(\omega) \quad , \qquad (4.39)$$

ii) $T_1 = T, T_2 = 0$ ("hot" particle and "cold" surface), when (4.31) reduces to

$$F_x = -\frac{3\hbar V}{2\pi z_0^{\ 5}}\int_0^\infty d\omega\frac{1}{\exp(\hbar\omega / k_B T)-1}\alpha''(\omega)\frac{d}{d\omega}\Delta''(\omega) \quad , \qquad (4.40)$$

iii) thermal equilibrium, $T_1 = T_2 = T$, when the drag force is determined from (4.28).

Furthermore, let us take a spherical particle with polarizability (4.23) . Both for the particle and surface we assume same dielectric function (C10), corresponding to conducting materials. Then, evidently, $\alpha(0) = R^3$ . In addition, we take (4.38) at $T_1 = 0$, $T_2 = T$, while in (4.29) we use parameters

$k_F = 0.95$, $\omega_F = 0.46$, $\omega_P = 0.61$ $a.u.$, relevant to the surface at $r_s = 2$ jellium ; $d \equiv z_0 = 0.5$ and $1\,nm$, $I(d) = 14$ (according to [40]) and $R = 1\,nm$ . Due to the dependence $F_x \sim z_0^{-3}$ in (4.38), the corresponding numerical data have additional relativistic factor, $\xi^2 = (2\pi\sigma z_0 / c)^2$ . In what follows we take $\xi^2 = 0.1$, referring to typical values of $\sigma = 10^{17}\,s^{-1}$, $z_0 = 15.3\,nm$ . Parameter $\xi$ determines role of retardation in the problem. For the data shown in Figs.7(1,b) we also assume $V = 1\,m/\sec$ .

Fig.6 compares the calculated dependencies of the normalized tangential forces ($F_x /(\hbar V R^3 / z_0^5)$) on parameter $a = 2\pi\sigma\hbar / k_B T$ . Lines 1 – 4 correspond to (4.28), (4.39), (4.40) and (4.38), respectively. Horisontal lines correspond to Eq. (4.29) at $z_0 = 0.5$ and $1\,nm$. We see that nonretarded thermal – induced drag FDF (lines 1,2,3) dominates at sufficiently small $a$. This needs either small $\sigma$, or high $T$. For high resistivity materials we have $a << 1$ even at very low temperatures of the order of 1 $K$ or less. In addition, one should note that Eq. (4.29) results in rapidly decreasing force at $z_0 > 0.5\,nm$ ($F_x \sim z_0^{-10}$), so the crossover value of $a$ increases with increasing $z_0$ (compare lines for $z_0 = 0.5$ and $1\,nm$) . Therefore, the quantum – induced FDF are smaller than the thermal ones in a broad practically important range of temperatures. Relativistic correction ( line 5) is small even for normal metals. Also, we see that the thermal –induced FDF (see lines 1, 2, 3) prove to be not very sensitive to the particle –surface temperature gradient.

Figs.7(a,b) demonstrate the force –distance dependencies. Lines 1 – 5 correspond to formulae (4.28), (4.29), (4.39), (4.40) and (4.38). In addition, lines 6 show the relativistic corrections according to paper [55] (Eq.(49) at $a = 1, \sigma = \sigma_t$). From the insert we see that the relativistic correction (at $a = 1$) being determined from Eq.(4.38) (line 5) dominates at about $z_0 \geq 60\,nm$, while that one from [55] –at about $z_0 \geq 20\,nm$. This is due to the different distance dependence ($F_x \propto z_0^{-3}$ in (4.38)

and $F_x \propto z_0^{-1}$ in [55]). The representative case ($\sigma = 10^{17}\, s^{-1}$) corresponds to normal metals, while for more resistive materials the relativistic corrections become negligibly small. Anyhow, at large separations of the order of $50 \div 100\, nm$ the absolute values of FDF prove to be very small irrespectively of the used materials and approximations.

Now let us compare calculations of the heat flow between a spherical probing particle and a flat surface. Quite recently this problem has been discussed in [46, 54]. In particular, as it was pointed out by Pendry [46], the local heating of a surface by an STM tip can be used for local modification of the surface because the heat flow via evanescent modes of the fluctuating electromagnetic field can be by about 9 orders of magnitude larger than that corresponding to black body radiation.

The authors of [54, 55] have reported on crucial influence of retardation effects for normal metals even at very small separations between the bodies: $1\, nm$ for two parallel semiinfinite bodies and $10\, nm$ for a small spherical particle and a flat surface. They claimed that the corresponding effect is dominated by $S-$wave contribution to the heat flow. Both Pendry [46] and Volokitin et. al. [54] have considered the simplest case $V = 0$, when the heat flux can be simply determined using the energy dissipation integral (see (3.15)). We also have considered this case in [91] using Eq. (4.37), which allows to analyze the retardation effect in "pure" form. The proper contribution is defined by the second term in figure brackets of Eq. (4.37), which contains both the contribution from $S-$wave (the term proportional to $\Delta_m''$), and that from $P$- wave (the term proportional to $\Delta_e'' / c^2$).

We have shown [91] that conclusions [54] are in error in view of the following reasons: 1) the contribution from $S-$wave is always smaller than that from $P-$wave; 2) in the range of separations from zero to several $\mu m$, retardation gives rise to smaller heat flux as compared to the nonretarded approximation. As seen from (4.37), the resulting heat flow consists of two terms, of which the first comes from evanescent modes ($k > \omega / c$), while the second one – from radiative modes ($k < \omega / c$). In particular, in thermal configuration "hot" particle - "cold surface", at $\eta = \dfrac{2 z_0 k_B T}{c \hbar} \ll 1$, using the same conditions, the contribution $\dot{Q}_r$ from radiative modes is given by [91]

$$\dot{Q}_{\rm r} \approx -36\frac{\hbar R^3 z_0}{\sigma^{3/2} c^4}\left(\frac{k_B T}{\hbar}\right)^{15/2} \tag{4.41}$$

The range of validity (4.41) is wide, because $\eta = 1$ at $z_0 = 4\,\mu m$, $\sigma \approx 10^{17}\,\sec^{-1}$ and $T = 300$ K. At $z_0 > 4\,\mu m$ the involved integral in (4.37) quickly decreases due to the oscillating function $\sin(2\tilde{q}_0 z_0)$. The contribution from evanescent modes we have calculated numerically. It is worthwhile to compare these results with the nonretarded heat flow which is given by (4.37) in the limit $c \to \infty$. In typical case for normal metals ($k_{\rm B} T / \hbar << 2\pi\sigma$) this results in [46, 54]

$$\dot{Q}_{\rm n\,r} = -\frac{3\pi}{40}\frac{\hbar R^3}{z_0^{\,3}\sigma^2}\left(\frac{k_{\rm B} T}{\hbar}\right)^4 \tag{4.42}$$

In opposite case of high resistivity materials and $k_{\rm B} T / \hbar >> 2\pi\sigma$, we obtain [73, 91]

$$\dot{Q}_{\rm nr} = -\frac{4}{5}k_{\rm B} T\frac{R^3}{z_0^{\,3}}\sigma \tag{4.43}$$

One sees that (4.43) agrees with simple estimation given in Section 2. Fig. 8 shows the dependence of the fraction $\dot{Q}_{\rm S} / \dot{Q}_{\rm P}$ vs. distance, where $\dot{Q}_{\rm S}$ and $\dot{Q}_{\rm P}$ correspond to the total (evanescent and radiative ) $S-$wave and $P$-wave contributions to the heat flow [91]. We see that relative $S-$wave contribution increases with increasing parameter $2\pi\sigma\hbar / k_{\rm B} T$, but it is smaller than that from $P-$wave.

Fig.9 compares the total reduced heat flux (in relation to $\dot{Q}_{\rm nr}$, (4.42)) at different separations and other parameters (lines 1-4). Line 5 is drawn according to work [54] (Eq. (33)) and corresponds to $S-$wave conribution predicted by these authors. We see that these results strongly overestimate the heating rate. This is caused by difference in the basic formulas (cf. (4.37) and (28) from Ref. [54]), and by oversimplified calculation of the involved integral over the wave numbers [54], as well.

So, finally, we conclude that retardation effect leads to smaller heat flow and is very small in many cases of practical importance. The corresponding corrections to the friction force are also small. This agrees with the results [46] and seems to be physically clear : retardation implies less effective interaction between the bodies Concerning the retardation and relativistic effects together, a question is : to what extent the nonrelativistic formulas are valid in calculations of the FDF and heat flow ?

The above results show that they are practically suitable in estimations of the upper bound of the heat flow up to the separations of the order of several $\mu m$, while in calculations of FDF – up to about $0.1 \mu m$. However, it is possible that more detailed analysis of dynamic regime using general equations (4.34) and (4.36) will result in somewhat different estimations.

### 4.7 Nonlinear velocity-resonance effects

It must be noted that some kind of resonance coupling is possible in the nonlinear velocity regime [73, 92]. Let the functions $\alpha''(\omega), \Delta''(\omega)$ to have the form

$$\alpha''(\omega) = \frac{\pi e^2 f_{0n}}{2m\omega_{0n}} \delta(\omega - \omega_{0n}),$$ (4.44)

$$\Delta''(\omega) = \frac{\pi \omega_s}{2} \delta(\omega - \omega_s),$$ (4.45)

where $\omega_{0n}$ and $f_{0n}$ are the atomic line –frequency and oscillator strength for the atomic transition $0 \rightarrow n$, $\omega_s$ is the surface excitation frequency (for instance, the surface plasmon frequency). Assuming the corresponding line –widths to be much less than the difference $|\omega_s - \omega_{0n}|$ and inserting (4.44), (4.45) into (4.26) and (4.32) yields

$$F_x = -\frac{\hbar e^2 f_{0n}}{2m} \frac{\omega_s}{\omega_{0n}} \frac{t^4 \left(K_0(t) + K_2(t)\right)}{\Delta\omega z_0^{\ 4}} \left[ \frac{1}{\exp(\hbar\omega_{0n}/k_B T_1) - 1} - \frac{1}{\exp(\hbar\omega_s/k_B T_2) - 1} \right]$$ (4.46)

$$\dot{Q} = -\frac{\hbar e^2 f_{0n}\omega_s}{2m} \frac{t^3 \left(K_0(t) + K_2(t)\right)}{\Delta\omega z_0^{\ 3}} \left[ \frac{1}{\exp(\hbar\omega_{0n}/k_B T_1) - 1} - \frac{1}{\exp(\hbar\omega_s/k_B T_2) - 1} \right]$$ (4.47)

where $\Delta\omega = |\omega_s - \omega_{0n}|$, $t \equiv 2\Delta\omega z_0 / V$, $K_{0,2}(t)$ are the modified Bessel functions.

The functions $F_x(t)$ and $\dot{Q}(t)$ proved to have maxima at $t = 3.25$ and $t = 2.1$, respectively. The coresponding relations $V = 0.62\Delta\omega z_0$ and $V = 0.95\Delta\omega z_0$ may be considered as the resonance dynamic conditions.

The directions of the tangential force and heat flow depend on temperature factors in the square brackets of (4.47). For a neutral atom, as $T_1 = 0$, the tangential force becomes accelerating (if $T_2 > 0$), and the heat flow is positive, too. Therefore, an atom is heated. We attribute this thermal effect to a some kind of "Lamb shift" of the atomic levels caused by atom–surface coupling. If the surface has low–frequency absorption peaks tuned to the temperature $T_2$, the corresponding effect seems to be observable. Really, assuming the following set of parameters :

$\omega_{0n} = 5\,eV$, $\omega_s = 0.1\,eV$, $T_1 = 0$, $T_2 = 600\,K$, $z_0 = 1\,nm$, $V = 3 \cdot 10^6\,m/s$, $f_{0n} = 1$, from (4.46) we get the energy increment $dE/dx = 0.2\,eV/\mu m$. Such an effect is really detectable using the modern time –off –flight spectroscopy technique. At the same time, the heat flow equals $3 \cdot 10^{-8}\,W$.

Also, it is interesting to compare the corresponding tangential and normal forces. Neglecting the velocity dependence of the vdW energy, we write it in the simple form [8]

$$U(z_0) = -\frac{\hbar\omega_{0n}\alpha(0)}{8z_0^{\,3}} \frac{\omega_s/\omega_{0n}}{1 + \omega_s/\omega_{0n}} \tag{4.48}$$

where $\alpha(0) = e^2 f_{0n}/m\omega_{0n}^{\,2}$ is the static atomic polarizability. Using (4.46) and (4.48) yields

$$K = \max\left|F_x/F_z\right| \approx 10\,\frac{\omega_{0n} + \omega_s}{\Delta\omega}\left|\frac{1}{\exp(\hbar\omega_{0n}/k_B T_1) - 1} - \frac{1}{\exp(\hbar\omega_s/k_B T_2) - 1}\right| \tag{4.49}$$

From (4.49) we obtain $K \gg 1$ if the temperature factor is of the order of 1. This result is not trivial, because usually, FDF are thought to be much smaller than ordinary vdW forces (see also Section 2).

## 4.8 Spatial dispersion effects

The spatial dispersion effects can be included in the theory if use is made of nonlocal surface –response functions (see Section 3.3 and Appendix C). In particular, for parallel movement of a charged particle, using (3.1), (3.22), (3.23) and taking the limit $V_\perp \to 0$ yields [86] ($V_{II} = V$)

$$F_x = -\frac{2(Ze)^2}{\pi V^2}\int_0^\infty dq_y \int_0^\infty d\omega\,\frac{\omega}{q}\exp(-2qz_0)\Delta''(q, \omega) \tag{4.50}$$

where $z_0$ is the distance from the surface, $q^2 = q_y^{\,2} + \omega^2/V^2$. In local case Eq.(4.50) reduces to (4.2) and (4.4). Therefore, a simple way to generalize the formulas for the charged particles which were obtained in the case of nonlocal response, is to make the replacement $\Delta(\omega) \to \Delta(q, \omega)$. This is appropriate in calculations both conservative and dissipative forces. However, in the case of constant and fluctuating dipoles one needs to get an additional evidence in favour of this modification [93].

In the case of moving constant dipole, the starting equations of SRM takes the form (cf. with (3.21)-(3.23))

$$-k^2 \phi_{\omega \mathbf{k}} = 8\pi^2 \left[ i\delta(\omega - \mathbf{q}\mathbf{V})\mathbf{k}\mathbf{d} + i\delta(\omega - \mathbf{q}\mathbf{V}')\mathbf{k}\mathbf{d} + \rho_s(\mathbf{q}, \omega) \right], \; z < 0;$$

$$\phi_{\omega \mathbf{k}} = \frac{8\pi^2 \rho_s(\mathbf{q}, \omega)}{\varepsilon(\mathbf{k}, \omega)k^2} \;, \quad z > 0; \qquad\qquad (4.51)$$

$$\rho_s(\mathbf{q}, \omega) = -2i \frac{\delta(\omega - \mathbf{q}\mathbf{V})}{(\pi / k + I_0)} \int_{-\infty}^{+\infty} dk_z \frac{\mathbf{k}\mathbf{d}}{k_z^2 + q^2} \qquad\qquad (4.52)$$

where $I_0$ is defined by (3.24) and $\mathbf{V} = \mathbf{V}'$ in the case of parallel motion. As it is seen from (4.52), the corresponding integral diverges at $d_z \neq 0$. This is principal shortcoming of SRM. For other projections of the dipole moment we have no problems, and (assuming $d_z = 0$) using (4.51), (4.52) yields (subtracting the Fourier-component of the unscreened potential, $\dfrac{4\pi \rho_s(\mathbf{q}, \omega)}{k^2}$)

$$\phi^{\mathrm{ind}}_{\omega \mathbf{q}}(z_0) = \frac{4\pi^2 i}{q} \Delta(q, \omega) \mathbf{d}\mathbf{q} \exp(-2qz_0)\delta(\omega - \mathbf{q}\mathbf{V}) \qquad\qquad (4.53)$$

Comparing (4.53) with the analogous expression in local case (see (A16)) we note their equivalence after the replacements $\mathbf{d}\mathbf{q} \to d_x q_x + d_y q_y - i d_z k_z$, $\Delta(q, \omega) \to \Delta(\omega)$. Therefore, Eq.(4.53) can be used in more general case when the dipole moment has an arbitrary orientation in space. Additionally, this is substantiated by calculation [86], where the authors found an exact solution of the problem (at $d_z \neq 0, d_x = d_y = 0$) using Eq.(3.25) for two separate bare charges situated at the distances $z$ and $(z + d_z)$ from the surface.

Analogously, the induced potential of the fluctuating dipole is given by [93]:

$$\phi^{\mathrm{ind}}_{\omega \mathbf{q}}(z_0) = \frac{2\pi i}{q} \Delta(q, \omega) \begin{pmatrix} d^{\mathrm{sp}}_x (\omega - \mathbf{q}\mathbf{V})q_x + d^{\mathrm{sp}}_y (\omega - \mathbf{q}\mathbf{V})q_y - \\ -i d^{\mathrm{sp}}_z (\omega - \mathbf{q}\mathbf{V})k_z \end{pmatrix} \exp(-2qz_0) \qquad (4.54)$$

where $\mathbf{d}^{\mathrm{sp}}(\omega - \mathbf{q}\mathbf{V})$ is the Fourier-transform of the fluctuating dipole moment $\mathbf{d}(t)$. The foregoing calculations of the conservative and dissipative forces completely repeat the corresponding ones in the local case. Therefore, all the basic formulas obtained in Sections 4.1 - 4.7 are generalized after the simple replacement $\Delta(\omega) \rightarrow \Delta(q, \omega)$.

In order to study an influence of the spatial dispersion on dissipative forces, let us consider a particle (bare charge, dipole molecule, spherical atom) moving above conducting surface with velocity $V << V_F$. Assuming, for simplicity, the asymptotic limit $z_0 >> 1/2q_{TF}$ , with $V_F$ and $q_{TF}$ being the Thomas-Fermi velocity and and wave-vector, one can write down the imaginary part of the surface response function in the form (see [86] and Appendix C)

$$\Delta''(q, \omega) = \omega \left( \frac{2}{\sqrt{3}\omega_P} \frac{q}{q_{TF}} \left( \ln(2q_{TF}/q) - \frac{1}{2} \right) + \frac{1}{2\pi\sigma} \right) \tag{4.55}$$

where $\sigma$ is the static conductance , $\omega_P$ is the bulk plasma frequency. Also, in this low –velocity and low –frequency limit, the imaginary part of the atomic polarizability is given by ( $\gamma_{0n}$ is the involved line-width and hereafter we use the atomic units )

$$\alpha''(\omega) \approx \omega \sum_n \frac{f_{0n}\gamma_{0n}}{\omega_{0n}^3} \tag{4.56}$$

Furthermore, inserting Eqs.(4.55), (4.56) into (4.2), (4.10), (4.27) yields [73]

i)    charged particle

$$F_x \approx -\frac{(Ze)^2 V}{16\pi\sigma z_0^3} - \frac{3}{2\pi} \frac{(Ze)^2 V}{(q_{TF}z_0)^4} \ln(0.692 q_{TF}z_0) \tag{4.57}$$

ii)    dipole molecule

$$F_x \approx -\frac{3}{64\pi\sigma} \frac{(3d_x^2 + d_y^2 + 4d_z^2)V}{z_0^5} -$$
$$-\frac{15}{8\pi} \frac{(3d_x^2 + d_y^2 + 4d_z^2)q_{TF}^2 V}{(q_{TF}z_0)^6} \ln(0.44 q_{TF}z_0) \tag{4.58}$$

iii)    neutral atom ( $2\pi\sigma > k_B T$  and  $V/z_0 << k_B T$ )

$$F_x = -\frac{3k_1 k_2}{2\pi} V k_B^2 (T_1^2 + T_2^2) \left[ \frac{5}{\sqrt{3}\omega_P} \frac{q_{TF}^5}{(q_{TF}z_0)^6} \ln(0.44 q_{TF}z_0) + \frac{1}{2\pi\sigma z_0^5} \right] \tag{4.59}$$

where $k_1 = 1.645$, $k_2 = \sum_n f_{0n}\gamma_{0n}/\omega_{0n}^3$.

From   (4.57)-(4.59) one sees that in addition to the ordinary Ohm's law dissipation term ( $F_x \sim 1/\sigma$ ), each of the expressions contains also the contribution from the electron –hole excitations. Eqs.(4.57), (4.58) completely agree with the ones obtained in [86].

For bare charges and dipole molecules, the energy loss (friction) due to electron-hole excitations dominates at $z_0 < 0.5 \; nm$ (for an aluminium –like surface, $r_s = 2$) [86]. For neutral atoms above the metal surface ($\sigma = 10^{17} \; \sec^{-1}$, $\omega_p = 10^{15} \; s^{-1}$, $q_{\mathrm{TF}} = 1 \; a.u.$) the Ohm's dissipation term in Eq. (4.59) is smaller than that from the electron –hole excitations at $z_0 < 5 \; nm$. To get a numerical estimation, one can assume $k_2 \approx \alpha(0) \langle \gamma_{0n} / \omega_{0n} \rangle \approx 0.01\alpha(0)$, where $\alpha(0)$ is the static polarizability. Then at

$T_1 = 0$, $T_2 = 300 \; K$, $\alpha(0) = 27 \; a.u.$ ($Xe$ atom), $z_0 = 10 \; a.u.$ and $V = 1 \; m/s$

we get from (4.59) $F_x = -0.4 \cdot 10^{-19} \; N$. For high resistivity materials ($\sigma = 10^{10} \div 10^{11} \; \sec^{-1}$) the second term in Eq.(4.59) strongly dominates and the friction force increases by 5 to 6 orders of magnitude.

Evidently, more features could be expected in the nonlinear – velocity regime ($V/z_0 > 2\pi\sigma$) and for other types of surface excitations, especially in the low-frequency range.

## 4.9 Structural effects

The electron density of a crystalline solid in the surface region is not uniform and manifests lattice periodicity in the lateral directions. So does the dielectric response function, thus giving rise to an anysotropy of the friction and energy loss. In general case, for a given trajectory of the moving particle above a surface, the energy loss can be written in the form

$$\frac{dE}{dx}(\mathbf{r}) = \left(\frac{dE}{dx}\right)_0 + \sum_{\mathbf{G}} \left(\frac{dE}{dx}\right)_{\mathbf{G}} \cdot \exp(\mathrm{i}\,\mathbf{G}\mathbf{r}) \qquad (4.60)$$

where $\left(\dfrac{dE}{dx}\right)_0$ corresponds to a uniform electron density, $\left(\dfrac{dE}{dx}\right)_{\mathbf{G}}$ is the Fourier-component corresponding to the two –dimensional reciprocal lattice vector $\mathbf{G}$, $\mathbf{r}$ is the particle position vector.

The proper generalization of the theory can be performed by analogy with that one which has been used in calculation of the energy loss of the channeling particles [94]. Thus, the structure –dependent surface response function will be defined by [95]

$$\Delta(\mathbf{q},\omega,\mathbf{G}) = \Delta(\mathbf{q},\omega) f(\mathbf{G}) \mathbf{e}(\mathbf{q}) \mathbf{e}(\mathbf{q}+\mathbf{G}) \qquad (4.61)$$

where $\Delta(\mathbf{q},\omega)$ is the surface response function (3.26), $f(\mathbf{G})$ is the normalized Fourier–component of the electron density, $\mathbf{e}(\mathbf{q})$ and $\mathbf{e}(\mathbf{q}+\mathbf{G})$ are the unit vectors in directions of vectors $\mathbf{q}$ and $\mathbf{q}+\mathbf{G}$. Evidently, the product $\mathbf{e}(\mathbf{q}) \mathbf{e}(\mathbf{q}+\mathbf{G})$ takes the form

$$\mathbf{e}(\mathbf{q}) \mathbf{e}(\mathbf{q}+\mathbf{G}) = \frac{q + G\cos\phi}{(q^2 + G^2 + 2qG\cos\phi)^{1/2}} \qquad (4.62)$$

where $\phi$ is the angle between $\mathbf{q}$ and $\mathbf{G}$.

Assuming radial symmetry of a typical atomic electron density distribution, $n(r)$, the Fourier–factor $f(\mathbf{G})$ will be given by

$$f(\mathbf{G}) = \frac{\int\limits_0^\infty dr\, r n\!\left(\sqrt{r^2 + z_0{}^2}\right) J_0(Gr)}{2\pi \int\limits_0^\infty dr\, r n\!\left(\sqrt{r^2 + z_0{}^2}\right)} \qquad (4.63)$$

where $J_0(x)$ is the Bessel function, $z_0$ being the distance to surface.

For instance, assuming $n(r) = \dfrac{Z\lambda^2}{4\pi a^2}\exp(-\lambda r/a)$ ($\lambda$ is the model parameter, and $a$ is the screening length), from (4.63) we get

$$f(G) = \frac{\lambda}{2\pi(\lambda^2 + G^2 a^2)^{1/2}}\exp\!\left(-z_0((\lambda^2 + G^2 a^2)^{1/2} - \lambda)\right) \qquad (4.64)$$

As the function $f(\mathbf{G})$ has the exponential factor, the contributions to the tangential force from large $\mathbf{G}$ will quickly tend to zero.

Finally, one needs to mention the results [43], where the authors have got a structure–dependent correction to the quantum–induced drag fluctuation force between two closely spaced solids, using the high–order quantum perturbation theory. The corresponding force was obtained to have the distance dependence $F \sim \exp(-2Gd)/d^4$, where $d$ is the spacing between the surfaces.

## 5. Conservative and dissipative forces between atomic particles and cylindrical surface

Study of FDF between the particles (nanoprobes) and curved surfaces is of considerable interest not only for the friction and energy loss problems, but also in applications related with transmission of particle beams through the nanotubes [24, 25] , and to investigating   adsorbing and frictional properties of the fullerene substance and porous structures [96].

Generally speaking, consideration of the fluctuation forces (conservative and dissipative) for curved surfaces involves additional difficulties of mathematical origin. Due to this, there is a comparatively small number of publications on this subject (see, for instance, reviews [97, 98] devoted to vdW forces). A problem of physical adsorption on the curved surfaces has been studied by Schmeits and Lucas [99]. Recently, using the method of conformal mapping, the vdW forces and heat flow for bodies terminated by surfaces with various curvatures (in the plane –parallel geometry) have been also calculated in [81].

For cylindrical surface, the dissipative forces (at zero temperature) were firstly considered by one of us  [100, 101]. In this work, we are reviewing more general results [102, 103], corrected with account of the recent considerations [56, 73] for the attractive and dissipative forces acting on nonrelativistic particles moving parallel to an axis of the convex/concave cylindrical surface at an arbitrary temperature.

We use the same restrictions and method of calculation as in the case of a particle and a flat surface. Fig.10a shows the case of convex cylindrical surface, Fig.1b – the case of concave surface which encloses polarizable medium. The assumptions of validity of the dipole approximation and neglecting retardation implies $r_0 << h << c / \omega_0$, where  $h = |R - a|$, $R$ and $a$ being the radial  distance from the cylinder axis and the cylinder radius, respectively, $r_0$ is the characteristic atomic radius, $\omega_0$ is the characteristic frequency. Eq.(3.1) is used in the case of a charged particle, while for dipoles, instead  of (3.2) – (3.4), we employ the equations for the tangential (dissipative) and normal (conservative) components of the force:

$$F_z = \left\langle \nabla_z (\mathbf{dE}) \right\rangle, \ \ F_R = \left\langle \nabla_R (\mathbf{dE}) \right\rangle \, , \tag{5.1}$$

Besides, we use obvious relations for the conservative potential

$$F_R = -\frac{\partial U}{\partial R}, \quad U = -1/2 \left\langle \mathbf{dE} \right\rangle \tag{5.2}$$

The needed solutions to the Poisson's equation for the electric potential are obtained in Appendix A, the necessary correlators of the dipole moments and electric fields – in Appendix D.

## 5.1 Charged particle

In the case of motion parallel to a symmetry axis of the convex cylindrical surface the components of the force are given by

$$F_R(R,V) = \frac{4(Ze)^2}{\pi} \sum_{n=0}^{\infty} \int_0^{\infty} dk \, k \, K_n(kR) K_n'(kR) \Delta_n'(kV) \,, \tag{5.3}$$

$$F_z(R,V) = -\frac{4(Ze)^2}{\pi} \sum_{n=0}^{\infty} \int_0^{\infty} dk \, k \, K_n^{\,2}(kR) \Delta_n''(kV) \qquad, \tag{5.4}$$

$$\Delta_n(\omega) = \frac{(\varepsilon(\omega)-1)I_n(ka)I_n'(ka)}{\varepsilon(\omega)I_n'(ka)K_n(ka) - I_n(ka)K_n'(ka)} \,, \tag{5.5}$$

where $K_n(x), I_n(x)$ are the cylindrical finctions of order n , the primed functions define the corresponding derivatives. For simplicity, we omit argument $ka$ in $\Delta_n(\omega)$ ; $\Delta_n'$ and $\Delta_n''$ define the proper real and imaginary parts. Using (5.3), the dynamic conservative potential takes the form

$$U(R,V) = -\frac{2(Ze)^2}{\pi} \sum_{n=0}^{\infty} \int_0^{\infty} dk \, K_n^{\,2}(kR) \Delta_n'(kV) \tag{5.6}$$

In the case of motion inside a cylindrical channel, one has to make the replacements $K_n(x) \leftrightarrow I_n(x)$ in (5.3) -(5.6).

## 5.2 Dipole molecule

For a dipole molecule moving parallel to a convex cylindrical surface we get

$$U(R,V) = -\frac{2}{\pi R^2} \sum_{n=0}^{\infty} \int_0^{\infty} dk \, K_n^{\,2}(kR) \Delta_n'(kV) \cdot$$
$$\cdot \left[ n^2 d_\phi^{\,2} + (kR)^2 d_z^{\,2} + (kR)^2 \Phi_n^{\,2}(kR) d_r^{\,2} \right] \tag{5.7}$$

$$F_z(R,V) = -\frac{4}{\pi R^2} \sum_{n=0}^{\infty} \int_0^{\infty} dk \left[ n^2 d_\phi^{\,2} + (kR)^2 d_z^{\,2} + (kR)^2 \Phi_n^{\,2}(kR) d_r^{\,2} \right] \cdot$$
$$\cdot \Delta_n''(kV) K_n^{\,2}(kR) \,, \tag{5.8}$$

where $d_\phi, d_r, d_z$ are the components of the dipole moments in the cylindrical coordinate system (see Fig.10a) , the function $\Phi_n(z)$ is given by

$$\Phi_n(z) \equiv \frac{d}{dz} \ln K_n(z) \tag{5.9}$$

Again, in the case of motion inside a cylindrical channel, one has to make the replacement $K_n(x) \leftrightarrow I_n(x)$.

### 5.3 Neutral spherical particle (ground state atom)

In the most general case, when the particle and surface have different temperatures $T_1$ and $T_2$, the resulting formulae have the form

$$U(R,V) = -\frac{\hbar}{\pi^2 R^2} \sum_{n=0}^{\infty} \iint d\omega dk \, K_n^{\,2}(kR)\left[n^2 + (kR)^2 + (kR)^2 \Phi_n^{\,2}(kR)\right] \cdot$$
$$\cdot \left\{ \begin{array}{l} \coth(\omega\hbar/2k_B T_1)\alpha''(\omega)\left[\Delta_n'(\omega+kV) + \Delta_n'(\omega-kV)\right] + \\ + \coth(\omega\hbar/2k_B T_2)\Delta_n''(\omega)\left[\alpha'(\omega+kV) + \alpha'(\omega-kV)\right] \end{array} \right\}$$

(5.10)

$$F_z(R,V) = -\frac{2\hbar}{\pi^2 R^2} \sum_{n=0}^{\infty} \iint d\omega dk \, k K_n^{\,2}(kR)\left[n^2 + (kR)^2 + (kR)^2 \Phi_n^{\,2}(kR)\right] \cdot$$
$$\cdot \left\{ \begin{array}{l} \coth(\omega\hbar/2k_B T_1)\,\alpha''(\omega)\left[\Delta_n''(\omega+kV) - \Delta_n''(\omega-kV)\right] + \\ + \coth(\omega\hbar/2k_B T_2)\Delta_n''(\omega)\left[\alpha''(\omega+kV) - \alpha''(\omega-kV)\right] \end{array} \right\}$$

5.11

Eq. (5.10) generalizes the known one for the static vdW potential of attraction of an atom to the cylindrical surface at $T$=0 . Really, taking $V$=0, $T_1 = T_2 = 0$ we get from (5.10)

$$U(R) = -\frac{2\hbar}{\pi^2 R^2} \sum_{n=0}^{\infty} \int_0^{\infty} dk \, K_n^{\,2}(kR)\left[n^2 + (kR)^2 + (kR)^2 \Phi_n^{\,2}(kR)\right]$$
$$\cdot \, \text{Im}\int_0^{\infty} d\omega \alpha(\omega)\Delta_n(\omega)$$

(5.12)

Turning the contour of integration in (5.12) over the angle $90^0$, so that it will coincide the upper imaginary half –axis, the frequency integral in Eq. (5.12) reduces to

$$\text{Im}\int_0^{\infty} d\omega \, \alpha(\omega)\,\Delta_n(\omega) = \int_0^{\infty} d\omega \, \alpha(i\omega)\,\Delta_n(i\omega) \,,$$

(5.13)

Therefore, with account of (5.13), Eq.(5.12) takes the form equivalent to [99, 104, 105]

$$U(R) = -\frac{\hbar}{\pi^2 R^2} \sum_{n=0}^{\infty} \int_0^{\infty} dk \, K_n^{\,2}(kR)\left[n^2 + (kR)^2 + (kR)^2 \Phi_n^{\,2}(kR)\right]$$
$$\cdot \, \frac{\alpha(i\omega)\,(\varepsilon(i\omega) - 1)I_n'(ka)I_n(ka)}{\varepsilon(i\omega)K_n(ka)I_n'(ka) - K_n'(ka)I_n(ka)}$$

(5.14)

The corresponding dynamic corrections to the potential, as in the case of flat surface, will contain only even powers of velocity. Moreover, from (5.11), to the linear - velocity order, the tangential force is given by

$$F_z(R,V) = -\frac{2\hbar V}{\pi^2 R^2} \sum_{n=0}^{\infty} \int_0^{\infty} dk\, k^2 K_n{}^2(kR)\, C_n(k) \left[ n^2 + (kR)^2 + (kR)^2 \Phi_n{}^2(kR) \right] \quad (5.15)$$

$$C_n(k) = \int_0^{\infty} d\omega \left\{ \left[ \coth\left(\frac{\omega\hbar}{2k_B T_1}\right) - \coth\left(\frac{\omega\hbar}{2k_B T_2}\right) \right] \left[ \begin{array}{l} \alpha''(\omega)\dfrac{d\Delta_n''(\omega)}{d\omega} - \\ \Delta_n''(\omega)\dfrac{d\alpha''(\omega)}{d\omega} \end{array} \right] + \right.$$
$$\left. + \alpha''(\omega)\Delta''(\omega)\left(-\frac{d}{d\omega}\right)\left[ \coth\left(\frac{\omega\hbar}{2k_B T_1}\right) + \coth\left(\frac{\omega\hbar}{2k_B T_2}\right) \right] \right\} \quad (5.16)$$

One sees, that tangential force has two terms, one of which has an arbitrary sign (the first term in the figure brackets of (5.16)), while the second one has a constant sign. At $T_1 = T_2$ the first contribution vanishes and we have the dissipative force, $F_z < 0$. The similar result we have got in the case of flat surface (see (4.31)).

A transition to the case of concave cylindrical surface is also performed using the replacement $K_n(x) \leftrightarrow I_n(x)$.

### 5.4 An atom moving parallel to a thin thread

In this case the general formula (5.11) is simplified, as $a << R$ and the main contribution to the integral over $k$ comes from $k \leq 1/R$. Then, at $a << R$ we get $ka << 1$ and $\Delta_n(\omega)$ can be expanded over $ka$. Using the well-known asymptotic relations [106]

$$I_n(z) \approx \frac{(z/2)^n}{\Gamma(n+1)} \ ,$$

$$K_n(z) \approx (-1)^{n+1} I_n(z)\ln(\gamma z/2) + \frac{(n-1)!}{2}(2/z)^n \ , \quad \gamma \approx 1.781 \ , \ z << 1$$

$$(5.17)$$

and taking into account (5.5) yields

$$\Delta_0(\omega) = \frac{(ka)^2}{2} \frac{\varepsilon(\omega) - 1}{1 - \varepsilon(\omega)\dfrac{(ka)^2}{2}\ln(\gamma\, ka/2)} \qquad 5.18)$$

$$\Delta_1(\omega) = \frac{(ka)^2}{2} \frac{\varepsilon(\omega) - 1}{\varepsilon(\omega) + 1} \qquad (5.19)$$

The next order functions $\Delta_n(\omega)$ (at $n \geq 2$) contain the additional small factor $(ka)^{n-1}$ and can be omitted in the sum (5.15). Moreover, in the case of a dielectric thread,

$$\varepsilon(\omega)\frac{(ka)^2}{2}\ln(\gamma\, ka\, /\, 2) << 1, \tag{5.20}$$

and the corresponding term in the denominator of (5.18) can be omitted, too. A further simplification (5.20) is straightforward, because the corresponding integrals are explicitly evaluated with account of the table integral [106]

$$\int_0^\infty dx\, x^{a-1} K_m(x) K_n(x) = \frac{2^{a-3}}{\Gamma(a)}\Gamma\left(\frac{a+m+n}{2}\right)\Gamma\left(\frac{a+m-n}{2}\right)\Gamma\left(\frac{a-m+n}{2}\right)\cdot$$

$$\cdot\, \Gamma\left(\frac{a-m-n}{2}\right), \quad \mathrm{Re}\, a > |\mathrm{Re}\, m| + |\mathrm{Re}\, n| \tag{5.21}$$

Using these results, to linear velocity order, Eq. (5.15) reduces to

$$F = -\frac{\hbar a^2 V}{R^7}\int_0^\infty d\omega\left\{0.33\left[\alpha''(\omega)\varepsilon''(\omega)\left(-\frac{d}{d\omega}\right)\left[\begin{matrix}\coth(\hbar\omega\,/\,2k_B T_1)+\\+\coth(\hbar\omega\,/\,2k_B T_2)\end{matrix}\right]\right]+\right.$$

$$+\left[\coth(\hbar\omega\,/\,2k_B T_1)-\coth(\hbar\omega\,/\,2k_B T_2)\right]\left(\alpha''\frac{d\varepsilon''}{d\omega}-\varepsilon''\frac{d\alpha''}{d\omega}\right)+$$

$$+1.1\left[\alpha''(\omega)\Delta''(\omega)\left(-\frac{d}{d\omega}\right)\left[\coth(\hbar\omega\,/\,2k_B T_1)+\coth(\hbar\omega\,/\,2k_B T_2)\right]+\right.$$

$$\left.\left.+\left[\coth(\hbar\omega\,/\,2k_B T_1)-\coth(\hbar\omega\,/\,2k_B T_2)\right]\left(\alpha''(\omega)\frac{d\Delta''(\omega)}{d\omega}-\Delta''(\omega)\frac{d\alpha''(\omega)}{d\omega}\right)\right]\right\} \tag{5.22}$$

At $T_1 = T_2 = T$ (5.22) takes the simplest form

$$F = -\frac{2\hbar a^2 V}{R^7}\int_0^\infty d\omega\, \alpha''(\omega)\left(0.33\,\varepsilon''(\omega)+1.1\Delta''(\omega)\right)\left(-\frac{d}{d\omega}\right)\coth\left(\frac{\hbar\omega}{2k_B T}\right) \tag{5.23}$$

For $dc$ conductors both terms in the denominator (5.18) must be taken into account, since $\varepsilon(\omega) \to 4\pi\sigma\mathrm{i}\,/\,\omega$ at $\omega \to 0$. Then at $\omega >> 2\pi\sigma(a\,/\,R)^2|\ln(\gamma a\,/\,2R)|$ from (5.15) we get

$$F \approx -8.3\frac{\hbar a^2\sigma V}{R^7}\int_0^\infty d\omega\left[\alpha''(\omega)\coth(\hbar\omega\,/\,2k_B T_1)+\omega\frac{d\alpha''(\omega)}{d\omega}\coth(\hbar\omega\,/\,2k_B T_2)\right] \tag{5.24}$$

I.e. the force –distance dependence proves to be the same as for the dielectric thread. One can show that at $T_1 = T_2 = 0$ both Eq.(5.22) and Eq.(5.24) result in zero drag force.

For an isolated atomic string, parameter $a$ can be interpreted as a length of characteristic electron density decay in radial direction from the string axis.

## 6. Dissipative interactions nanotip-flat surface and between two flat surfaces

### 6.1 An additive approximation

In order to strictly calculate FDF between a nanoprobe with arbitrary shape and a plane (or curved) surface in the framework of the above developed theory, one must determine equilibrium fluctuation spectrum of the electromagnetic field in the gap between contacting bodies. The involved geometrical restrictions and resultant mathematical complexity are due to principal features of the fluctuation – induced forces : they cannot be obtained from a pairwise sum of two -body potentials.

Fortunately, in the case of conservative vdW forces, the assumption of additive interaction between individual particles is a close approximation, which allows one to correctly calculate the dependence of the forces upon the spacing between the solids: only the interaction constant is affected by this approximation [107, 108]. For a convex nanoprobe and a flat surface, this constant, as calculated in additive approximation, is no more than 5 to 20 % in error and can be additionally corrected using the effective renormalization [108]. As a suitable working hypothesis, the FDF are assumed to be additive, too [48, 54, 55, 59, 109, 110].

We assume that the probe has the form of a paraboloid of revolution which is defined by the canonical equation $z = d + (x^2 + y^2)/2R$, where $z$ is measured from the sample surface, $d$ is the minimum distance between the surface and the probe's apex, and $R$ is the probe's curvature radius. Using the Clausius-Mossotti relation (being justified for materials with cubic symmetry of lattice), the atomic polarizability is expressed in terms of the dielectic function of the probe's material,

$$\alpha''(\omega) = \frac{3}{4\pi N} \mathrm{Im} \frac{\varepsilon_1(\omega) - 1}{\varepsilon_1(\omega) + 2}, \qquad (6.1)$$

where $N$ is the volume concentration of atoms, $\varepsilon_1(\omega)$ is the corresponding dielectric function. In what follows, the dielectric functions of the probe and the surface under investigation are labeled by indices "1" and "2", respectively. In the case of a small spherical particle, instead of (6.1) one can use (4.23). After substituting Eq. (6.1) into

Eq. (4.31) and integrating over the probe's volume, the resultant lateral force is given by

$$F_x = -\frac{3}{16\pi} \frac{\hbar R V}{d^3} J(\varepsilon_1(\omega), \varepsilon_2(\omega)) \qquad (6.2)$$

where the structure of the overlap integral $J(\varepsilon_1(\omega), \varepsilon_2(\omega))$ is identical to the quotient ones in Eq. (4.31) with the replacement $\alpha''(\omega)$ by $\left(\frac{3}{4\pi N} \widetilde{\Delta}''(\omega)\right)$,

$$\widetilde{\Delta}''(\omega) = \operatorname{Im} \frac{\varepsilon_1(\omega) - 1}{\varepsilon_1(\omega) + 2} \qquad (6.3)$$

In addition, it must be noted that in derivation of Eq. (6.2) the upper limit of integration over the tip height was taken to be infinity due to the large aspect ratio. At $\hbar\omega << k_B T$, with account of (4.31), (6.1), (6.3), Eq. (6.2) reduces to

$$F_x = -\frac{3}{16\pi} \frac{k_B T R V}{d^3} J(\varepsilon_1(\omega), \varepsilon_2(\omega)) \qquad (6.4)$$

$$J(\varepsilon_1(\omega), \varepsilon_2(\omega)) = \begin{cases} \int_0^\infty d\omega \dfrac{\widetilde{\Delta}_1''(\omega)\Delta_2''(\omega)}{\omega^2} & , \ T_1 = T_2 = T \qquad (a) \\[2mm] \int_0^\infty \dfrac{d\omega}{\omega} \widetilde{\Delta}_1''(\omega) \dfrac{d}{d\omega}\Delta_2''(\omega) & , \ T_1 = T, \ T_2 = 0 \qquad (b) \\[2mm] \int_0^\infty \dfrac{d\omega}{\omega} \Delta_2''(\omega) \dfrac{d}{d\omega}\widetilde{\Delta}_1''(\omega) & , \ T_1 = 0, \ T_2 = T \qquad (c) \end{cases}$$

(6.5)

At $\hbar\omega >> k_B T$, the overlap integral $J(\varepsilon_1(\omega), \varepsilon_2(\omega))$ can be evaluated for simple dielectric functions, too. At arbitrary temperatures of the tip and sample (or when the spatial dispersion effects are taken into account) we should use general Eqs. (4.26), (4.27).

Obviously, an additivity approximation must be worse for normal metals, when the screening effects of an electron gas should be important. In this case we can profit using the known analytical formula for the polarizability of metal spherical cluster of radius $R$, obtained in the Thomas –Fermi approximation [111]

$$\alpha(\omega) = R^3 \left\{ 1 + \frac{3(1 - \kappa R \coth(\kappa R))}{2\big(\varepsilon(\omega) - 1\big)(1 - \kappa R \coth(\kappa R)) + \varepsilon(\omega)(\kappa R)^2} \right\} \qquad (6.6)$$

where $\kappa^2 = \dfrac{4}{\pi} \dfrac{(9\pi/4)^{1/3}}{r_s}$, $r_s$ is the jellium parameter (in atomic units), $\varepsilon(\omega)$ is the dielectric function. It is not difficult to show that the resonance absorption frequency corresponding to (6.6) coincides the bulk plasma frequency, $\omega_P$, while that one corresponding to (4.23) with the Drude function (C7) – the Mie frequency, $\omega_P/\sqrt{3}$. Also, $\alpha''(\omega) \propto R^2$ up to small $R$ of the order of $1\,nm$, and in addition, the proper peak values are lower than in the Drude model. Still more drastical difference we get in the low - frequency range. In this case, from (4.24) we get $\alpha''(\omega) \approx \dfrac{2R^3}{\omega_P \tau} \dfrac{\omega}{\omega_P}$, while from (6.6), respectively

$$\alpha''(\omega) \approx \frac{3R^3}{\omega_P \tau} \frac{\omega}{\omega_P} \frac{\kappa R \coth(\kappa R) - 1}{(\kappa R)^2 - 2(\kappa R \coth(\kappa R) - 1)} \qquad (6.7)$$

As, typically, $\kappa R \gg 1$, the polarizability (6.6a) is by $2\kappa R/3$ times lower than from (4.24). We shall turn to the numerical estimation of these issues in Section 6.4.

As the dominating contribution to the integral $J(\varepsilon_1(\omega), \varepsilon_2(\omega))$ in (6.5) comes from the frequency ranges with a noticeable crossover of the absorption bands of the contacting materials, let us discuss below several possible cases.

## 6.2 Low-frequency absorption mechanisms

The corresponding dielectric functions are defined by Eqs.(C10), (C12), with the proper parameters $\sigma_{1,2}$ $\varepsilon_{1,2}$ and $\tau_{1,2}$ (for instance, in the case of mica surface, $\tau = 10^{-10} \div 10^{-9}\,\text{sec}$, $\varepsilon \approx 6$), the subscripts $i = 1, 2$ indicate the tip and sample, respectively.

Using (C10), (C12), integrals (6.5) reduce to the similar form . So, for a contact of two conductors , we get (see Appendix E)

$$\int_0^\infty d\omega \frac{\widetilde{\Delta}_1''(\omega)\Delta_2''(\omega)}{\omega^2} = \frac{3}{4(2\sigma_1 + 3\sigma_2)} \qquad (6.8)$$

$$\int_0^\infty \frac{d\omega}{\omega} \widetilde{\Delta}_1''(\omega) \frac{d}{d\omega} \Delta_2''(\omega) = \frac{9\sigma_2}{4(2\sigma_1 + 3\sigma_2)^2} \qquad (6.9)$$

$$\int_0^\infty \frac{d\omega}{\omega} \Delta_2''(\omega) \frac{d}{d\omega} \widetilde{\Delta}_1''(\omega) = \frac{3\sigma_1}{2(2\sigma_1 + 3\sigma_2)^2} , \qquad (6.10)$$

Eqs. (6.4)- (6.10) are valid at $k_B T / 2\pi\hbar \gg \max(\sigma_1, \sigma_2)$. Maximum values of (6.9), (6.10) correspond to $\sigma_2 = 2\sigma_1 / 3$. Then, taking account of (6.4) yields

$$F_x = -C_{a,b,c} \frac{k_B TRV}{\sigma_1 d^3} \tag{6.11}$$

where $C_{a,b,c} = 0.009,\ 0.022,\ 0.022$ correspond to (6.5 ,a,b,c), respectively. At $k_B T / 2\pi\hbar \ll \min(\sigma_1, \sigma_2)$, similarly, from (4.31) we get

$$F_x = -D_{a,b,c} \frac{k_B^2}{\hbar} \frac{V R T^2}{d^3 \sigma_1 \sigma_2} \tag{6.12}$$

where $D_{a,b,c} = 0.0075, 0.0037, 0.0037$ correspond to (6.5,a,b,c). Eq.(6.11) is seen to give much larger result than (6.12) due to the additional small factor $k_B T / 2\pi\hbar\sigma$. In general case, the corresponding force –temperature dependence is scaled as $T^\lambda$, $1 \le \lambda \le 2$, if the involved dielectric functions are independent of temperature.

Assuming the typical parameters of the the tip of scanning probe microscope, $R = 30\ nm$, $d = 0.5\ nm$, $T = 300\ K$, $V = 1\ m/s$ and $\sigma_1 = \sigma_2 = 10^{17}\ \text{sec}^{-1}$ (normal metals) from (6.12) we get $F \sim 10^{-23}\ (N)$, and therefore the drag force is extremely small. But, for amorphous carbon or metal –insulator composites ($\sigma_1 = 10^{10}\ \text{sec}^{-1}$) from (6.11) we get $F \sim 1\ pN$, while for contacts between still more resistive materials (and dielectrics), the frictional force may be of the order of $1\ nN$ - the typical value for contact SPM mode.

For a contact between normal metals, a noticeable contribution to the friction can be related also with nonlocal contribution to the dielectric response .Using (4.59), one can simply show that $F_x \propto VRT^2 / d^6$. At $d = 0.4 \div 0.5\ nm$ this contribution is larger than (6.12), but, nevertheless, it is very small, too. This agrees with the results which we have got in Section 4.6.

## 6.3 High-frequency absorption mechanisms

Additional contributions to the tangential force can be associated with other absorption mechanisms, relevant to absorption bands in the infrared or optical range of spectra. They are determined from (6.2) or (6.4), in dependence of the relation between $k_B T$ and $\hbar\omega_0$, with $\omega_0$ being the characteristic frequency. The proper frequency integrals can be evaluated for simple dielectric response functions, such as Lorentzians and Drude functions (see (C7), (C14)). The integrals analogous to (6.6)-(6.10) again reduce to the similar form (see (F4), (F10), (F11)). For instance, for two Drude conductors at $T_1 = T$, $T_2 = 0$, $k_B T / \hbar >> \max(\omega_{P1}, \omega_{P2})$ we obtain (see Appendix F)

$$F_x = -\frac{3}{16\pi} \frac{k_B T R V \sqrt{3}}{d^3 \omega_{P1}} G_2 \left( \frac{\sqrt{3}}{\omega_{P1} \tau_1}, \frac{\sqrt{2}}{\omega_{P2} \tau_2}, \frac{\omega_{P2}}{\omega_{P1}} \sqrt{\frac{3}{2}} \right),$$
(6.13)

$$G_2(x, y, z) = \int_0^\infty \frac{dt}{(1 - t^2 z^2)^2 + x^2 z^2 t^2} \frac{d}{dt} \frac{t}{(1 - t^2)^2 + y^2 t^2}$$
(6.14)

In comparison with formulas corresponding to the low-frequency absorption dielectric response, the functions (F11) and (6.14) may have an arbitrary sign, while (F10) is positive in any case. For instance, $G_2 < 0$ at $\omega_{P1} / \sqrt{3} > \omega_{P2} / \sqrt{2}$, and therefore the tangential force accelerates a "hot" tip . On the contrary, at $\omega_{P1} / \sqrt{3} < \omega_{P2} / \sqrt{2}$ we have an accelerative force on a "cold" tip. For homogenious contacts (at arbitrary temperatures) or heterogenious ones (at equal temperatures), the tangential force is always frictional.

Our numerical estimates show that $G_{1,2}(x, y, z) = 10 \div 10^3$ under significant overlap of the absorption peaks of contacting materials, otherwise the resulting values prove to be negligibly small. However, as we have noted in Section 4.7, if the proper spectra are not tuned, it is still possible a resonance tip–sample coupling under the corresponding conditions.

For the above set of SPM parameters, $G_{1,2} \sim 10^3$ and $\omega_{P1} = 10^{13}$ sec$^{-1}$ (this may be typical for doped semiconductors like germanium), from (6.13), (6.14) we get $F \sim 10$ $pN$. In order to get the contribution from still higher frequencies, one should use Eq. (4.31), but even if the resonance frequency is close to the Wien frequency, $k_B T / \hbar$, the resulting force will be small due to the exponential cutting factors in (4.31). Therefore, the above obtained estimates seem to be maximal ones.

## 6.4 Remarks on heat flow between an STM tip and a flat surface

The general results obtained in Section 4.6 can be used when analysing process of heating of a surface by an STM tip. We have seen that nonretarded approximation is valid up to the separations from the surface of the order of several $\mu m$ for normal metals and much larger for semimetals , therefore it is sufficient to use Eqs. (4.33) and (4.47). Then, by analogy with (6.1), (6.2), for a paraboloidal tip, using (4.33) yields

$$\dot{Q} = -\frac{3\hbar R}{4\pi d} J_1(\widetilde{\Delta}_1{}''(\omega), \Delta_2{}''(\omega)) - \frac{9\hbar}{16\pi^2}\frac{RV^2}{d^3} J_2(\widetilde{\Delta}_1{}''(\omega), \Delta_2{}''(\omega)) \qquad (6.15)$$

where the structure of the integrals $J_{1,2}(\widetilde{\Delta}_1{}''(\omega), \Delta_2{}''(\omega))$ is identical to (4.33) with the replacements $\alpha''(\omega) \to \widetilde{\Delta}_1{}''(\omega)$, $\Delta''(\omega) \to \widetilde{\Delta}_2{}''(\omega)$. Furthermore, if use is made of the dielectric function (C10), from (6.15) we get

1) $\min k_B(T_1, T_2)/2\pi\hbar >> \max(\sigma_1, \sigma_2)$

$$\dot{Q} = -\frac{3\pi}{2} k_B(T_1 - T_2)\frac{R}{d}\frac{\sigma_1\sigma_2}{(2\sigma_1 + 3\sigma_2)} - \frac{9}{8\pi^2}\frac{RV^2}{d^3}\frac{k_B(T_1 - T_2)\sigma_1}{(9\sigma_2{}^2 + 4\sigma_1{}^2)} \qquad (6.16)$$

2) $\max k_B(T_1, T_2)/2\pi\hbar << \min(\sigma_1, \sigma_2)$

$$\dot{Q} = -\frac{3\pi}{160}\frac{R}{d}\frac{k_B{}^4(T_1{}^4 - T_2{}^4)}{\hbar^3\sigma_1\sigma_2} \qquad (6.17)$$

Eq.(6.17) slightly differs from that one obtained in [55, 80] by a numerical factor of the order of 1.

In the general case of arbitrary temperatures, the corresponding frequency integrals $J_{1,2}(\widetilde{\Delta}_1{}''(\omega), \Delta_2{}''(\omega))$ need to be calculated numerically. It is seen that in (6.17) the velocity - dependent term is absent, while in (6.16) it may be significant, because its ratio to the first term is of order $(V/\sigma d)^2$ and rapidly increases with increasing resistivity.

At $T_1 = 300\,K$, $T_2 = 0$, $R/d = 30$ and $\sigma = 9\cdot10^9\,\sec^{-1}$ (silicon) from (6.16) we get $\dot{Q} = -1\cdot10^{-9}\,W$. Then, assuming the effectively heated surface area to be of the order of $d^2 = 10^{-18}\,m^2$, the corresponding heat flux $d^2Q/dtdS$ per unit surface area equals $1\cdot10^9\,W/m^2$ - i.e. much larger than in the case of black body radiation intesity - $4.6\,W/m^2$. At the same time, for a contact between normal metals, ($\sigma = 10^{17}\ \text{sec}^{-1}$) at same conditions, from (6.17) we get $\dot{Q} = -4.5 \cdot 10^{-13}\ W$ and $d^2 Q / dt dS = 4.5 \cdot 10^5\ W / m^2$.

It is interesting to compare the results for metals with those obtained using polarizability (6.7), when approximating an STM tip by a spherical particle with radius $R$. In this case the condition $\max k_B (T_1, T_2) / 2\pi \hbar << \min(\sigma_1, \sigma_2)$ is fulfilled and using (4.33) yields

$$\dot{Q} = -\frac{\pi}{40}\left(\frac{R}{R+d}\right)^3 \frac{k_B^{\ 4}(T_1^{\ 4} - T_2^{\ 4})}{\hbar^3 \sigma_1 \sigma_2} \frac{\kappa R \coth(\kappa R) - 1}{(\kappa R)^2 + 2 - 2\kappa R \coth(\kappa R)} \qquad (6.18)$$

The fraction of (6.18) to (6.17) vs. $d$, $R$ is represented in Fig.11 from which we conclude that screening effect leads to drastic decrease of the polarizability and the heating rate for metal particles. This again confirms that more effective heating regime is to be expected for high –resistivity materials.

New features may be specific for dynamic regime. As it follows from (4.47), a direction of the heat flow, $\dot{Q}$, depends on $T_1 / \omega_0$ and $T_2 / \omega_s$. Moreover, as we have seen in Section 4.7, $|\dot{Q}| = \max$ at $V = 0.95 z_0 |\omega_0 - \omega_s|$. This equation represents the "resonance heating" condition. Because the wave –vectors of the order $k \sim 1 / z_0$ dominates integral (4.47), this implies that the resonance is tuned to these wave –vectors. The same conclusion holds for the resonance force, Eq. (4.46).

In principle, the nonlinear –velocity effects can be observed in the SPM modulation mode at velocities of $1 \div 100\ m / s$ and distances to the surface of the order of $1\ nm$, if the characteristic frequency ranges in the interval $10^{10} \div 10^{12}\ \text{sec}^{-1}$ [73]. For faster particles, the corresponding distances can be larger, proportional to the velocity.

### 6.5 Friction of flat surfaces

Due to nonrelativistic scope of this paper, we leave aside the retardation effects. However, as we have shown in Section 4.6, they can be neglected at separations in the range 1 to $\sim 30\ nm$ which are of prime interest here.

If use is made of an additive approximation, a transition to friction force between two semiinfinite bodies with smooth flat surfaces is trivial [73]: the integration over the tip volume is replaced by integration over the distance $z_0$ (in the interval $(0, \infty)$) and the result must be multiplied by the surface area, $S$. The formulae for the shear stress between the bodies separated by the distance $d$ are obtained from those ones for a paraboloidal tip and a flat surface by myltiplying on factor $3/2\pi R d$. For instance, Eqs.(6.2), (6.4), (6.13) take the form (in our notation, the frictional stress is taken with minus sign)

$$F/S = -\frac{9\hbar V}{32\pi^2 d^4}J(\varepsilon_1(\omega),\varepsilon_2(\omega)),\tag{6.19}$$

$$F/S = -\frac{9}{32\pi^2}\frac{k_{\mathrm{B}}TV}{d^4}J(\varepsilon_1(\omega),\varepsilon_2(\omega)),\quad \hbar\omega << k_{\mathrm{B}}T\tag{6.20}$$

$$F/S = -\frac{9}{32\pi^2}\frac{k_{\mathrm{B}}TV\sqrt{3}}{d^4\omega_{\mathrm{P1}}}G_2\left(\frac{\sqrt{3}}{\omega_{\mathrm{P1}}\tau_1},\frac{\sqrt{2}}{\omega_{\mathrm{P2}}\tau_2},\frac{\omega_{\mathrm{P2}}}{\omega_{\mathrm{P1}}}\sqrt{\frac{3}{2}}\right),\tag{6.21}$$
$$T_1 = T, T_2 = 0, k_{\mathrm{B}}T/\hbar >> \max(\omega_{\mathrm{P1}},\omega_{\mathrm{P2}})$$

where the overlap integrals have same meaning (see (6.5), (6.13) and Appendix F). In particular, for two conductors with dielectric response (C10) we get

$$F/S = -0.0043\frac{k_{\mathrm{B}}TV}{d^4\sigma_1}, k_{\mathrm{B}}T/2\pi\hbar >> \max(\sigma_1,\sigma_2), T_1 = T_2 = T\tag{6.22}$$

$$F/S = -0.0036\frac{k_{\mathrm{B}}T^2V}{\hbar d^4\sigma_1\sigma_2}\;,\quad k_{\mathrm{B}}T/2\pi\hbar << \min(\sigma_1,\sigma_2), T_1 = T_2 = T\tag{6.23}$$

Eq.(6.23) is in close resemblance with the corresponding one in [44, 45], where the authors also presented  more general expressions for frictional stress in terms of the reflectivity factors (see also [41]).

However, in addition to the results [44, 45], corresponding to  the nonretarded case, formulae (6.17) - (6.19)  predict some new effects:  i) at different temperatures of the bodies a moving one can be accelerated ; ii)  heat may flow from cold to hot body;  iii) the temperature dependence of the frictional stress strongly depends on contacting materials; in particular, the larger friction is expected  for homo- and heterocontacts between semiconductors and dielectrics with the characteristic dependence  $F/S \sim T$.  For instance, at

$$T = 300\,K,\;\; \omega_{\mathrm{P}} = 10^{13}\,s^{-1}\;,\; d = 1\,nm, G_2 = 10^3\,, V = 1\,m/s\;,$$

from Eq. (6.21) we get  $F/S = -2\cdot10^4\,N/m^2$. This value is by 12 orders of magnitude larger than that obtained from (6.23) for normal metals. However, even this value is much less than adhesional shear stress in an atomically dense contact ($10^8\,N/m^2$ ).

More important contribution to the friction force between normal metals comes from the nonlocal part of the dielectric response function. Using approximation (C2)-(C5), and assuming $\Delta_1''(q,\omega) = \Delta_2''(q,\omega) = \widetilde{\Delta}_1''(q,\omega)$, $\omega << \omega_s$, after the proper integration of the linearized Eq.(4.27) we get [73] (the atomic units are used here)

$$F/S \approx -2.16\, k_B^2 (T_1^2 + T_2^2) \frac{1}{\omega_P^4 q_{TF}^2} \frac{V}{d^5} \begin{bmatrix} \ln^2(1.21 q_{TF}) + \ln^2(2d) + \\ + 2\ln(1.21 q_{TF})\ln(2d) - \\ 3.74\ln(0.89 q_{TF} d) \end{bmatrix} \qquad (6.24)$$

Assuming $T_1 = T_2 = 300\,K$, $\omega_P = 9\,eV$, $q_{TF} = 1.07\,a.u.$, $d = 1\,nm$ and $V = 1\,m/s$, from (6.24) we get $F/S = -3.4 \cdot 10^{-4}\quad N/m^2$. This is a factor $\sim 10^4$ larger than we get from (6.19).

The authors of [55] noted that formulae for the frictional stress between semi-infinite bodies can be used in calculations of the friction force between a small particle and a surface making use a replacement $4\pi N\alpha \rightarrow 2R_P = 2\Delta_2(\omega)$ and the Deryagin's approximation [112]

$$F = 2\pi \int\limits_{-\infty}^{+\infty} dr\, r\, \sigma\big(L(r)\big) \qquad (6.25)$$

where $\sigma\big(L(r)\big)$ is the frictional stress between flat surfaces, $L(r)$ is the tip-sample separation as function of the distance $r$ from the tip symmetry axis. On the other hand, making use of (4.30) with the replacement $\alpha \rightarrow R_P/2\pi N$, and integrating over the volume of the moving semiinfinite body, we can get the expression for $F/S$ which agrees with that one obtained in [41] in a quite different way (see Section 2):

$$F/S = \frac{\hbar}{\pi^3} \int\limits_0^\infty k_x dk_x \int\limits_0^\infty dk_y \, \exp(-2k\,z_0) \int\limits_0^{k_x V} d\omega\, R_{P2}''(\omega) R_{P1}''(\omega - k_x V) \qquad (6.26)$$

From (6.24), for a homogenious contact in the limit $V \rightarrow 0$, using the dielectric function (C10) yields

$$F/S = -\frac{5}{2^9 \pi^4} \frac{\hbar V^3}{z_0^6 \sigma^2} \qquad (6.27)$$

Therefore, an extension of the theory of the particle –surface friction in the case of flat –flat geometry, with the help of an additive approximation, as well as the different way, when we start from the problem of friction between two semi-infinite bodies and, using (6.25), calculate the friction force between a sperical (paraboloidal) particle and a semi –infinite body, proves to give very similar results in the nonretarded case. This is strong argument in favour of an additive approximation.

## 7. Discussion of several experiments

### 7.1 Measuring of dissipative forces in the modulation mode of scanning probe microscopes

Quantitative measurements of dynamic interactions between the nanoprobes and the slabs have been performed in several experiments using the scanning probe microscopes [47, 113-118]. So, the conservative forces in normal modulation mode SPMs were studied in [113-115]. Gotsmann et.al. [116-118] studied both conservative and dissipative forces. Also, Dorofeyev et. al. [47] observed Brownian motion of the damped cantilever.

As the dissipative forces are of prime interest here, we shall restrict our discussion by the results obtained in [116, 117, 47]. However, it is worthwhile noting that in real experimental situation different mechanisms contribute to damping of vibrating nanoprobe. To date, there is no clear distinction between them and much work must be done in order to reach clearer understanding of the involved processes. For instance, in modulation mode (tapping mode) of SPMs, an important role may have the energy dissipation due to a breaking/forming of the adhesion bonds - a mechanism being more characteristic for the contact mode [109, 119]. Other possibilities have been discussed in [41,50, 109]. In order to diminish signals which are not related with FDF, we have proposed to use parallel vibration geometry [120], when the SPM tip vibrates at a fixed controllable distance from the surface and an additional contribution to the damping of adhesive nature should be much less.

In [116], the dissipative forces were measured in the case of a silicon probe oscillating along normal to mica surface in UHV ($T = 300~K$). The cantilever had stiffness $k = 40~N/m$, the curvature radius of the tip was $R= 20~nm$, the set point amplitude $A = 32~nm$, the quality factor $Q$=22815, and the ground frequency $f$= 296.6 $kHz$. From this work, the mean dissipation power was equal to about 0.1 and 0.02 $pW$ at the minimum approach distances to the surface of 0.1 and 0.5 $nm$, respectively. Therefore, the corresponding distance dependence is scaled as $\overline{P} \propto h^{-1}$. In addition, the internal dissipation power of free vibrations is estimated to be $P^{(i)} = \pi k A^2 f / Q = 1.7~pW$.

Yet we have not obtained the formula for the dissipative force under normal motion of the nanoprobe, and in order to get a numerical estimation, we assume the corresponding force to be twice that for parallel motion, as in the cases of bare charges (see Section 4.1). Furthermore, for the silicon -mica contact, we can use the results of Section 6.2 and Appendix F. At typical parameters $\varepsilon = 6$ , $\tau = 10^{-9}\ s$ , $\sigma = 0.01\,\Omega^{-1}\,m^{-1}$ , the corresponding force is $F = -C\dot{z}/z^3$ , with $C = 6 \cdot 10^{-38}\ J \cdot m^2 \cdot s$ . Assuming the simple harmonic cantilever movement $z(t) = A\cos(2\pi f\,t)$ , $\dot{z}(t) = -2\pi\,fA\sin(2\pi f\,t)$ , the mean dissipation power is given by

$$\overline{P}(d) = f\int_0^{1/f} F(t)\dot{z}(t)dt = 4\pi\,f^2 A^{-1}C\int_0^{\pi}\frac{\sin^2(x)}{(d/A+\cos(x))^3}dx = \frac{2\pi^2 f^2 C}{A\big((d/A)^2 - 1\big)^{3/2}} \quad (7.1)$$

where $d$ is the cantilever support distance. The minimal approach distance is $h = d - A$ , and at $h << A$ we get from (7.1)

$$\overline{P}(h) = \sqrt{\frac{A}{2}}\pi^2 f^2 C / h^{3/2} \tag{7.2}$$

Then, at $h = 0.1$ and $h = 0.5\ nm$ we get $\overline{P} = 6.6 \cdot 10^{-3}$ and $0.6 \cdot 10^{-3}\quad pW$ , respectively, i.e. less than experimental values by 2 orders of magnitude. To reach concordance with this experiment, one should assume the relaxation time to be about $\tau = 10^{-7}$ sec , while the surface conductance must be smaller by 2 order of magnitude. The corresponding material parameters seem to be somewhat problematic in the case of pure silicon and mica at room temperature. However, one should bear in mind that the used value of constant $C$ defines only a part of the dissipative force due to absorption in the low frequency range of electromagnetic spectra.

In the case of parallel nanoprobe vibration, the mean dissipation power is given by ( $x(t) = A\cos(2\pi ft)$ )

$$\overline{P}(h) = \frac{C}{h^3}f\int_0^{1/f}\dot{x}^2 dt = \frac{2\pi^2 A^2 f^2 C}{h^3} \tag{7.3}$$

Therefore, at $h = 0.5\ nm$ we get $\overline{P} \approx 0.9\ pW$ and evidently, the fluctuation –induced dissipation would be more intensive.

In another experiment [117], the authors have measured conservative and dissipative interactions between an aluminium tip and an Au(111) surface in UHV. The used set of parameters is:

$$f = 267.2 \, kHz \, , \, A = 24 \, nm \, , \, k = 40 \, N \, / \, m, \, R = 35 \, nm \, , \, Q = 19050.$$

In this experiment, the measured dependence of the dissipation force vs. distance is in good agreement with the law $F \sim z_0^{-3}$, and an analytical fit of the damping coefficient gives $C \approx 8 \cdot 10^{-35} \, N \cdot m^2 \cdot s$, with the experimental error of about $\pm 40\%$. Then, according to (7.2, the mean dissipation power is $0.55 \pm 0.22 \, pW$ at $h = d\text{-}A = 0.5 \, nm$.

Unfortunately, the expected theoretical estimates prove to be in much worse agreement with the experimental ones. So, from Eq. (6.9), we get only $C = 10^{-46}$ to $10^{-45}$ $N \cdot m^2 \cdot s$. With account of nonlocal contribution to the dielectric response, the constant $C$ can be larger by several orders of magnitude, but, nevertheless, the dissipation power seems to be negligibly small. In addition, the friction mediated by the electron exchange mechanism [109] is very small, too.

In paper [47], the authors have studied Brownian motion of an aluminium coated nanotip near a surface of gold in UHV. The measured damping coefficient $\gamma / m_{eff}$ at a distance to the snap – on point of about 5 $nm$ was equal to 150 $s^{-1}$ ($m_{eff} = 10^{-8}$ to $10^{-10} \, g$). The corresponding friction force (1.5 to 0.015 $nN$) is again very large, so that it can unlikely be attributed to the FDF. This follows both from our calculations [73], and from [50], too. However, the authors of [47] have a different opinion. Also, Persson and Volokitin [50, 54, 80] believe that these experimental data could be explained using the mechanism of the thermally activated point – like defect flipping.

Besides, we can consider a possible role of the resonance coupling effect (see Section 4.7). At the above experimental conditions, the maximal tip velocity is estimated to be $V_m = 2\pi \, fA = 0.04 \, m / s$. Then at $z_0 = 0.5 \, nm$, the resonance dynamic condition implies $\Delta\omega = V_m / 0.62 \, z_0 = 1.3 \cdot 10^8 \, \sec^{-1}$. Such frequencies may be, in principle, characteristic for surface acoustic phononic modes, but, nevertheless, the corresponding mechanism needs to be elaborated in more details, because it necessitates noticeable electron – phonon coupling.

On the whole, we see that the above discussed SPM experiments still do not provide reliable and consistent information on FDF, thus giving rise to further experimental and theoretical studies.

**7.2 Sliding friction of adsorbates (quartz crystal microbalance experiments)**

The quartz crystal microbalance (QCM) technique was adopted for friction measurements in 1986-1988 by Widom and Krim [121]. QCM consists of a single crystal of quartz oscillating in transverse shear motion with a quality factor of about $10^5$. Adsorption onto the microbalance produces the shifts in both the frequency $f_0$ and the quality factor $Q$. The characteristic slip time $\tau$ and friction parameter $\eta$ (the shear stress per unit velocity) are determined by [21, 121]

$$\delta(Q^{-1}) = 4\pi\tau\delta f_0 \ , \ \eta = \rho/\tau \qquad (7.4)$$

where $\rho$ is mass of the adsorbate per unit area. In typical experiments [121, 122], friction is studied of inert gas films (mono- and bilayers) on Ag/Au(111) surfaces, being characterized by $\tau = 2$ to $3\,ns$ . Also, when probing the friction forces, the quartz crystal resonators may be used in another arrangement [123].

To date, there is no well –accepted theoretical interpretation of the QCM results. So, Krim and coworkers guess that more preferable mechanism of friction is phononic one [124,125], while other authors [38, 68, 126, 127] argue that on systems like Xe on Ag, the electronic mechanisms dominate.

As far as the electronic friction is concerned, Persson and Volokitin [38], using the high –order perturbation theory, have proposed Eq. (4.29) for the drag friction force of physisorbed atoms on a metal surface, which is seen to be independent of temperature. Note that these authors and Liebsch [40] have employed the following relation between the drag friction force $F_x$ and the friction coefficient $\eta$: $F_x = -M\,\eta\,V$ , with $M$ being the adsorbate mass, therefore, in this notation $\eta = 1/\tau$ .

Using (4.29) and the time-dependent local density approximation, Liebsch obtained $\eta = 3.4 \cdot 10^8\ s^{-1}$ at $d = 0.24\,nm$ [40]. Persson [68] has found $\eta = 0.4 \cdot 10^8\ s^{-1}$ using the same $d$ and making use a simpler model of the surface, while from the corresponding QCM observations it follows $\eta = 8 \cdot 10^8\ s^{-1}$ [128]. One can note a reasonable agreement between the experiment and the theory, but the dependence $F_x \sim d^{-10}$ seems to be extremely sensitive to small variations of the separation. Note, that from data on molecular beam scattering [129], the minimum distance between the adsorbed *Xe* atom and the ion cores on an Ag(111) surface is estimated to be 0.355 *nm* , while $d = 0.24\,nm$ defines the distance from the jellium edge. Evidently, since we do not control $d$ independently, the theoretical estimations cannot be accepted with a confidence. For instance, an obvious shortcoming of the theory is due to its inability to describe a dependence of the slip time on the film coverage.

Another point of ambiguity of the electronic friction models is related with the recent results [55], where the authors obtained a new formula for the friction coefficient (see Eq.(42) Ref. [55]) . According to this one, the slip time is given by

$$\tau = \frac{16}{3\pi} \frac{M z_0^8}{\hbar \alpha(0)^2} \left( \frac{4\pi \hbar \sigma}{k_B T} \right)^2 \qquad (7.5)$$

Then in the case where the temperature is tuned to the conductivity, so as to give maximum friction for $Xe$ physisorption, at

$$M = 2.2 \cdot 10^{-22} \, g, \ \alpha(0) = 4 \cdot 10^{-24} \, cm^3, \ z_0 = 0.24 \, nm \ \text{and} \ \frac{4\pi \hbar \sigma}{k_B T} = 1$$

from (7.5) we get $\tau = 2.4 \, ns$ . In drastic contrast to (4.29), formula (7.5) is very sensitive to the temperature and surface resistivity. Moreover, the authors of [55] have obtained nonlocal contribution to the slip time (also temperature –dependent), but it has proven to be a factor $\sim 10^{-4}$ smaller than estimated in Ref. [38].

It is worthwhile noting in this turn, that in this situation we should also take into account an important linear – order  (in polarizability and electric field) contributions to the friction force, which are determined by Eq. (4.27) or other ones like (4.28), (4.31) or   (4.59), when the dependence  $\eta(z_0)$  is less strong ( $\eta \sim z_0^{-5}$ ), being temperature-dependent, and , what is more important – the slip time depends on  the structure of the adsorbed film via the dielectric response function.

For instance, the solid $Xe$ film is known to have strong exciton absorption bands and those ones due to intraband transitions [130]. On the other hand, the solidified film is characterized by new phononic modes, and when cooling  (below the freezing point), this can lead to a positive contribution to the tangential force (under sufficient electron-phonon coupling), thus giving rise to smaller friction force if the film temperature is lower than that of the surface (see  Eq.(4.31)). Such an effect can be, in principle, responsible for the small friction of solidified incommensurate films [131], because any structural transformations caused by heat transfer influence the dielectric response and, therefore, will modify FDF. However, we do not clearly understand the role of  heating effects related with evanescent fields in these 2D-systems.

In view of what has been said above, we claim that in general, large work is still needed to reach more clarity when relating theoretical models with QCM observations.

Finally, we shall briefly touch on the recent QCM experiments on superconducting surfaces [132]. It was found that friction of the solid nitrogen film drops abruptly (by about two times as compared to the normal state value) below the transition temperature (7.2 K) of lead. Evidently, this requires significant role of the electronic friction processes in this system [133]. The authors note that their theory cannot explain this experiment, because the electronic sliding friction decreases continuously when the system is cooled below $T_c$, in a way correlating with the fraction of electrons in the superconducting condensate. In our opinion, this conclusion follows from oversimplified character of the used model, attributing the electronic friction exclusively with the *dc* -resistivity. However, in these systems an essential role may have the electron –phonon coupling [132]. Really, the phonon frequencies ($10^{12}$ to $10^{13}$ sec$^{-1}$ ) exactly coincide the energy gap of lead: $\omega_c = 4k_B T_c / \hbar = 3.8 \cdot 10^{12}$ sec$^{-1}$ . If, in normal state, there is a noticeable contribution to the interaction from this frequency range, then below $T_c$ we must observe a drop of the friction. In favour of such a possibility one can mention the experimental results on the energy dependence of the density states, $N(\omega)$ for lead [134], where the nonmonotonous character of the function $N(\omega)$ at $\omega > 2k_B T_c$ is clearly seen.

Still more intriguing results are to be expected when studying the electronic friction on surfaces of high temperature superconductors, where a great variety of interesting features of the involved dielectric and absorption properties has been observed (see, for instance, [135] and references therein).

Close to these problems are those related with the frictional drag between parallel 2D- electron systems [23, 136, 137]. Some comments on this matter were made in [43, 80] but, on the whole, the problem is still open for discussion.

### 7.3 Experiments on passage of neutral atomic beams near a surface

As far as conservative force is concerned, one can mention earlier papers [138-140], where the authors have studied dynamic vdW forces. So, the authors of [138, 139] have found somewhat weaker force than theoretically expected one in experiments with neutral alkali atoms scattered from metal cylinders, whereas Arnold et.al. [140] have observed stronger vdW attraction between semiconductor plates in dynamic regime.

In view of the general results (see (4.19), (4.22)), the velocity –dependent effects may become more significant for higher beam energies. In this relation, we must note that velocity effects are expected to be important not also in vdW interactions, but also in repulsive interatomic interactions [141].

Amirav and Cardillo [142,143] have performed experiments on scattering of a beam of hyperthermal xenon atoms (2-15 eV energy) from the surface of Ge or In-P $p$-$n$ junction. They were able to estimate the product of the electron-hole pair-excitation probability, which was found to be of the order of 0.2 per $Xe$ atom at 9 eV. The corresponding theoretical calculations using the self –energy approach [33] show that the long –range vdW coupling gives only a small contribution: about $10^{-5}$ per atom at normal incidence to the total electron –hole pair excitation probability which quickly goes down with increasing the incidence angle.

In addition to the experiments discussed in Sections 7.1, 7.2, where we usually are dealing with small velocities ($V < 1\,m/s$ ), measurements of FDF can be performed at much greater velocities of the order of Fermi velocity of metals [144] .We have done theoretical estimation of the expected drag force and stopping power for a well - collimated neutral beam of helium atoms passing in close vicinity to a flat surface of aluminium [144]. A possible experimental arrangement is similar to that has been exploited when measuring transmission of electrons through thin microchannels in the metal foils (of 20-200 nm in diameter) [26] or, for instance, when the atomic beam moves above a metal (semiconductor) bar deposited on a substrate, the height of a single bar must be greater than the diameter of the beam.

Unfortunately, in these calculations [144] the temperature factors have not been taken into account correctly. In view of the present day results, in this case we dot not expect a noticeable contribution from the surface plasmon coupling (for normal metal substrates). However, there might be expected similar effects due to a resonance with the low –frequency surface excitations like surface plasmons in doped semiconductors and polaritons in dielectric materials. Also, the resonance stopping effects can be observed for hot neutral molecules and clusters above a cold substrate (see Eq.(4.46)). The higher is the temperature $T_1$, the greater number of inner states of the incident particles will satisfy the frequency – tuning conditions, thus leading to a stronger particle –surface coupling, as well.

Moreover, one should bear in mind a possibility for the low – frequency nonlocal contribution to the dissipative force, Eq.(4.59). For instance, for a neutral helium atom with the velocity of $3 \cdot 10^6 \ m/s$, passing above a metal, assuming the same conditions as those given after (4.46), the estimated energy loss will be of the order of $0.1 \ eV/\mu m$. Much larger stopping power is expected for atomic beams above poor conductors and dielectrics.

Evidently, in real experimental situation, some part of the output beam will be ionized, but atoms which remain in a neutral state, would be characterized by positive or negative energy increments differing from those of the most outgoing ions. Consequently, the total energy distribution spectrum of the particles must reveal these features.

Concluding this discussion, we guess that rapid progress in development of experimental technique presents a great challenge for further studying and using the results of measurements of the fluctuation forces when probing the atomic –scale properties of various materials in the nearest future.

## 8. Summary and conclusions

Using general methods of classical electrodynamics and fluctuation electromagnetic theory, we have considered a lot of problems related with conservative and dissipative interactions of moving particles and nanoprobes near flat and cylindrical surfaces with account of different dielectric properties and the spatial dispersion effects of materials. In the case of a neutral particle and a surface, of prime interest is the van der Waals coupling. We have used minimal number of assumptions and obtained explicit nonrelativistic expressions for the involved normal / lateral forces and heating rates of the particles in the regime where retardation can be neglected , i.e. when the particle- surface separation does not exceed, approximately 20 to 40 $nm$.

It is very important that all the obtained results for both conservative and dissipative interactions retrieve the known from literature results in the static and dynamic regimes for different types of particles. In addition, as it follows from our relativistic analysis, the obtained nonrelativistic formulae for FDF exactly coincide with the relativistic ones in the limit $c \rightarrow \infty$. This is a new principal point, because to date, no such an agreement has been reached in numerous works of other authors. In our opinion, a lot of descrepancies between the results obtained by several research groups to date, have arised via usage of very different theoretical approachs, on the one hand, and the lack of understanding of fundamental relations between the basic physical quantities, on another hand. For the first time, we have formulated these relations for both relativistic and nonrelativistic case.

In this paper, we have restricted consideration of interactions in the molecule-surface systems by dipole approximation. However, the same method is applicable for consideration of quadrupole – surface interaction, and other high order multipole terms of the energy expansion [145].

An important point is that a moving particle may not only lose energy, but, in definite cases, the energy may be picked up. The physical reason is that a composite particle has the internal states of freedom and so the energy exchange with the surface is not so trivial as in the case of interaction of bare charge and dipole molecule. Due to this, some resonance effects may appear. The resonances are characterized by somewhat different velocities with concern of the tangential force and heat flow.

The spatial dispersion of the surface excitations gives rise to still more complex picture of the interactions, because the corresponding dispersion relations are crucially dependent on geometry of the contacting bodies, resulting in specific dependencies of the fluctuation forces on distance and material properties. In addition, any structural changes of the interfaces may significantly alter the dielectric properties and the interactions.

We have discussed several recent experimental results where the fluctuation dissipative forces could be probed in SPM and QCM measurements, but, in our opinion, the present day status of the theory and experiment still does not allow to compare them in an unambiguous way. Nevertheless, the methods of SPMs make it possible to probing even extremely small forces of the order of $10^{-17} N$ [171]. Most of the numerical estimates of FDF in this paper are larger by several orders of magnitude. We guess that such forces can be studied more successfully if use is made of lateral modulation mode SPMs, or by measuring transmission energy spectra of neutral atomic/molecular beams passing through nanochannels (for instance, in metal foils and nanotubes). These experiments present a challenge for future investigation.

Among theoretical problems, which are of prime interest, are the problems of normal to surface motion for a neutral particle, the spatial dispersion effects for different types of surface excitations and near field structure of fluctuating field close to curved surfaces, relation between electromagnetic and phononic processes, heating effects and, in general, a more detailed elaboration of the relativistic theory.

### Acknowledgements

We are very grateful to P.M. Echenique and J.I. Juaristi for their helpful comments with concern of the plasma dispersion effects and sending us their paper reprints. Comments and communications from B.N.J. Persson are also greatly acknowledged, as well as submission of the paper reprints from him, A.I. Volokitin, M.Stoneham, B.Gotsmann, I. Dorofeyev, H.Fuchs, F.Fortsmann.

### Appendix A . Solution to the Poisson's Equation

### i) Flat surface

When calculating the induced electric field of a surface created by moving nonrelativistic particle (see Fig.4) the problem reduces to a solution of the involved Poisson's equation for the electric potential. The corresponding Fourier-transformed equation (over the variables $x,y,t$) is

$$\left(\frac{d^2}{dz^2} - k^2\right)\phi_{\omega k}(z) = A\delta(z - z_0) + B\delta'(z - z_0)$$
$$k^2 = k_x^2 + k_y^2$$

(A1)

For the sake of maximal generality, the coefficients $A$, $B$ are taken as arbitrary ones. Then the general solution to (A1) is applicable to different cases (charged particle, dipole molecule and fluctuating dipole).

General solution to the homogenious equation $\left(\frac{d^2}{dz^2} - k^2\right)\phi_{\omega k}(z) = 0$ reads

$$\phi_{\omega k}(z) = C_1 \exp(-kz) + C_2 \exp(kz)$$

(A2)

where the physical meaning have only the solution with $C_1 = 0$ at $z < 0$ and $C_2 = 0$ at $z > 0$.

In order to find out particular solution of (A1), one has to obtain the corresponding Green's function, satisfying the equation

$$\left(\frac{d^2}{dz^2} - k^2\right) G(z, z') = \delta(z - z') \tag{A3}$$

From (A3) we get

$$G(z, z') = -\frac{1}{2\pi} \int_{-\infty}^{+\infty} \frac{\exp(ik_z(z - z'))}{k_z^2 + k^2} dk_z = -\frac{1}{2k} \exp(-k|z - z'|) \tag{A4}$$

Then, following general method for solution of the inhomogeneous differential equation [147], we have

$$\phi_{\omega k}(z) = \int_{-\infty}^{+\infty} G(z, z') \left[A\delta(z' - z_0) + B\delta'(z' - z_0)\right] dz' =$$
$$= \exp(-k|z - z_0|)\left[-A/2k + B/2 \, sign(z - z_0)\right] \tag{A5}$$

where we used the known relations

$$\int_{-\infty}^{+\infty} f(x)\delta'(x) dx = -\int_{-\infty}^{+\infty} f'(x)\delta(x) dx \quad,$$

$$\frac{d|x|}{dx} = sign(x)$$

Summing (A2) and (A5), the general solution of (A1) is

$$\phi_{\omega k}(z) = C_1 \exp(-kz) + C_2 \exp(kz) + \exp(-k|z - z_0|)\left[-A/2k + B/2 \, sign(z - z_0)\right] \tag{A6}$$

Coefficients $C_1$ and $C_2$ are determined from the boundary conditions of continuity of the electric potential and normal component of electric induction at $z=0$:

$$\phi_{\omega k}(+0) = \phi_{\omega k}(-0) \tag{A7}$$

$$d\phi_{\omega k}(+0)/dz = \varepsilon(\omega) \, d\phi_{\omega k}(0)/dz \tag{A8}$$

From (A6)-(A8) we get

$$C_1 = C_2 = \Delta(\omega)(A/2k + B/2)\exp(-kz_0) \tag{A9}$$

where $\Delta(\omega) = \dfrac{\varepsilon(\omega) - 1}{\varepsilon(\omega) + 1}$ is the surface response function.

Substituting (A9) into (A6) yields (at $z>0$)

$$\phi_{\omega k}(z) = \Delta(\omega)(A/2k + B/2)\exp(-k(z + z_0)) +$$
$$+ \exp(-k|z - z_0|)\left[-A/2k + B/2 \, sign(z - z_0)\right] \tag{A10}$$

Eq.(A10) contains the induced potential of the surface (the first term) and the self potential of the particle (the second term). Therefore, the induced potential is given by

$$\phi^{ind}_{\omega k}(z) = \Delta(\omega)(A/2k + B/2)\exp(-k(z + z_0)) \tag{A11}$$

Now, let us consider the special cases.

***1.Charged particle.*** The Poisson equation for a moving particle with the charge $Ze$ reads

$$\Delta\phi = -4\pi\,Ze\,\delta(x - Vt)\delta(y)\delta(z - z_0) \tag{A12}$$

The corresponding equation for Fourier amplitude $\phi_{\omega k}$ takes the form

$$(d^2/dz^2 - k^2)\phi_{\omega k}(z) = -8\pi^2 Ze\,\delta(\omega - k_x V)\delta(z - z_0) \tag{A13}$$

Comparing Eq.(A1) with (A13), we have

$$A = -8\pi^2 Ze\,\delta(\omega - k_x V)\,,\; B = 0\,,$$

then (A11) reduces to

$$\phi^{\text{ind}}_{\omega k}(z) = -\frac{4\pi^2}{k}Ze\,\delta(\omega - k_x V)\Delta(\omega)\exp(-k(z + z_0))$$

(A14)

***2.Dipole molecule.*** The Poisson equation takes the form

$$(d^2/dz^2 - k^2)\phi_{\omega k}(z) = 8\pi^2\,\delta(\omega - k_x V)((ik_x d_x + ik_y d_y)\delta(z - z_0) +$$
$$+ d_z\delta'(z - z_0)) \tag{A15}$$

Evidently, (A15) reduces to (A1) at

$$A = 8\pi^2\,\delta(\omega - k_x V)(ik_x d_x + ik_y d_y)\,,\; B = 8\pi^2\,\delta(\omega - k_x V)d_z\qquad,$$

then from (A11) w get

$$\phi^{\text{ind}}_{\omega k}(z) = \frac{4\pi^2}{k}\delta(\omega - k_x V)\Delta(\omega)(ik_x d_x + ik_y d_y + kd_z)\exp(-k(z + z_0)) \tag{A16}$$

***3.Neutral spherically symmetric particle (ground state atom).*** The Poisson equation is similar to (A15), but the components of the dipole moment are replaced by the Fourier-transformed amplitudes of the spontaneous dipole moment

$$(\frac{d^2}{dz^2} - k^2)\phi_{\omega k}(z) = 4\pi\,\delta(z - z_0)(ik_x d^{sp}_x(\omega - k_x V) + ik_y d^{sp}_y(\omega - k_x V)) +$$
$$+ 4\pi\,d^{sp}_z(\omega - k_x V)\delta'(z - z_0) \tag{A17}$$

Comparing (A17) with (A1) yields

$$A = 4\pi(ik_x d^{sp}_x(\omega - k_x V) + ik_y d^{sp}_y(\omega - k_x V))\,,$$
$$B = 4\pi\,d^{sp}_z(\omega - k_x)$$

and finally, from (A11) we get

$$\phi^{\text{ind}}_{\omega k}(z) = \frac{2\pi}{k}\Delta(\omega)(ik_x d^{\text{sp}}_x(\omega - k_x V) + ik_y d^{\text{sp}}_y(\omega - k_x V) +$$

$$+ kd^{\text{sp}}_z(\omega - k_x V))\exp(-k(z + z_0)) \tag{A18}$$

Comparing Eqs.(A16) with (A18) we see that when passing from fluctuating dipole $\mathbf{d}^{\text{sp}}(t)$ to the constant dipole $\mathbf{d}$ one needs to replace $\mathbf{d}^{\text{sp}}(\omega - k_x V)$ by $2\pi\,\delta(\omega - k_x V)\mathbf{d}$ .

## ii) Cylindrical surface

At first , consider an auxiliary problem of interaction between the resting point charge and a cylindrical surface (see Fig.10). The involved Poisson's equation (using cylindrical coordinate system) is written by

$$\left(\frac{\partial^2}{\partial r^2} + \frac{1}{r}\frac{\partial}{\partial r} + \frac{1}{r^2}\frac{\partial^2}{\partial\phi^2} + \frac{\partial^2}{\partial z^2}\right)\Phi(r,\phi,z) = -\frac{4\pi Ze}{R}\delta(r - R)\delta(\phi)\delta(z) \tag{A19}$$

Chosing the form of solution to be

$$\Phi(r,\phi,z) = \sum_{n=-\infty}^{n=+\infty}\int_{-\infty}^{+\infty}\frac{dk}{2\pi}u_n(r)\exp(ikz)\frac{\exp(in\phi)}{2\pi} \tag{A20}$$

and introducing (A20) into (A19) yields

$$\cdot\left(\frac{d^2}{dr^2} + \frac{1}{r}\frac{d}{dr} - \frac{n^2}{r^2} - k^2\right)u_n(r) = -\frac{2Ze}{R}\delta(r - R) \tag{A21}$$

After integrating Eq.(A21) over $r$ in the range $(R - \varepsilon, R + \varepsilon)$ and taking the limit $\varepsilon \to 0$ , we get the following boundary conditions at $r = R$ :

$$u_n(R+0) - u_n(R-0) = 0,$$
$$du_n(R+0)/dr - du_n(R-0)/dr = -2Ze/R \tag{A22}$$

Therefore, the electric penential is continuous function at the particle location point ( $r = R$ ), while the involved derivative has the finite break depending on the charge value.

Introducing the dimensionless variable $\xi = kr$ , Eq.(A21) (at $r \neq R$ ) takes the form of the modified Bessel equation, of which the general solution is given by linear combination of the cylindrical functions, $I_n(\xi)$ and $K_n(\xi)$ [115]. It is natural way to separate the integration range in (A21) on the domains

$0 \le r \le a$ , $a < r \le R$ and $r > R$ . In these domains the finite solutions are of the form:

$$u_n(r) = D_n I_n(kr) , \; 0 \le r \le a ;$$
$$u_n(r) = A_n K_n(kr) + B_n I_n(kr) , \; a < r \le R ; \qquad (A23)$$
$$u_n(r) = C_n K_n(kr) , \; r > R$$

Assuming $\varepsilon$ to be the static dielectric permittivity of the cylinder, the conventional continuity conditions at $r = a$ yield:

$$u_n(a+0) - u_n(a-0) = 0,$$
$$du_n(a+0)/dr - \varepsilon \, du_n(a-0)/dr \qquad (A24)$$

Coefficients $A_n, B_n, C_n, D_n$ are explicitly determined from (A22) and (A24). Solving the corresponding system of linear equations allows to write down the final expression for $u_n(r)$ :

a) $0 \le r < a$

$$u_n(r) = 2Ze\left(-\frac{(\varepsilon-1)I'_n(ka)I_n(ka)K_n(kR)K_n(kr)}{\varepsilon I'_n(ka)K_n(ka) - I_n(ka)K'_n(ka)} + I_n(kr)K_n(kR)\right) \qquad (A25)$$

b) $r > R$

$$u_n(r) = 2Ze\left(-\frac{(\varepsilon-1)I'_n(ka)I_n(ka)K_n(kR)K_n(kr)}{\varepsilon I'_n(ka)K_n(ka) - I_n(ka)K'_n(ka)} + I_n(kR)K_n(kr)\right) \qquad (A26)$$

From (A25) and (A26) we obtain the Green's function of the involved radial equation

$$\left(\frac{d^2}{dr^2} + \frac{1}{r}\frac{d}{dr} - \frac{n^2}{r^2} - k^2\right)G_n(r,r') = \delta(r-r') ,$$

being adjusted to the bondary conditions:

$$G_n(r,r') = \begin{cases} r'\big[\Delta_n K_n(kr)K_n(kr') - I_n(kr')K_n(kr)\big], \, r > r' \\ r'\big[\Delta_n K_n(kr)K_n(kr') - I_n(kr)K_n(kr')\big], \, a < r < r' \end{cases} \qquad (A27)$$

$$\Delta_n = \frac{(\varepsilon-1)I'_n(ka)I_n(ka)}{\varepsilon I'_n(ka)K_n(ka) - I_n(ka)K'_n(ka)} \qquad (A28)$$

Coming down to the case of moving particle, one should bear in mind that the expansion (A20) will contain an additional Fourier –frequency integral, while in Eqs. (A27), (A28) the static dielectric permittivity is replaced by the dynamic one - $\varepsilon(\omega)$ .

a) *A moving charged particle*. In this case the involved Poisson's equation reads

$$\left(\frac{\partial^2}{\partial r^2}+\frac{1}{r}\frac{\partial}{\partial r}+\frac{1}{r^2}\frac{\partial^2}{\partial \phi^2}+\frac{\partial^2}{\partial z^2}\right)\Phi(r,\phi,z,t)=-\frac{4\pi Ze}{R}\delta(r-R)\delta(\phi)\delta(z-Vt) \quad \text{(A29)}$$

Furthermore, Eq.(A20) is replaced by

$$\Phi(r,\phi,z,t)=\sum_{n=-\infty}^{n=+\infty}\int_{-\infty}^{+\infty}d\omega\int_{-\infty}^{+\infty}\frac{dk}{2\pi}u_n(r)\frac{\exp(in\phi)}{2\pi}\frac{\exp(i(kz-\omega t)}{2\pi} \quad \text{(A30)}$$

From (A29), (A30) we obtain the proper Fourier –component of the potential, $u_n(\omega k,r)$:

$$\left(\frac{d^2}{dr^2}+\frac{1}{r}\frac{d}{dr}-\frac{n^2}{r^2}-k^2\right)u_n(\omega k,r)=-\frac{8\pi^2 Ze}{R}\delta(r-R)\delta(\omega-kV) \quad \text{(A31)}$$

Performing convolution of the right-hand side of Eq. (A31) with the Green's function (A27), the Fourier –component of the induced potential takes the form

$$u^{ind}{}_n(\omega k,r)=-8\pi^2 Ze\,\delta(\omega-kV)\Delta_n(\omega)K_n(kR)K_n(kr) \quad \text{(A32)}$$

**b)** ***A moving dipole molecule.*** The Poisson's equation $\Delta\Phi=4\pi\,\text{div}\,\mathbf{P}$ reads

$$\left(\frac{\partial^2}{\partial r^2}+\frac{1}{r}\frac{\partial}{\partial r}+\frac{1}{r^2}\frac{\partial^2}{\partial \phi^2}+\frac{\partial^2}{\partial z^2}\right)\Phi(r,\phi,z,t)=$$

$$=4\pi\left\{\begin{array}{l}\dfrac{1}{r}\dfrac{\partial}{\partial r}\delta(r-R)\delta(\phi)\delta(z-Vt)d_r+\\[2mm]\dfrac{1}{r}\dfrac{\partial}{\partial \phi}\dfrac{\delta(r-R)}{r}\delta(\phi)\delta(z-Vt)d_\phi+\\[2mm]\dfrac{\partial}{\partial z}\dfrac{\delta(r-R)}{r}\delta(\phi)\delta(z-Vt)d_z\end{array}\right\} \quad \text{(A33)}$$

Taking the Fourier transform of the right –hand side of Eq. (A33) over variables $\omega,k$, and Fourier series over $n\phi$, and substituting (A31) into (A33) yields

$$\left(\frac{d^2}{dr^2}+\frac{1}{r}\frac{d}{dr}-\frac{n^2}{r^2}-k^2\right)u_n(\omega k,r)=8\pi^2\delta(\omega-kV)\times$$

$$\left(\frac{1}{r}\delta'(r-R)d_r+\frac{in}{r^2}\delta(r-R)d_\phi+\frac{ik}{r}\delta(r-R)d_z\right) \quad \text{(A34)}$$

Finally, after convolution of the right –hand side of Eq.(A34) with (A27), the involved Fourier –component of the induced potential is given by

$$u^{\text{ind}}{}_n(\omega k, r) = 8\pi^2 \delta(\omega - kV) \Delta_n(\omega) \left\{ -K_n(kr)K'_n(kR)|k|d_r + \right.$$

$$\left. + \frac{in}{R} K_n(kR)K_n(kr)d_\phi + ikK_n(kR)K_n(kr)d_z \right\} \tag{A35}$$

**c) _Neutral spherically-symmetric particle (ground-state atom)._** Eq.(A33) must be modified by replacing the projections of the constant dipole moment by the time – dependent ones. By analogy with derivation of (A33), we can write (see also text after (A18)):

$$\left( \frac{d^2}{dr^2} + \frac{1}{r}\frac{d}{dr} - \frac{n^2}{r^2} - k^2 \right) u_n(\omega k, r) = 4\pi \left( \frac{1}{r}\delta'(r - R)d^{sp}{}_{\text{r}}(\omega - kV) + \right.$$

$$\left. + \frac{in}{r^2} \delta(r - R)d^{sp}{}_\phi(\omega - kV) + \frac{ik}{r}\delta(r - R)d^{sp}{}_z(\omega - kV) \right) \tag{A36}$$

(A36) The corresponding Fourier –component of the induced potential is given by

$$u_n(\omega k; r) = 4\pi\Delta_n(\omega) \left\{ \begin{array}{l} -K_n(kr)K'_n(kR)|k|d_{\text{r}}{}^{sp}(\omega - kV) + \\ + \dfrac{in}{R} K_n(kr)K_n(kR)d_\phi{}^{sp}(\omega - kV) + \\ + ikK_n(kr)K_n(kR)d_z{}^{sp}(\omega - kV) \end{array} \right\} \tag{A37}$$

Using (A37), the induced electric field is given by

$$\mathbf{E}^{\text{ind}}(r, \phi, z, t) = \frac{1}{(2\pi)^3} \sum_{n=-\infty}^{\infty} \iint \mathbf{E}_{\omega k}{}^{\text{ind}}(r) e^{in\phi} e^{i(kz - \omega t)} \, d\omega dk \tag{A38}$$

$$E_{\text{r}}{}^{\text{ind}}(\omega k; r) = -\frac{d}{dr} u_n(\omega k; r) \tag{A39}$$

$$E_\phi{}^{\text{ind}}(\omega k; r) = -\frac{in}{r} u_n(\omega k; r) \tag{A40}$$

$$E_z{}^{\text{ind}}(\omega k; r) = -ik u_n(\omega k; r) \tag{A41}$$

In the case of a particle moving inside a cylindrical channel, the calculations are completely the same and result in the modified Eqs. (A32), (A35), (A37)-(A41) after the replacement $K_n(x) \leftrightarrow I_n(x)$.

### B. Relation between the rate of work of electromagnetic field in different reference frames

Consider a moving nonmagnetic particle (having zero magnetic moment, $\mathbf{m} = 0$, in its rest frame), a surface and the corresponding reference frames ($K'$) and ($K$), respectively (Fig.4). Using the Lorentz transformations for the density current, charge density and electric field amplitude yields

$$j_x{}' = \frac{j_x - \rho V}{\sqrt{1 - V^2/c^2}}$$

$$j_y{}' = j_y, j_z{}' = j_z \qquad\qquad (B1)$$

$$\rho' = \frac{\rho - V j_x/c^2}{\sqrt{1 - V^2/c^2}}$$

$$E_x{}' = E_x$$

$$E_y{}' = \frac{E_y - VH_z/c}{\sqrt{1 - V^2/c^2}}, \qquad\qquad (B2)$$

$$E_z{}' = \frac{E_z + VH_y/c}{\sqrt{1 - V^2/c^2}}$$

With account of $d^3r = d^3r'\sqrt{1 - V^2/c^2}$ , (B1), (B2) we get

$$\int \langle \mathbf{j}'\mathbf{E}'\rangle d^3r' = \frac{1}{1 - V^2/c^2}\left\{ \int \langle \mathbf{j}\mathbf{E}\rangle d^3r - F_x V \right\}, \qquad\qquad (B3)$$

$$F_x = \int \langle \rho E_x \rangle d^3r + \frac{1}{c}\int \langle [\mathbf{j}\mathbf{H}]_x \rangle d^3r \qquad\qquad (B4)$$

where $F_x$ is the tangential force in the $K$-system. Evidently, in the limit $c \to \infty$ Eq. (B3) reduces to (3.15).

By definition, $\mathbf{j}' = \mathbf{j}'_e + \mathbf{j}'_m = \dfrac{\partial \mathbf{P}'}{\partial t'} + c\,\mathrm{rot}\,\mathbf{M}'$ , $\mathbf{P}' = \delta(x')\delta(y')\delta(z')\mathbf{d}'$ ,

$\mathbf{M}' = \delta(x')\delta(y')\delta(z')\mathbf{m}'$ ,

Therefore, in the $K'$-frame of the nonmagnetic particle ($\mathbf{m}' = 0$, $\mathbf{j}'_m = c\,\mathrm{rot}\,\mathbf{m}' = 0$) we get

$$\int \langle \mathbf{j}'\mathbf{E}'\rangle d^3r' = \int \left\langle \frac{\partial \mathbf{P}'}{\partial t'}\mathbf{E}' \right\rangle d^3r' = \int \langle \delta(x')\delta(y')\delta(z')\dot{\mathbf{d}}'\mathbf{E}'\rangle d^3r' = \langle \dot{\mathbf{d}}'\mathbf{E}'\rangle \qquad (B5)$$

In addition, from the Lorentz transformation for the parallel and perpendicular components of the polarization $\mathbf{P}$, assuming $\mathbf{M}' = 0$,

$$\mathbf{P}_{\mathrm{II}} = \mathbf{P}'_{\mathrm{II}}, \quad \mathbf{P}_\perp = \frac{\mathbf{P}'_\perp + \dfrac{1}{c}\mathbf{V}\times\mathbf{M}'}{\sqrt{1 - V^2/c^2}},$$

and using the identity $\delta(\alpha x) = \delta(x)/|\alpha|$ , we get the relations

$$d_x(t) = d'_x(t')\sqrt{1 - V^2/c^2}$$
$$d_y(t) = d'_y(t')$$
$$d_z(t) = d'_z(t')$$

(B6)

Using (B5), (B6) and relation $dt' = dt\sqrt{1 - V^2/c^2}$ yields

$$\int \langle \mathbf{j'E'} \rangle d^3 r' = \langle \mathbf{\dot{d}'E'} \rangle = \frac{1}{1 - V^2/c^2} \left\{ \langle \mathbf{\dot{d}E} \rangle - \frac{V}{c} \langle [\mathbf{\dot{d}H}]_x \rangle \right\}$$

(B7)

Finally, from (B3), (B5), (B7) we get

$$\int \langle \mathbf{jE} \rangle d^3 r = F_x V + \langle \mathbf{\dot{d}E} \rangle - \frac{V}{c} \langle [\mathbf{\dot{d}H}]_x \rangle \equiv F_x V + dQ/dt ,$$

(B8)

with $dQ/dt$ being the heat flow. It must be emphasized that all the quantities in (B8) are expressed in the laboratory frame.

## C. Surface dielectric response functions

1) approximation for the low –velocity case of atomic particles ($V << V_{\mathrm{F}}$) [148]

($\theta(x)$ is the unit step –function, the atomic units are used here and in (3.26)-(3.30))

$$\varepsilon(k, \omega) = 1 + \frac{\omega_{\mathrm{P}}^2}{V_{\mathrm{F}}^2 k^2 [1 - i\pi\omega\,\theta(2k_{\mathrm{F}} - k)/2kV_{\mathrm{F}}]/3 - \omega(\omega + i\gamma)}$$

(C1)

Eq. (C1) generalizes the Lindhard's hydrodynamic model [149] with the plasmon propagation velocity $V_{\mathrm{F}}/\sqrt{3}$, $\omega_{\mathrm{P}}$ is the bulk plasma frequency, $V_{\mathrm{F}}$ is the Fermi velocity. The term proportional to $\omega V_{\mathrm{F}} k$ in denominator describes the damping due to electron –hole excitations. The term containing $\gamma$ describes "frictional" damping of collective states. If Eq. (C1) is expanded in a power series in $\omega$ , it agrees with the small frequency expansion of $\varepsilon^{\mathrm{L}}(k, \omega)$, being the Lindhard dielectric function, to first order. The presence of $\theta(x)$ accounts for the fact that $\mathrm{Im}(1/\varepsilon^{\mathrm{L}}(k, \omega))$ vanishes at $k > 2k_{\mathrm{F}}$, because the small energy particle –hole excitations are forbidden for the momentum transfer much larger than $2k_{\mathrm{F}}$. In the low –frequency limit, using (3.24), (C1) yields [86]

$$\Delta(q, \omega) = \frac{1 - q(A + B + C)}{q(A + B + C) + 1}$$

(C2)

$$A = \frac{1}{(q^2 + q^2_{\mathrm{TF}})^{1/2}}$$

(C3)

$$B = \frac{-\mathrm{i}\omega\, q_{TF}}{4\,V_F}\left[\frac{2q^2{}_{TF}+q^2}{q^2{}_{TF}(q^2+q^2{}_{TF})^{3/2}}\ln\!\left(\frac{(q^2+q^2{}_{TF})^{1/2}+q_{TF}}{(q^2+q^2{}_{TF})^{1/2}-q_{TF}}\right)-\frac{2}{(q^2+q^2{}_{TF})q_{TF}}\right] \quad \text{(C4)}$$

$$C = \frac{-\mathrm{i}\omega\,\gamma\, q^2{}_{TF}}{s^2}\left[\left(\frac{(q^2+q^2{}_{TF})^{1/2}-q}{q^4{}_{TF}q(q^2+q^2{}_{TF})^{1/2}}\right)-\frac{1}{2q^2{}_{TF}(q^2+q^2{}_{TF})^{3/2}}\right] \quad \text{(C5)}$$

where $s = V_F/\sqrt{3}$ , $q_{TF} = \omega_P/s$ .

Also, it is interesting to get the small wave –vector expansion (3.26) without restrictions on frequency. In this case we obtain [73]

$$\Delta(q,\omega) = \frac{\omega_s{}^2}{\omega_s{}^2-\omega^2-\mathrm{i}\omega\gamma}\left(1-\frac{\mathrm{i}}{3}\frac{\omega V_F}{\omega^2-\omega_s{}^2+\mathrm{i}\omega\gamma}q\ln(4q_F/q)\right) \quad \text{(C6)}$$

where $q_F$ is the Fermi wave –vector . By definition,
$V_F = q_F = \pi q_{TF}{}^2/4$, $q_{TF} = \sqrt{3}\,\omega_P/V_F$ , $\gamma = \omega_P{}^2/4\pi\,\sigma$ , with $\sigma$ being the static conductance.

2) local Drude approximation

$$\varepsilon(\omega) = 1 - \frac{\omega_P{}^2}{\omega(\omega+\mathrm{i}/\tau)} \quad \text{(C7)}$$

where $\tau$ is the relaxation time for electron scattering. Substituting (C7) into (3.27) yields ( $\omega_s = \omega_P/\sqrt{2}$ )

$$\Delta(\omega) = \frac{\omega_s{}^2(\omega_s{}^2-\omega^2)}{(\omega_s{}^2-\omega^2)^2+\omega^2/\tau^2}+\mathrm{i}\frac{\omega_s{}^2\,\omega/\tau}{(\omega_s{}^2-\omega^2)^2+\omega^2/\tau^2} \quad \text{(C8)}$$

3) "weak" nonlocal approximation [150]

$$\Delta(q,\omega) = \frac{\varepsilon(\omega)-1}{\varepsilon(\omega)+1}\left(1+\frac{2\varepsilon(\omega)}{\varepsilon(\omega)+1}qd(\omega)+...\right) \quad \text{(C9)}$$

where $\varepsilon(\omega)$ is the Drude function (3.31), $d(\omega)$ is the shift of the centroid of the screening charge in the metal. In the limit $\omega\to 0$ , $q << \omega/V_F$ the imaginary part of (3.33) takes the form [40]

$$\Delta''(q,\omega) = 2\,I(q)\frac{\omega}{\omega_P}\frac{q}{q_F},$$

where $I(q)$ is weakly dependent function of $q$ of the order of 1. This function is close to that one in (4.55).

4) low –frequency local dielectric response for good conductors

$$\varepsilon(\omega) = 1+\frac{4\pi\,\sigma\,\mathrm{i}}{\omega} \quad , \quad \text{(C10)}$$

$$\Delta''(\omega) = \frac{2\pi\sigma\omega}{\omega^2 + 4\pi^2\sigma^2} \tag{C11}$$

5) Debye appproximation for dielectrics

$$\varepsilon(\omega) = 1 + \frac{\varepsilon(0) - 1}{1 - i\omega\tau} \tag{C12}$$

$$\Delta''(\omega) = \frac{2(\varepsilon(0) - 1)\omega/\tau}{\omega^2 + \left(\frac{\varepsilon(0) + 1}{2\tau}\right)^2} \tag{C13}$$

where $\tau$ is the dipole relaxation time, $\varepsilon(0)$ is the static dielectric permittivity .

6) Lorentzian absorption line

$$\varepsilon_i(\omega) = \varepsilon_{i\infty} + \frac{\left(\varepsilon_{i0} - \varepsilon_{i\infty}\right)\omega_{0i}^2}{\left(\omega_{0i}^2 - \omega^2\right) - i\gamma_i\omega} \tag{C14}$$

where the subscript "i" denotes type of material , $\varepsilon_{i0}$ and $\varepsilon_{i\infty}$ are the corresponding static and high –frequency values of the dielectric permittivity, $\omega_{0i}$ and $\gamma_i$ are the line frequency and width.

## D. Fluctuation-dissipation theorem and correlators of physical quantities

According to the general results of the fluctuation electromagnetic theory, the spectral density of the symmetrized correlation function of the equilibrium fluctuating electromagnetic field is related with the retarded photon Green's function [8]

$$\left(E^{sp}_i(\mathbf{r})E^{sp}_k(\mathbf{r}')\right) = \frac{i}{2}\coth\frac{\omega\hbar}{2k_BT}\frac{\omega^2}{c^2}\left[D_{ik}(\omega,\mathbf{r},\mathbf{r}') - D^*_{ki}(\omega,\mathbf{r}',\mathbf{r})\right] \tag{D1}$$

In an isotropic medium without of the spatial dispersion, the retarded Green's function $D_{ik}(\omega,\mathbf{r},\mathbf{r}')$ satisfies equation [8]

$$\left(rot_{ik}rot_{kl} - \frac{\omega^2}{c^2}\varepsilon(\omega)\delta_{il}\right)D_{lm}(\omega,\mathbf{r},\mathbf{r}') = -4\pi\hbar\delta_{im}\delta(\mathbf{r} - \mathbf{r}') \tag{D2}$$

At the same time, as it follows from the Maxwell equations, the electric field created by a neutral particle with the dipole moment $\mathbf{d}(t)$, being located at a space point $\mathbf{r}'$, obeys the equation [8]

$$\left(rot\,rot - \frac{\omega^2}{c^2}\varepsilon(\omega)\right)\mathbf{E}(\omega,\mathbf{r}) = 4\pi\frac{\omega^2}{c^2}\mathbf{d}(\omega)\delta(\mathbf{r} - \mathbf{r}') \tag{D3}$$

Comparing (D2) and (D3) one sees that function $D_{lm}(\omega,\mathbf{r},\mathbf{r}')$ at fixed $m$ and $\mathbf{r}'$

( being the incoming parameters) exactly equals the electric field produced by a point –like dipole at $\mathbf{r}'$:

$$d_1(\omega) = -\frac{\hbar c^2}{\omega^2} \delta_{1m} \tag{D4}$$

Due to this, we may not solve Eq.(D2), while instead, assuming the nonrelativistic case ( therefore, Eq. (D2) again reduces to the Posson equation) and to use the previously obtained solutions (A18) and (A37) at $V=0$ (for a solid surface in rest). Now let us consider the special cases.

### a) *Flat surface*

Due to homogeneity over coordinates $x$, $y$, the retarded Green function $D_{ik}(\omega, \mathbf{r}, \mathbf{r}')$ can be written by

$$D_{ik}(\omega, \mathbf{r}, \mathbf{r}') = D_{ik}(\omega, x - x', y - y', z, z') =$$
$$\iint \frac{dk_x}{2\pi} \frac{dk_y}{2\pi} D_{ik}(\omega \mathbf{k}, z, z') \exp\left[i\left(k_x(x - x') + k_y(y - y')\right)\right] \tag{D5}$$

where the corresponding Fourier transform is given by

$$D_{ik}(\omega \mathbf{k}, z, z') =$$
$$\iint d(x - x') d(y - y') D_{ik}(\omega, x - x', y - y', z, z') \exp\left[-i\left(k_x(x - x') + k_y(y - y')\right)\right] \tag{D6}$$

Making use of the Fourier transform of Eq.(D1) yields

$$\left(E^{sp}{}_i(z) E^{sp}{}_k(z')\right) = \frac{i}{2} \coth \frac{\omega \hbar}{2k_B T} \frac{\omega^2}{c^2} \left[D_{ik}(\omega \mathbf{k}, z, z') - D^*{}_{ki}(\omega \mathbf{k}, z, z')\right] \tag{D7}$$

Furthermore, bearing in mind remarks after Eq.(D4) and expressing $D_{ik}(\omega \mathbf{k}, z, z')$ with account of (D4) and (A18), we get

$$D_{xm}(\omega \mathbf{k}, z, z') = -ik_x \phi^{ind}{}_{\omega \mathbf{k}}(z) =$$
$$= ik_x \frac{2\pi}{k} \frac{\hbar c^2}{\omega^2} \Delta(\omega)\left(ik_x \delta_{xm} + ik_y \delta_{ym} + k\delta_{zm}\right) \exp(-k(z + z')) \tag{D8}$$

Assuming $m = x, y, z$, from (D8) we obtain

$$D_{xx}(\omega \mathbf{k}, z, z') = = -\frac{2\pi}{k} k_x{}^2 \left(\frac{\hbar c^2}{\omega^2}\right) \Delta(\omega) \exp(-k(z + z')) \tag{D9}$$

$$D_{xy}(\omega \mathbf{k}, z, z') = D_{yx}(\omega \mathbf{k}, z, z') = -\frac{2\pi}{k} k_x k_y \left(\frac{\hbar c^2}{\omega^2}\right) \Delta(\omega) \exp(-k(z + z')) \tag{D10}$$

$$D_{x\,z}(\omega\mathbf{k},z,z') = -D_{z\,x}(\omega\mathbf{k},z,z') = \frac{2\pi}{k}\,\mathrm{i}\,k_x k\left(\frac{\hbar c^2}{\omega^2}\right)\Delta(\omega)\exp(-k(z+z')) \tag{D11}$$

In the same way, we can write down other components of $D_{x\,m}(\omega\mathbf{k},z,z')$:

$$D_{y\,y}(\omega\mathbf{k},z,z') = -\frac{2\pi}{k}\,k_y{}^2\left(\frac{\hbar c^2}{\omega^2}\right)\Delta(\omega)\exp(-k(z+z'))$$

$$D_{z\,z}(\omega\mathbf{k},z,z') = -\frac{2\pi}{k}\,k^2\left(\frac{\hbar c^2}{\omega^2}\right)\Delta(\omega)\exp(-k(z+z'))$$

$$D_{x\,z}(\omega\mathbf{k},z,z') = -2\pi\,k_x\left(\frac{\hbar c^2}{\omega^2}\right)\Delta(\omega)\exp(-k(z+z'))$$

$$D_{y\,z}(\omega\mathbf{k},z,z') = -D_{z\,y}(\omega\mathbf{k},z,z') = \frac{2\pi}{k}\,\mathrm{i}\,k_y k\left(\frac{\hbar c^2}{\omega^2}\right)\Delta(\omega)\exp(-k(z+z'))$$

$$\tag{D12}$$

One can see that Eqs.(D9)-(D12) have no singularities at $z = z'$, because in our derivation of $D_{i\,m}(\omega\mathbf{k},z,z')$ we used the induced electric field (A18), which does not contain the self–field of the fluctuating dipole. Therefore, Eqs.(D9)-(D12) determine "normalized" Green's functions which are needed when calculating spectral densities of the surface fluctuation field at the particle location point $z = z' = z_0$. Then, making use of (D1) and (D12) yields

$$\left(E^{s\,p}{}_x(z_0)E^{s\,p}{}_x(z_0)\right)_{\omega\mathbf{k}} = \hbar\coth\frac{\omega\hbar}{2k_B T}\frac{2\pi}{k}k_x{}^2\exp(-2kz_0)\,\mathrm{Im}\,\Delta(\omega) \tag{D13}$$

and, analogously,

$$\left(E^{s\,p}{}_y(z_0)E^{s\,p}{}_y(z_0)\right)_{\omega\mathbf{k}} = \hbar\coth\frac{\omega\hbar}{2k_B T}\frac{2\pi}{k}k_y{}^2\exp(-2kz_0)\,\mathrm{Im}\,\Delta(\omega)$$

$$\left(E^{s\,p}{}_z(z_0)E^{s\,p}{}_z(z_0)\right)_{\omega\mathbf{k}} = \hbar\coth\frac{\omega\hbar}{2k_B T}\frac{2\pi}{k}k^2\exp(-2kz_0)\,\mathrm{Im}\,\Delta(\omega)$$

$$\tag{D14}$$

Assuming the case of stationary fluctuations, the correlator of two physical quantities can be related with the involved spectral density [151]

$$\left\langle A_\omega B_{\omega'}\right\rangle = 2\pi\,\delta(\omega+\omega')\left(AB\right)_\omega \tag{D15}$$

In our case, this leads to

$$\left\langle \mathbf{E^{sp}}_{\omega\mathbf{k}}(z_0)\mathbf{E^{sp}}_{\omega'\mathbf{k}'}(z_0)\right\rangle = (2\pi)^3\,\delta(\omega+\omega')\delta(\mathbf{k}+\mathbf{k}')\left(\mathbf{E^{sp}}(z_0)\mathbf{E^{sp}}(z_0)\right)_{\omega\mathbf{k}} =$$

$$= (2\pi)^3\,\delta(\omega+\omega')\delta(k_x+k_x')\delta(k_y+k_y')\times$$

$$\times\left[\left(E_x^{sp}(z_0)E_x^{sp}(z_0)\right)_{\omega\mathbf{k}} + \left(E_y^{sp}(z_0)E_y^{sp}(z_0)\right)_{\omega\mathbf{k}} + \left(E_z^{sp}(z_0)E_z^{sp}(z_0)\right)_{\omega\mathbf{k}}\right]$$

(D16)

Finally, substituting (D13) into (D16) yields:

$$\left\langle \mathbf{E^{sp}}_{\omega\mathbf{k}}(z_0)\mathbf{E^{sp}}_{\omega'\mathbf{k}'}(z_0)\right\rangle = 2(2\pi)^4\,k\,\exp(-2kz_0)\hbar\coth\frac{\omega\hbar}{2k_\mathrm{B}T}\times$$

$$\times\,\mathrm{Im}\,\Delta(\omega)\delta(\omega+\omega')\delta(k_x+k_x')\delta(k_y+k_y')$$

(D17)

## b) *Cylindrical surface*

By analogy with the case of flat surface, expressing the Fourier transform of the retarded Green's function $D_{ik}(\omega\mathbf{k},r,R)$ via the electric field of the dipole problem (see Eq. (D4)), and assuming $V=0$ in Eq. (A37) we get

$$D_{rm}(\omega k,r,R) = -\frac{d}{dr}\Phi^{in}_{\omega k}(r) =$$

$$= \frac{2\hbar c^2}{\omega^2}\sum_{n=-\infty}^{n=\infty}\Delta_n(\omega)\Big\{-k^2 K_n'(kr)K_n'(kR)\delta_{rm} + \frac{in|k|}{R}K_n'(kr)K_n(kR)\delta_{\phi m} +$$

$$ik|k|K_n'(kr)K_n(kR)\delta_{zm}\Big\}\exp(in\phi)$$

(D18)

Taking m = r in Eq. (D18) yields

$$D_{rr}(\omega k,r,R) = -\frac{2\hbar c^2}{\omega^2}\sum_{n=-\infty}^{n=\infty}\Delta_n(\omega)k^2 K_n'(kr)K_n'(kR)\exp(in\phi)$$

(D19)

By analogy with that, after simplifications we obtain two other diagonal components of the Green's function

$$D_{\phi\phi}(\omega k,r,R) = -\frac{2\hbar c^2}{\omega^2}\sum_{n=-\infty}^{n=\infty}\Delta_n(\omega)\frac{n^2}{R^2}K_n(kr)K_n(kR)\exp(in\phi)$$

(D20)

$$D_{zz}(\omega k,r,R) = -\frac{2\hbar c^2}{\omega^2}\sum_{n=-\infty}^{n=\infty}\Delta_n(\omega)k^2 K_n(kr)K_n(kR)\exp(in\phi)$$

(D21)

In this case, the Fourier –transformed Eq.(D1) takes the form

$$\left(E_i^{sp}(r)E_k^{sp}(R)\right)_{\omega k} = \frac{i}{2}\coth\frac{\omega\hbar}{2k_\mathrm{B}T}\frac{\omega^2}{c^2}\Big[D_{ik}(\omega,r,R) - D^*_{ki}(\omega,R,r)\Big],$$

$$i,k = r,\phi,z$$

(D22)

Eq.(D22) differs from (D7) in two aspects: i) now the wave vector is one-dimensional; 2) the coordinate $z$ is replaced by $r$ - the distance to the cylinder axis (see Fig.10). Subsituting (D20)-(D21) into (D22) and taking $r=R$, $\phi=0$ (the particle location point) yields

$$\left(E_r^{\ sp}(R)E_r^{\ sp}(R)\right)_{\omega \mathbf{k}} = 2\hbar \coth\frac{\omega\hbar}{2k_B T}\sum_{n=-\infty}^{n=\infty}\mathrm{Im}\,\Delta_n(\omega)k^2 K'_n(kR)^2 \tag{D23}$$

$$\left(E_\phi^{\ sp}(R)E_\phi^{\ sp}(R)\right)_{\omega \mathbf{k}} = 2\hbar \coth\frac{\omega\hbar}{2k_B T}\sum_{n=-\infty}^{n=\infty}\mathrm{Im}\,\Delta_n(\omega)\frac{n^2}{R^2} K_n(kR)^2 \tag{D24}$$

$$\left(E_z^{\ sp}(R)E_z^{\ sp}(R)\right)_{\omega \mathbf{k}} = 2\hbar \coth\frac{\omega\hbar}{2k_B T}\sum_{n=-\infty}^{n=\infty}\mathrm{Im}\,\Delta_n(\omega)k^2 K_n(kR)^2 \tag{D25}$$

Next, taking account of the relationship between the spectral density and correlator [151] (cf. with (D16)), we have

$$\left\langle \mathbf{E}^{sp}_{\omega \mathbf{k}}(R)\mathbf{E}^{sp}_{\omega' \mathbf{k}'}(R)\right\rangle = (2\pi)^2\,\delta(\omega+\omega')\delta(k+k')\times$$

$$\times\left[\left(E_r^{\ sp}(R)E_r^{\ sp}(R)\right)_{\omega \mathbf{k}} + \left(E_\phi^{\ sp}(R)E_\phi^{\ sp}(R)\right)_{\omega \mathbf{k}} + \left(E_z^{\ sp}(R)E_z^{\ sp}(R)\right)_{\omega \mathbf{k}}\right] \tag{D26}$$

Finally, substituting (D23)-(D25) into (D26) yields

$$\left\langle \mathbf{E}^{sp}_{\omega \mathbf{k}}(R)\mathbf{E}^{sp}_{\omega' \mathbf{k}'}(R)\right\rangle = (2\pi)^2\,\delta(\omega+\omega')\delta(k+k')\frac{2\hbar}{R^2}\coth\frac{\omega\hbar}{2k_B T}\times$$

$$\times\sum_{n=-\infty}^{n=\infty}\mathrm{Im}\,\Delta_n(\omega)K_n(kR)^2(n^2+(kR)^2+(kR)^2\Phi^2_n(kR)) \tag{D27}$$

with $\Phi_n(x)=d\ln K_n(x)/dx$. Moreover,

$$\Delta_n(\omega)=\frac{(\varepsilon(\omega)-1)I'_n(ka)I_n(ka)}{\varepsilon(\omega)\,I'_n(ka)K_n(ka)-I_n(ka)K'_n(ka)} \tag{D28}$$

### b) Cylindrical channel (concave surface)

According to Appendix A, all the formulae which are analogous to Eqs.(D27), (D28), can be obtained after the replacements

$$K_n(x)\leftrightarrow I_n(x),\ \Phi_n(x)\rightarrow \Psi_n(x)=d\ln I_n(x)/dx\ .$$ Then we get

$$\left\langle \mathbf{E}^{sp}_{\omega \mathbf{k}}(R)\mathbf{E}^{sp}_{\omega' \mathbf{k}'}(R) \right\rangle = (2\pi)^2 \, \delta(\omega + \omega')\delta(k + k')\frac{2\hbar}{R^2}\coth\frac{\omega \hbar}{2k_B T} \times$$

$$\times \sum_{n=-\infty}^{n=\infty} \mathrm{Im}\,\widetilde{\Delta}_n(\omega)I_n(kR)^2\left(n^2 + (kR)^2 + (kR)^2\,\Psi^2_n(kR)\right) \tag{D29}$$

$$\widetilde{\Delta}_n(\omega) = \frac{(\varepsilon(\omega) - 1)K'_n(ka)K_n(ka)}{\varepsilon(\omega)\,K'_n(ka)I_n(ka) - K_n(ka)I'_n(ka)} \tag{D30}$$

## E. Mathematical details of the calculation of the particle -surface interaction

In the process of calculation of correlators like $\left\langle \mathbf{d}^{sp}\mathbf{E}^{in} \right\rangle, \left\langle \mathbf{d}^{in}\mathbf{E}^{sp} \right\rangle, \left\langle \dot{\mathbf{d}}^{sp}\mathbf{E}^{in} \right\rangle, \left\langle \dot{\mathbf{d}}^{in}\mathbf{E}^{sp} \right\rangle$ the following integrals are encountered (for a flat surface)

$$J_1 = \int_{-\infty}^{+\infty}d\omega\int_{-\infty}^{+\infty}dk_x k_x k \exp(-2kz_0)\Delta(\omega - k_x V)\alpha''(\omega)\coth\frac{\omega \hbar}{2k_B T} \tag{E1}$$

$$J_2 = \int_{-\infty}^{+\infty}d\omega\int_{-\infty}^{+\infty}dk_x k_x k \exp(-2kz_0)\alpha(\omega - k_x V)\Delta''(\omega)\coth\frac{\omega \hbar}{2k_B T} \tag{E2}$$

$$J_3 = \int_{-\infty}^{+\infty}d\omega\omega\int_{-\infty}^{+\infty}dk_x k \exp(-2kz_0)\Delta(\omega - k_x V)\alpha''(\omega)\coth\frac{\omega \hbar}{2k_B T} \tag{E3}$$

$$J_4 = \int_{-\infty}^{+\infty}d\omega\int_{-\infty}^{+\infty}dk_x k \exp(-2kz_0)(\omega - k_x V)\alpha(\omega - k_x V)\Delta''(\omega)\coth\frac{\omega \hbar}{2k_B T} \tag{E4}$$

In order to simplify the above correlators, we use analytical properties of the incoming functions $\alpha(\omega), \varepsilon(\omega)$: namely, parity of their real parts and unparity of the imaginary ones [152]. It is common matter to show that the function $\Delta(\omega)$ also satisfies these conditions. As an example, consider the calculation of $J_1$. Let us define the auxiliary function

$$f(k,\omega) = k\exp(-2kz_0)\alpha''(\omega)\coth\frac{\omega \hbar}{2k_B T} \quad, k = \sqrt{k^2_x + k^2_y} \text{ . Evidently, } f(k,\omega) \text{ is an}$$

even function of $\omega$ and $k_x$.  At first, we write Eq.(E1) in the form

$$\int_{-\infty}^{+\infty}d\omega\int_{-\infty}^{+\infty}dk_x f(k,\omega)k_x \Delta(\omega - k_x V) =$$

$$= \int_{-\infty}^{+\infty}dk_x\int_0^{+\infty}d\omega f(k,\omega)k_x\left(\Delta'(\omega - k_x V) + \mathrm{i}\Delta''(\omega - k_x V)\right) +$$

$$+ \int_{-\infty}^{+\infty}dk_x\int_{-\infty}^0 d\omega f(k,\omega)k_x\left(\Delta'(\omega - k_x V) + \mathrm{i}\Delta''(\omega - k_x V)\right) \tag{E5}$$

Making use of the transposition $\omega \to -\omega$ in the second integral (E5) and taking account of parity of the functions $f(k, \omega)$, $\Delta'(\omega)$ and unparity of $\Delta''(\omega)$, yields

$$
\begin{aligned}
J_1 &= \int_{-\infty}^{+\infty} dk_x \int_0^{+\infty} d\omega\, f(k, \omega)\{k_x\big(\Delta'(\omega - k_x V) + \mathrm{i}\Delta''(\omega - k_x V)\big) + \\
&+ k_x\big(\Delta'(\omega + k_x V) - \mathrm{i}\Delta''(\omega + k_x V)\big)\} = \\
&= \int_0^{+\infty} d\omega \int_0^{+\infty} dk_x\, f(k, \omega)\{k_x\big(\Delta'(\omega - k_x V) + \mathrm{i}\Delta''(\omega - k_x V)\big) + \\
&+ k_x\big(\Delta'(\omega + k_x V) - \mathrm{i}\Delta''(\omega + k_x V)\big)\} + \\
&+ \int_0^{+\infty} d\omega \int_{-\infty}^0 dk_x\, f(k, \omega)\{k_x\big(\Delta'(\omega - k_x V) + \mathrm{i}\Delta''(\omega - k_x V)\big) \\
&+ k_x\big(\Delta'(\omega + k_x V) - \mathrm{i}\Delta''(\omega + k_x V)\big)\}
\end{aligned}
\tag{E6}
$$

Furthermore, transposing $k_x \to -k_x$ in the second integral (E6) and making use of parity of the function $f(k, \omega)$ over $k_x$ yields

$$
\begin{aligned}
J_1 &= \int_0^{+\infty} d\omega \int_0^{+\infty} dk_x\, f(k, \omega)\{k_x\big(\Delta'(\omega - k_x V) + \mathrm{i}\Delta''(\omega - k_x V)\big) + \\
&+ k_x\big(\Delta'(\omega + k_x V) - \mathrm{i}\Delta''(\omega + k_x V)\big) - \\
&- k_x\big(\Delta'(\omega + k_x V) + \mathrm{i}\Delta''(\omega + k_x V)\big) - \\
&- k_x\big(\Delta'(\omega - k_x V) - \mathrm{i}\Delta''(\omega - k_x V)\big)\}
\end{aligned}
\tag{E7}
$$

Finally, from (E7) we get

$$
\begin{aligned}
J_1 &= -2\mathrm{i} \int_0^{+\infty} d\omega \int_0^{+\infty} dk_x\, k_x k \exp(-2k z_0) \alpha''(\omega) \coth\frac{\omega \hbar}{2 k_{\mathrm{B}} T} \cdot \\
&\cdot \{\Delta''(\omega + k_x V) - \Delta''(\omega - k_x V)\}
\end{aligned}
\tag{E8}
$$

Integrals $J_2, J_3, J_4$ are calculated similarly to (E8).

Also, it is worthwhile to get a relation between the induced dipole moment $\mathbf{d}^{\mathrm{in}}(t)$ and the surface fluctuation field $\mathbf{E}^{\mathrm{sp}}(t)$. The corresponding linear integral relation is

$$
\mathbf{d}^{\mathrm{in}}(t) = \int_0^\infty \alpha(\tau) \mathbf{E}^{\mathrm{sp}}(t - \tau)\, d\tau
\tag{E9}
$$

Inserting (E9) into the Fourier integral for $\mathbf{E}^{\mathrm{sp}}(\mathbf{r}, t)$ which is taken at the particle location point $(Vt, 0, z_0, t)$, yields

$$
\mathbf{d}^{\mathrm{in}}(t) = \iiint \frac{d\omega\, dk_x dk_y}{(2\pi)^3} \alpha(\omega - k_x V) E_{\omega \mathbf{k}}^{\mathrm{sp}}(z_0) \exp\big(\mathrm{i}(k_x V - \omega) t\big)
\tag{E10}
$$

## F. Frequency overlap integrals

### i) dc conductors and Debye dielectrics

Let us consider the integrals (6.5)

$$J_1(\varepsilon_1(\omega), \varepsilon_2(\omega)) = \int_0^\infty \frac{d\omega}{\omega^2} \tilde{\Delta}_1''(\omega) \Delta_2''(\omega) \qquad , \tag{F1}$$

$$J_2(\varepsilon_1(\omega), \varepsilon_2(\omega)) = \int_0^\infty \frac{d\omega}{\omega} \tilde{\Delta}_1''(\omega) \frac{d}{d\omega} \Delta_2''(\omega) \qquad , \tag{F2}$$

$$J_3(\varepsilon_1(\omega), \varepsilon_2(\omega)) = \int_0^\infty \frac{d\omega}{\omega} \Delta_2''(\omega) \frac{d}{d\omega} \tilde{\Delta}_1''(\omega) \tag{F3}$$

which appear in calculations of the tangential force. Evidently, (F3) reduces to (F1) and (F2) after integration by parts:

$$\int_0^\infty \frac{d\omega}{\omega} \Delta_2''(\omega) \frac{d}{d\omega} \tilde{\Delta}_1''(\omega) = \int_0^\infty \frac{d\omega}{\omega^2} \Delta_2''(\omega) \tilde{\Delta}_1''(\omega) - \int_0^\infty \frac{d\omega}{\omega} \tilde{\Delta}_1''(\omega) \frac{d}{d\omega} \Delta_2''(\omega) \tag{F4}$$

In the case of homo- and heterocontacts between the $dc$ – conductors and Debye dielectrics with dielectric functions (C10), (C12), the functions $\tilde{\Delta}_1''(\omega)$ and $\Delta_2''(\omega)$ have the similar form $A_{ij} \dfrac{\omega}{\omega^2 + a_{ij}^2}$ , the subscripts $i = 1$ , $2$ and $j = c$ , d denote tip/surface and type of material (conductor/ dielectric), respectively . Parametrs $A_{ij}$ and $a_{ij}$ are listed in Table 1, where $\sigma_{1,2}$ are the tip/surface static conductivities, $\varepsilon_{1,2}$ - the tip/surface static dielectric permittivitties, and $\tau_{1,2}$ - the corresponding relaxation times.

Table 1

Parameters of dielectric response functions

| | $a_{ij}$ $j = c$ | $a_{ij}$ $j = d$ | $A_{ij}$ $j = c$ | $A_{ij}$ $j = d$ |
|---|---|---|---|---|
| $i = 1$ tip | $4\pi\sigma_1 / 3$ | $(\varepsilon_1 + 2) / 3\tau_1$ | $4\pi\sigma_1 / 3$ | $(\varepsilon_1 - 1) / 3\tau_1$ |
| $i = 2$ surface | $2\pi\sigma_2$ | $(\varepsilon_1 + 1) / 2\tau_2$ | $2\pi\sigma_2$ | $(\varepsilon_2 - 1) / 2\tau_2$ |

In order to calculate (F1) - (F3), we use the table integrals [106]

$$\int_0^\infty \frac{dx}{x^2 + a^2} \frac{1}{x^2 + b^2} = \frac{\pi}{2ab(a + b)} \tag{F5}$$

$$\int_0^\infty \frac{dx}{x^2 + a^2} \frac{d}{dx} \frac{x}{x^2 + b^2} = \frac{\pi}{2a(a + b)^2} \tag{F6}$$

With account of (F5), (F6) and the data from Table 1 we get

$$J_1(\varepsilon_1(\omega), \varepsilon_2(\omega)) = \frac{\pi A_{1j} A_{2j}}{2a_{1j} a_{2j}(a_{1j} + a_{2j})} \tag{F7}$$

$$J_2(\varepsilon_1(\omega), \varepsilon_2(\omega)) = \frac{\pi A_{1j} A_{2j}}{2a_{1j}(a_{1j} + a_{2j})^2} \tag{F8}$$

$$J_3(\varepsilon_1(\omega), \varepsilon_2(\omega)) = \frac{\pi A_{1j} A_{2j}}{2a_{2j}(a_{1j} + a_{2j})^2} \tag{F9}$$

## ii) *Drude conductors and dielectrics with Lorentzian absorption lines*

Using the dielectric functions (C7) and (C14), the functions $\tilde{\Delta}_1''(\omega)$ and $\Delta_2''(\omega)$ in (6.5) reduce to the form

$$\frac{B_{ij}\omega}{\left(\omega_{ij}^2 - \omega^2\right)^2 + \gamma_{ij}^2 \omega^2},$$

where the subscripts i=1,2 denote the tip and surface, j corresponds to Drude (D) and Lorentzian (L) dielectric functions. Parameters $\omega_{ij}$ and $B_{ij}$ are listed in Table 2, where $\gamma_{1D}$, $\gamma_{2D}$ are the corresponding inverse Drude relaxation times; $\gamma_{1L}$, $\gamma_{2L}$ - the line widths of the Lorentzians; $\omega_{P1}$ and $\omega_{P2}$ - are the bulk plasma frequencies; $\varepsilon_{01}$, $\varepsilon_{02}$ - are the static dielectric permeabilities; $\varepsilon_{\infty1}$, $\varepsilon_{\infty2}$ are the proper high frequency limits of the dielectric permeabilities.

Table 2

Parameters of the frequency overlap integrals between Drude conductors and dielectrics with Lorentzian lines

|  | $\omega_{ij}$ $j = D$ | $\omega_{ij}$ $j = L$ | $B_{ij}$ $j = D$ | $B_{ij}$ $j = L$ |
|---|---|---|---|---|
| $i = 1$ tip | $\omega_{P1}/\sqrt{3}$ | $\omega_{01}\sqrt{\dfrac{(\varepsilon_{01}+2)}{(\varepsilon_{\infty1}+2)}}$ | $\omega_{P1}{}^2\gamma_1/3$ | $3\omega_{1L}{}^2\gamma_{1L}\dfrac{(\varepsilon_{01}-\varepsilon_{\infty1})}{(\varepsilon_{01}+2)(\varepsilon_{\infty1}+2)}$ |
| $i = 2$ surface | $\omega_{P2}/2$ | $\omega_{02}\sqrt{\dfrac{(\varepsilon_{02}+1)}{(\varepsilon_{\infty2}+1)}}$ | $\omega_{P2}{}^2\gamma_2/2$ | $2\omega_{2L}{}^2\gamma_{2L}\dfrac{(\varepsilon_{02}-\varepsilon_{\infty2})}{(\varepsilon_{02}+1)(\varepsilon_{\infty2}+1)}$ |

Using the data from Table 2, integrals (F1), (F2) are reduced to

$$J_1(\varepsilon_1(\omega),\varepsilon_2(\omega)) = \frac{B_{1j}B_{2j}}{\omega_{1j}{}^4\omega_{2j}{}^3}\, G_1\left(\frac{\gamma_{1j}}{\omega_{1j}},\frac{\gamma_{2j}}{\omega_{2j}},\frac{\omega_{2j}}{\omega_{1j}}\right) \qquad (F10)$$

$$J_2(\varepsilon_1(\omega),\varepsilon_2(\omega)) = \frac{B_{1j}B_{2j}}{\omega_{1j}{}^4\omega_{2j}{}^3}\, G_2\left(\frac{\gamma_{1j}}{\omega_{1j}},\frac{\gamma_{2j}}{\omega_{2j}},\frac{\omega_{2j}}{\omega_{1j}}\right) \qquad (F11)$$

$$G_1(x,y,z) = \int_0^\infty \frac{dt}{(1-t^2z^2)^2 + x^2z^2t^2}\,\frac{1}{(1-t^2)^2 + y^2t^2} \qquad (F12)$$

$$G_2(x,y,z) = \int_0^\infty \frac{dt}{(1-t^2z^2)^2 + x^2z^2t^2}\,\frac{d}{dt}\frac{t}{(1-t^2)^2 + y^2t^2} \qquad (F13)$$

For two materials with $\gamma_{1j} = \gamma_{2j}$ and $\omega_{1j} = \omega_{2j}$ we get

$$G_2(x,y,z) \equiv 1/2 \int_0^\infty dt /((1-t^2)^2 + x^2 t^2)^2 > 0, \text{ otherwise (F13) has an arbitrary sign.}$$

The function (F12) is positive in any case.

FIGURES:

FIG.1

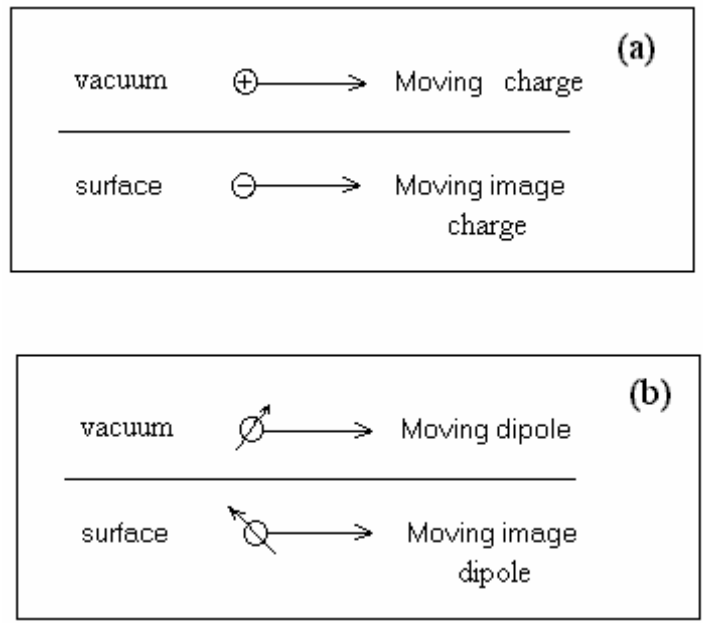

FIG.2

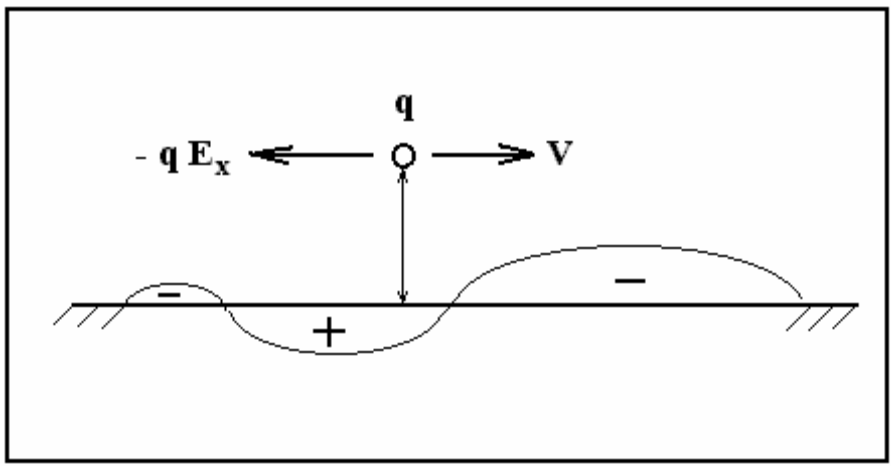

FIG.3

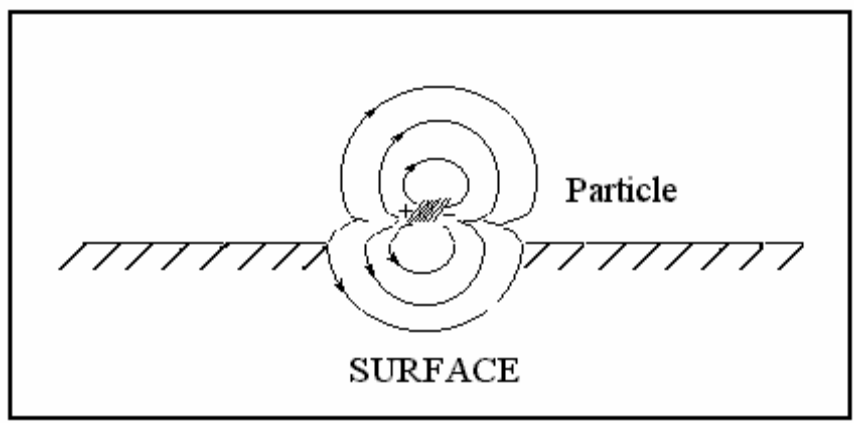

FIG.4

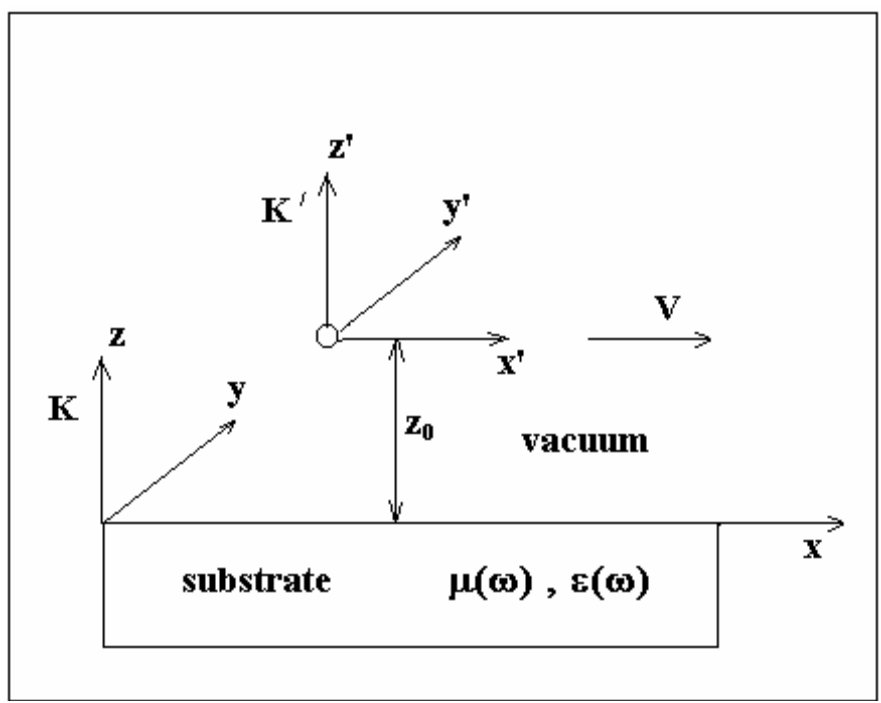

FIG.5

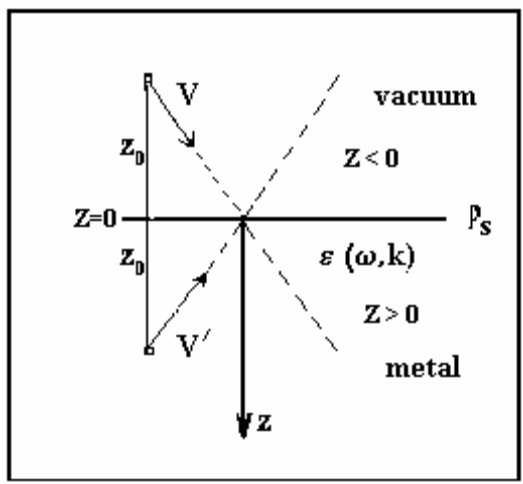

FIG.6

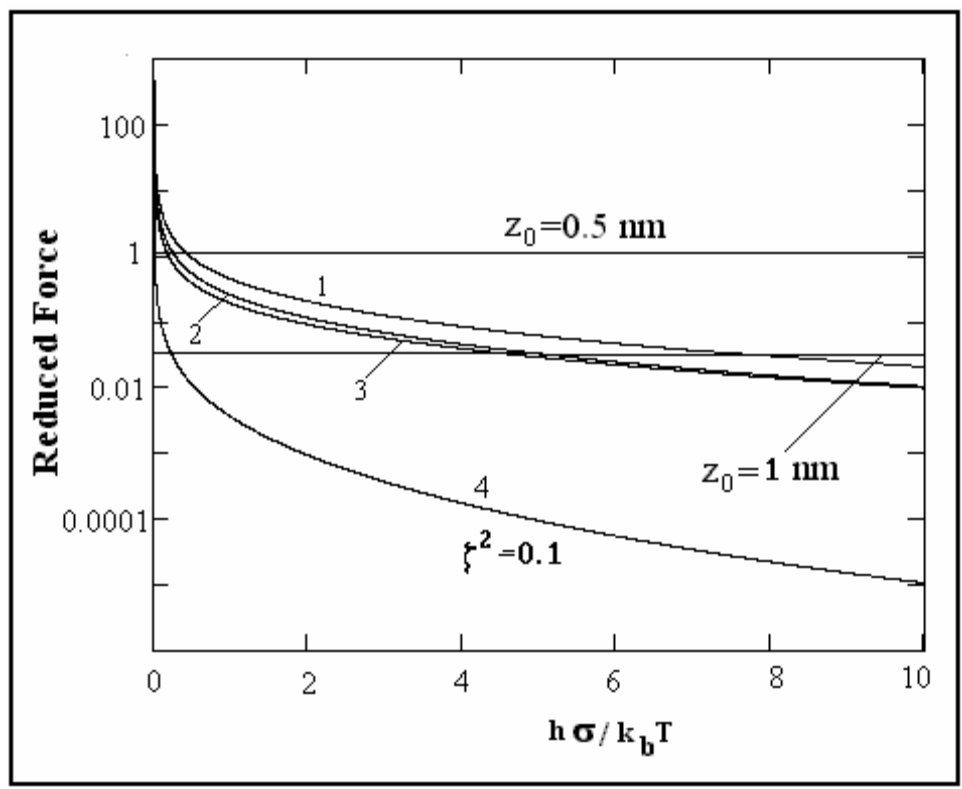

FIG.7a

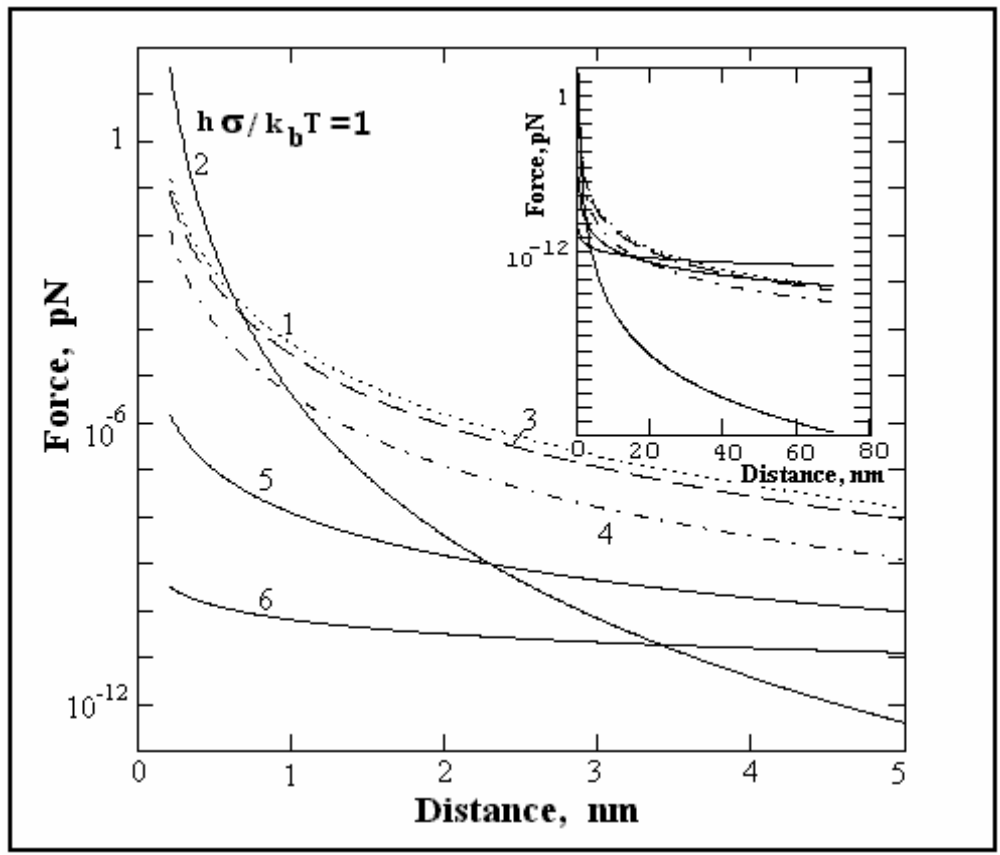

FIG.7b

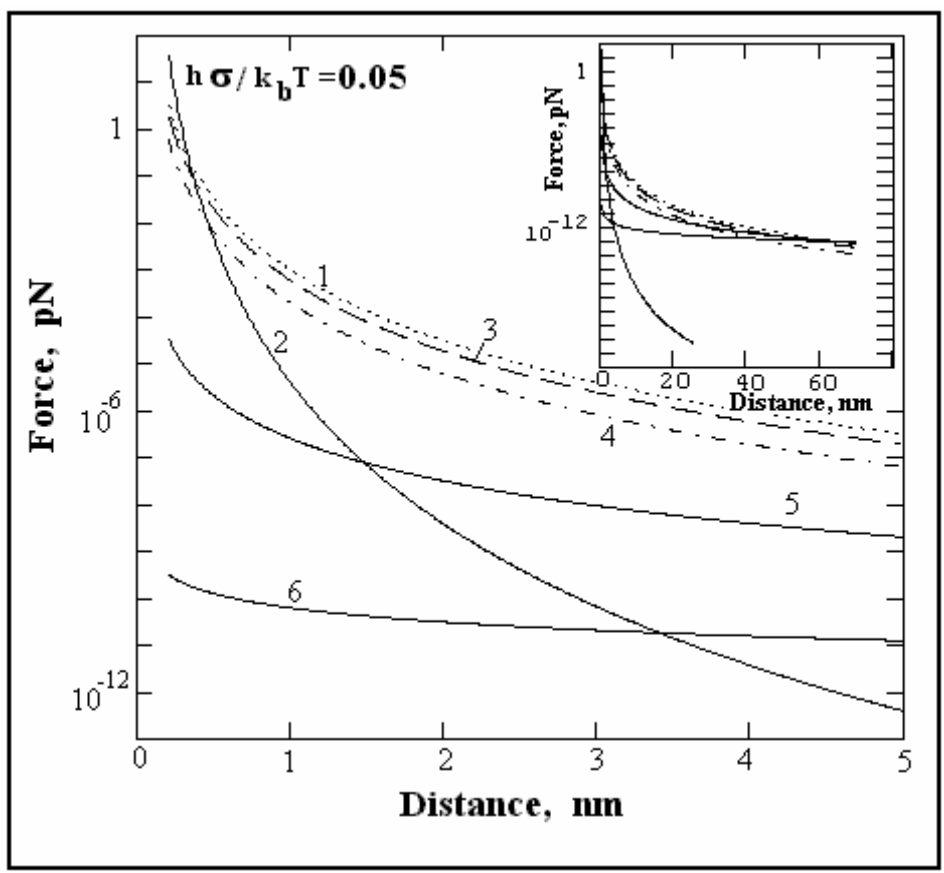

FIG.8

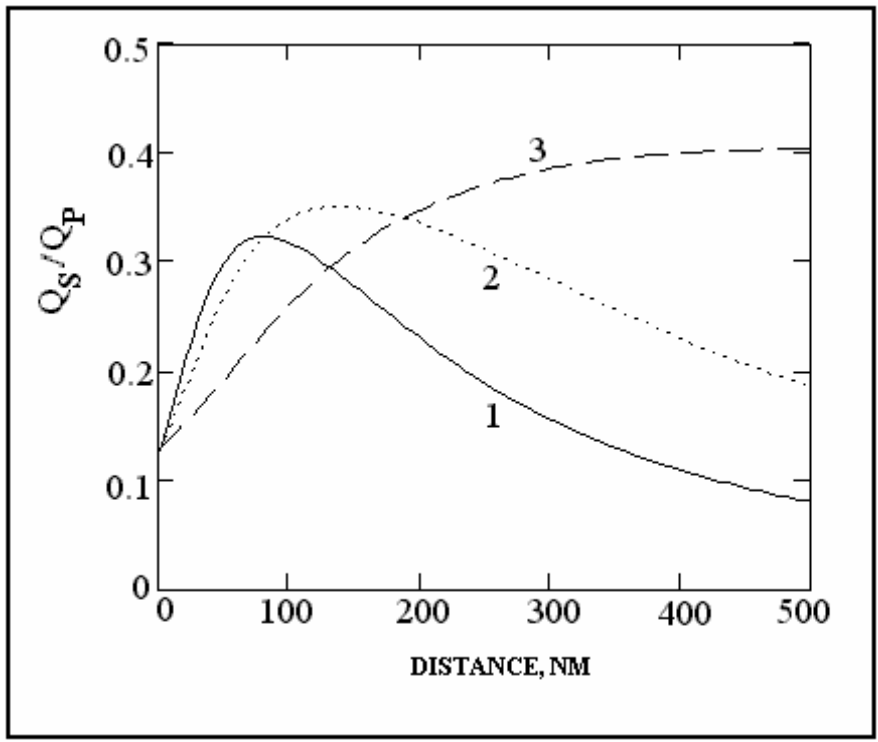

FIG.9

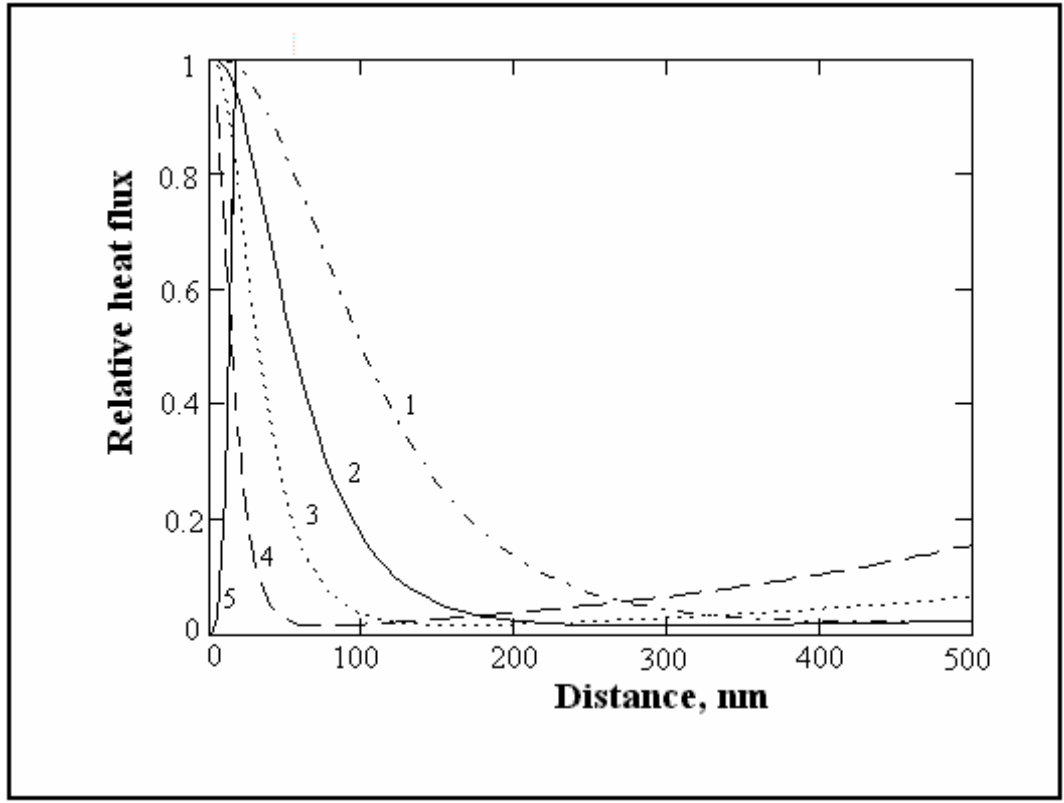

FIG.10

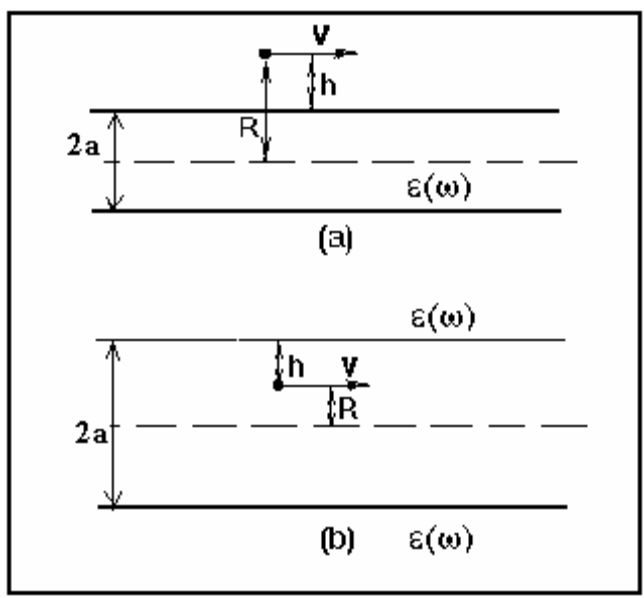

FIG.11

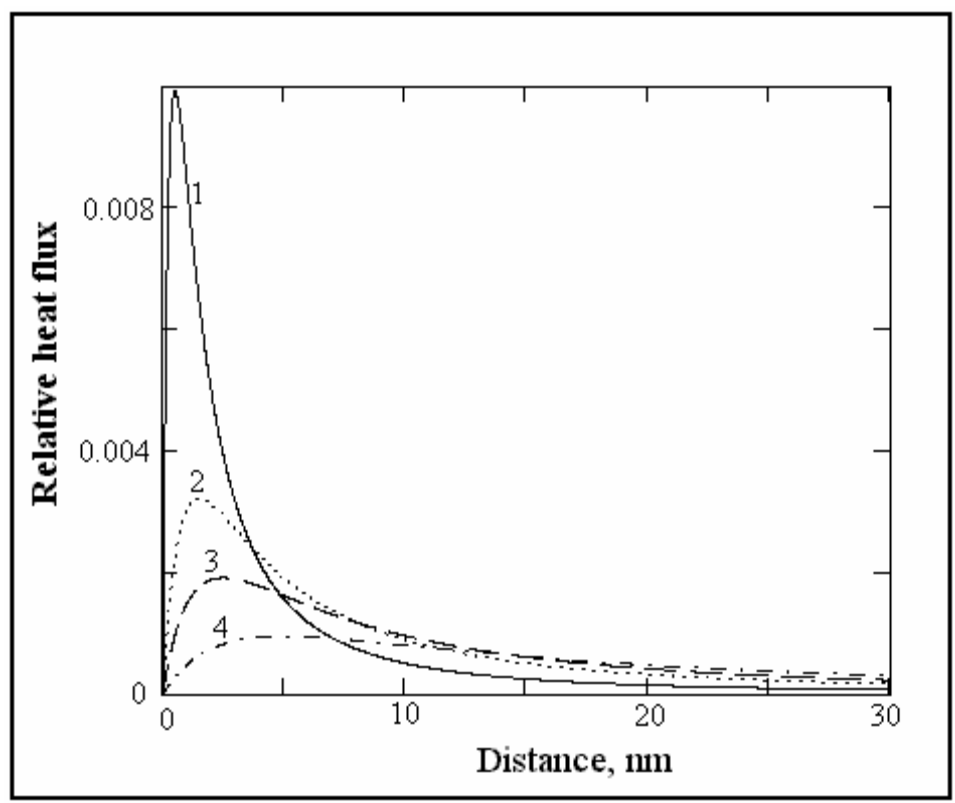

FIGURE CAPTIONS:

Fig.1(a,b)  Moving charge (a),  moving dipole (b) and their "images" in polarizable medium. The moving image charges give rise to Ohm's low heating within the metal.

Fig.2 The induced charge density near the moving charge (according to [26]).

Fig.3 Moving neutral particle and a substrate. The figure illustrates a thermal (quantum) fluctuation, giving rise to a temporal charge imbalance and an electric field. The electric field penetrates into substrate where it creates the electron-hole pairs and other excitations.

Fig.4 The rest frame of the moving particle ($K'$) and the laboratory frame ($K$) related with  resting solid surface.

Fig.5 Schematic representation of the charges used in the specular reflection model: particle of charge $Ze$ (or dipole moment $\mathbf{d}$) being reflected at t = 0 from the surface and its specular reflection; $\rho_s$ is a fictitious surface charge density fixed by the boundary conditions.

Fig.6 Normalized tangential force  ($F_x / (\hbar V R^3 / z_0^5)$) vs.  $a = 2\pi\sigma\hbar / k_B T$ . Lines 1 – 4  correspond to (4.29), (4.39), (4.40) and (4.38), respectively. Horizontal lines were calculated from  (4.30).  Relativistic  correction  (line  4)  corresponds  to $\xi^2 = (2\pi\sigma z_0 / c)^2 = 0.1$ .

Fig.7(a,b) Tangential force vs. distance to surface : (a)  $a = 2\pi\sigma\hbar / k_B T = 1$; (b) $a = 2\pi\sigma\hbar / k_B T = 0.05$.  The  calculations  correspond  to  the  particle  velocity $V = 1\, m / \sec$ .

Fig.8 Dependence  $\dot{Q}_S / \dot{Q}_P$   vs. distance to surface, where  $\dot{Q}_S$  and  $\dot{Q}_P$  denote the contributions from  $S$ -  and $P$ - polarized waves (according to Ref. [91]).

Fig.9. The distance dependence of ratio of the heat flow (4.37) to the nonretarded one - (4.42). Lines  1 – 4  correspond to :  1 -  $\sigma = 10^{16}\, s^{-1}$ ; $T = 300K$  ;  2 - $\sigma = 10^{17}\, s^{-1}$ ; $T = 100K$ ;  3 -  $\sigma = 10^{17}\, s^{-1}$ ; $T = 300K$ ;4 -  $\sigma = 5 \cdot 10^{17}\, s^{-1}$ ; $T = 300K$ . Line 5 is calculated according to Eq.(33) from Ref. [55]. Both the particle and substrate materials are assumed to have same conductance. The particle and substrate temperatures are assumed to be  $T_1 = T$ , $T_2 = 0$  (according to Ref. [91]).

Fig.10(a,b) Geometry of the particle – cylindrical surface interaction: (a)  particle moving parallel to generatrix of convex cylindrical surface; (b)  particle in  cylindrical

channel (near a concave surface). Axis $z$ of coordinate system is directed along the symmetry axis of the cylinders.

Fig.11 The ratio of the heat flow calculated from (6.16) to that one calculated from (6.15) vs. distance $d$ between a spherical metal particle and a flat metal surface ($r_s = 2\,a.u.$). Lines $1-4$ correspond to $R = 1, 3, 5, 10\,nm$, respectively.